\documentclass{aa} 

\newcommand{\changes}[0]{\textsc{chang-es }}
\newcommand{\sdsspos}[0]{$\textrm{RA} = 10^{\textrm{h}}52^{\textrm{m}}41.256^{\textrm{s}} \; \textrm{Dec} = +36^{\circ}40'08.95''$}

\usepackage{array}
\newcolumntype{M}[1]{>{\centering\arraybackslash}m{#1}}
\usepackage{multirow}
\usepackage{subcaption}

\usepackage{graphicx}
\usepackage{txfonts}
\usepackage{color}

\begin{document} 

   \title{Multi-epoch variability of AT~2000ch (SN~2000ch) in NGC~3432}
   \subtitle{A radio continuum and optical study\footnote{Table A.1. is only available in electronic form at the CDS via anonymous ftp to cdsarc.u-strasbg.fr}}

   \author{Ancla Müller
          \inst{1},
          Vanessa Frohn
          \inst{1}\fnmsep\inst{2},
          Lukas Dirks
          \inst{1},
          Michael Stein 
          \inst{1},
          Björn Adebahr
          \inst{1},
          \\
          Dominik J. Bomans
          \inst{1,3,4},
          Kerstin Weis\inst{1},
          \and
          Ralf-Jürgen Dettmar\inst{1,4}
          }

   \institute{Ruhr University Bochum (RUB), Faculty of Physics and Astronomy, Astronomical Institute, Universit\"atsstr. 150, 44801 Bochum, Germany\\
              \email{amueller@astro.rub.de}
         \and
             Argelander-Institut für Astronomie (AIfA), Universität Bonn, Auf dem H\"ugel 71, 53121 Bonn, Germany 
        \and 
            RUB Research Department Plasmas with Complex Interactions
        \and
            Ruhr Astroparticle and Plasma Physics Center (RAPP Center) 
             }

   \date{Received XXX, 2021; accepted XXX}

 
  \abstract
   {AT~2000ch is a highly variable massive star and supernova imposter in NGC~3432 first detected in 2000. It is similar and often compared to SN~2009ip, and it is therefore expected to undergo a core-collapse supernova (SN) ---{a SN imposter of similar brightness}--- in the near future.}
   {We characterize the long-term variability of AT~2000ch in the radio and optical regimes with archival data reaching back to the year 1984. We use these newly reduced observations in addition to observations in the literature to restrict the mass-loss rates of AT~2000ch at multiple epochs based on different approaches, and to infer the general properties of its circumstellar nebula with respect to the detected radio brightness.}
    {We extend the known optical light curve of AT~2000ch up {to the beginning of 2022} by performing point spread function (PSF) photometry on archival data from the Palomar Transient Factory and the Zwicky Transient Facility.
    We reduced archival radio continuum observations obtained with the Very Large Array using standard calibration and imaging methods and complemented these with pre-reduced \changes observations as well as observations obtained with the Westerbork Synthesis Radio Telescope and LOw Frequency ARray.
    For the analysis of AT~2000ch, we consider the optical light curve and color evolution, its radio continuum brightness at different frequencies and times, and the corresponding spectral indices. We estimated mass-loss rates and optical depths based on radio continuum brightnesses and H$\alpha$ fluxes.}
    {We report {two} newly detected outbursts of AT~2000ch similar to those found in the 2000s and 13 re-brightening events, of which at least four are not conclusively detected because of insufficient sampling of the light curve. The dates of all outbursts {and significant, well-sampled re-brightening events} are consistent with a period of {$\sim 201 \pm 12\,$days} over a total time-span of two decades. Such a behavior has never been found for any SN imposter, especially not for candidate SN~2009ip analogs. {During 2010 to 2012 and 2014 to 2018, we only have a few detections, which is insufficient to come to any conclusion as to a possible less eruptive phase of the transient.} We find steady dimming after the most recent re-brightening events and possible evidence of porosity in the circumstellar envelope, suggesting AT~2000ch may currently be in transition to a state of relative calm. We identified a second, unrelated source at a projected distance of $\sim 23\,$pc ($\sim0.5^{\prime\prime}$) that has contaminated the optical measurements of AT~2000ch {at its minimum luminosity} over the last two decades probably  on a $5\%-10\,\%$ level, but this does not affect our overall findings {and is negligible during re-brightening}.
    We are able to restrict the mass-loss rate of AT~2000ch to range between several $10^{-6}\,\textrm{M}_{\odot}/\textrm{yr}$ and several $10^{-5}\,\textrm{M}_{\odot}/\textrm{yr}$. The fresh ejecta appear to be optically thick to radio continuum emission at least within the first $\sim 25\,$days after {significant re-brightening}.}
   {We suggest that other SN imposter and probably also candidate SN~2009ip-analogs at comparable distances emit radio continuum fluxes on the order of a few to several tens of microJanskys at 6\,GHz. Deep and frequent continuum surveys in the radio range are needed to study these kinds of objects in a broader context.}


\titlerunning{Multi-epoch variability of AT~2000ch}
\authorrunning{M\"uller et al.}
    \maketitle


\section{Introduction}
    
    Over approximately the last decade, large-scale automated surveys have monitored the sky for optical transients (e.g., Catalina Real-Time Transient Survey; \citeauthor{drake2009} \citeyear{drake2009}, Master Robotic Net; \citeauthor{lipunov2010} \citeyear{lipunov2010}, Zwicky Transient Facility, ZTF; \citeauthor{bellm2014} \citeyear{bellm2014}, or All-Sky Automated Survey for Supernovae; \citeauthor{shappee2014} \citeyear{shappee2014}). Supernovae (SNe) in particular have taken a place of interest in these searches, because their frequency, characteristics, and physical mechanisms have profound implications for many astrophysical fields from galaxy evolution to cosmology. The large-scale and long-running surveys have also expanded our knowledge of all kinds of rare (and more frequent) transients \citep[e.g.,][]{Perley2020}.
    However, some massive stars can exhibit a sudden photometric brightening of a similar scale to core-collapse SNe before the end of their lifetime, while displaying spectral features associated with SNe of Type IIn or Type Ibn \citep{Schlegel1990, Pastorello2008, Hoss2017, Pastorello2018}, resulting in misclassified SNe (e.g., SN~1954J \citep{tammann1968, smith2001}, SN~1997bs \citep{vandyk2000}, SN~2002kg {\citep{weis2005, Maund2006}}, and SN~2009ip \citep{smith2010, foley2011}). Modern SN surveys yield these so-called SN imposters as inadvertent byproducts, but these observations can be used to explore stellar evolution models and stellar wind feedback with the surrounding interstellar medium.
    
    Supernova imposters are usually likened to the variability of luminous blue variables (LBVs), which are known to spontaneously increase in brightness by $\Delta m \geq 2$\,mag in optical broadband photometry in so-called giant eruptions \citep{LBVreview}. This episode of variability probably occurs on timescales of centuries or longer and is accompanied by significant mass loss. Historic examples include the
outburst of P Cyg around the turn of the 16th century \citep{degroot1988} and the outburst of $\eta$ Carinae in the 19th century, which resulted in the formation of the Homunculus nebula.
    A second mode of variability attributed to LBVs is the S Dor cycle, which consists of a semi-regular increase in brightness of $1-2\,\textrm{mag}$ at optical wavelengths on timescales of decades. The bolometric magnitude remains largely constant during both quiescence and outburst, but the visual maximum is instead caused by an expanded, cooler atmosphere shifting the peak of the spectral energy distribution (SED) from the {ultraviolet} (UV) to the visual regime. Spectra of LBVs evolve over the course of the S Dor cycle, with their appearance ranging from cool supergiants of spectral types A or F at visual maximum to hotter B-type supergiants or Of/WN stars at minimum light. Common features are emission lines of H, He\,I, and Fe\,II, often displaying P Cygni profiles \citep{LBVreview}. Both modes are characterized by luminosities near or above the Eddington limit and considerable mass loss, but so far there is no conclusive evidence that both phenomena can be consistently traced back to a singular physical mechanism or class of star \citep{Davidson2020}. For a more detailed description of LBVs, we refer the reader to the reviews of \citet{LBVreview} and \citet{weis2020}.
    
    Some imposters exhibit a more complex pattern of variability. SN~2009ip, for example, underwent three eruptive episodes between 2009 and 2012 during which the progenitor star experienced erratic brightening and dimming over $\Delta m_R \sim 4\,\textrm{mag}$ in the R-band on timescales of weeks \citep{Pastorello2013}. Its final outburst reached an absolute magnitude of $M_V = -17.7\,\textrm{mag}$, comparable to what could be expected from a true core-collapse SN, although definitive evidence for the nature of this outburst has not been uncovered \citep{fraser2013}.  This behavior is not unique among massive star transients, and indeed there are several stellar transients that show sudden brightness increases before becoming a SN weeks to years later; for example, SN~2005gl \citep{Gal-Yam2007, Gal-Yam2009}, SN~2015bh \citep{Elias-Rosa2016, Thoene2017}, LSQ13zm \citep{Tartaglia2016}, SN~2010bt \citep{Elias-Rosa2018a}, and {AT~2016jbu} {\citep{Kilpatrick2018,Brennan2022,Brennan2022_2}}. The sudden brightness variations may be an important mechanism in the increased mass loss in late evolutionary phases \citep{Groh2013} and the creation of the dense circumstellar environments leading to SN IIn and Ibn.
    
    There are other classes of variable objects currently being debated that also produce strong variations in the optical magnitude. For instance, the mass flow between binary objects and a merger process can explain variable light curves, with major outbursts expected to also be visible in the X-ray regime \citep[e.g.,][]{Zickgraf,Kashi2010,Bilinski2017}. Another example is pulsational pair instability in the case of very massive stars, which produces mass losses above 1\,M$_\odot$ during an individual outburst \citep[e.g.,][]{Woosley2007}. Several transients were linked to potential stellar merger events like SN~Hunt~248 \citep{Mauerhan2018} or AT~2017jfs \citep{Pastorello2019}. These events are now put into the box of intermediate luminosity transients and seem to span a large range in masses of the progenitor stars. Another intriguing new class of transients are the recently detected fast blue transients similar to AT~2018cow \citep{Prentice2018}; these are most probably likened to an accretion-powered jet from a direct collapse black hole \citep{Perley2021}. Following this interpretation, constraining previous activity of the massive progenitor star, especially in the radio, will help us to understand the evolutionary state of the progenitor and its circumstellar environment.
    
    Supernova imposters are expected to be visible in radio continuum, but only a few dedicated studies have been performed so far \citep[e.g.,][]{White,Chandra,Hancock}.
    Generally, an object can produce continuum emission due to radiation of individual particles that can be divided into a thermal (Bremsstrahlung) and a nonthermal (synchrotron) contribution \citep{Wilson2009}.
    For example, HII regions are often identified using Bremsstrahlung with a typical electron temperature of \(T\approx 10^4\)\,K~\citep{Condon2016}, while the magnetic-field characteristics of galaxies are often probed by thermally corrected, synchrotron spectra. SN imposters with a mass-loss rate of about $1 \times 10^{-6}$\,M$_\odot$/yr or higher are expected to produce bright radio continuum emission due to the interaction of strong stellar winds with the circumstellar medium (e.g., \citet{Slysh}) and should therefore be detectable (e.g., \citet{Horesh,Krauss}).
    
    There exist a few blind surveys searching for radio transients \citep{Bannister, Bower, Gal} and also some individual studies \citep{Anderson2019, Jaeger} identifying several radio SNe, but none could be associated with known SN imposters from the optical regime. 
    Nevertheless, the most famous eruptive stellar object, $\eta$ Carinae, is found to be very bright in radio continuum \citep[0.7\,Jy,][]{White}.
    The radio continuum is mainly of thermal origin. The morphology and extent match the {infrared (IR; dust)} and visible (gas) emission of the Homunculus nebula around the star. These latter authors compared it to an ultracompact HII region ---that nevertheless was not formed by the remnant of the formation of the star but rather by stellar material that was ejected during its giant eruption around 1843. Thereafter, \citet{White2005} investigated several observations of 8.6\,GHz emission to study the radio light curve of $\eta$ Car. These authors found significant differences in brightness and morphology, but no radial-velocity changes during the 5.5\,yr cycle attributed to the binary system.
    The also well-known SN~2009ip could not be detected in the radio range using different radio interferometers, but the flux density was found to be less than 66\,$\mu$Jy \citep{Hancock}, even though strong winds are expected based on different tracers. This minor continuum emission can be explained by a free-free optical depth of greater than 1, reducing the measurable free-free emission effectively \citep{Ofek2013a}. Generally, the radio continuum signatures of SN Ib or SN II are also found to agree with the aforementioned reasoning \citep[see e.g.,][and cites therein]{Krauss}. The radio signal alone cannot  therefore provide a clear identification of the stellar object but can be used to obtain reasonable estimates of stellar-wind parameters if the continuum emission is purely thermal. A combined study of the optical and radio light curve can reveal the origin of the variability and the stellar evolution through such evolutionary stages.
    
    The present study is concerned with a complementary analysis of the optical and radio variability of the imposter AT~2000ch in NGC~3432, which was first discovered by \citet{discovery} and described in more detail by \citet{Wagner2004} and \citet{Pastorello2010}. AT~2000ch is found to show optical variations of several magnitudes within a few days that appear about every 200\,days between 2008 and 2009. Such behavior is compared to imposters like SN~2009ip, though AT~2000ch has not (yet) undergone the potentially ``final'' extremely bright outburst. {In 2013, additional data were published showing magnitude variations but no major outburst \citep{Pastorello2013}.} In Section \ref{sec:hist}, we summarize the analyses of AT~2000ch performed so far. In this work, we reduced various data sets of AT~2000ch recorded in the radio range (first detected during the magnetic field studies of galaxy NGC~3432 by \citet{Miskolczi2017}) and complemented the optical light curve. The data used throughout this study and their reduction are presented in Sections \ref{sec:obs} and \ref{sec:red}, respectively. We present our variability results in both frequency regimes in Section \ref{sec:res}, studying  light curve and color of AT~2000ch, as well as its spectral index during different phases and epochs. In Section \ref{sec:disc}, we discuss the evolution of AT~2000ch over two decades and the origin of the variability in the radio range, and provide constraints on the mass-loss rate at different times. We conclude our findings in Section \ref{sec:conc}.
    

\begin{figure*}[ht]
        \centering
        \includegraphics[width=1.0\textwidth]{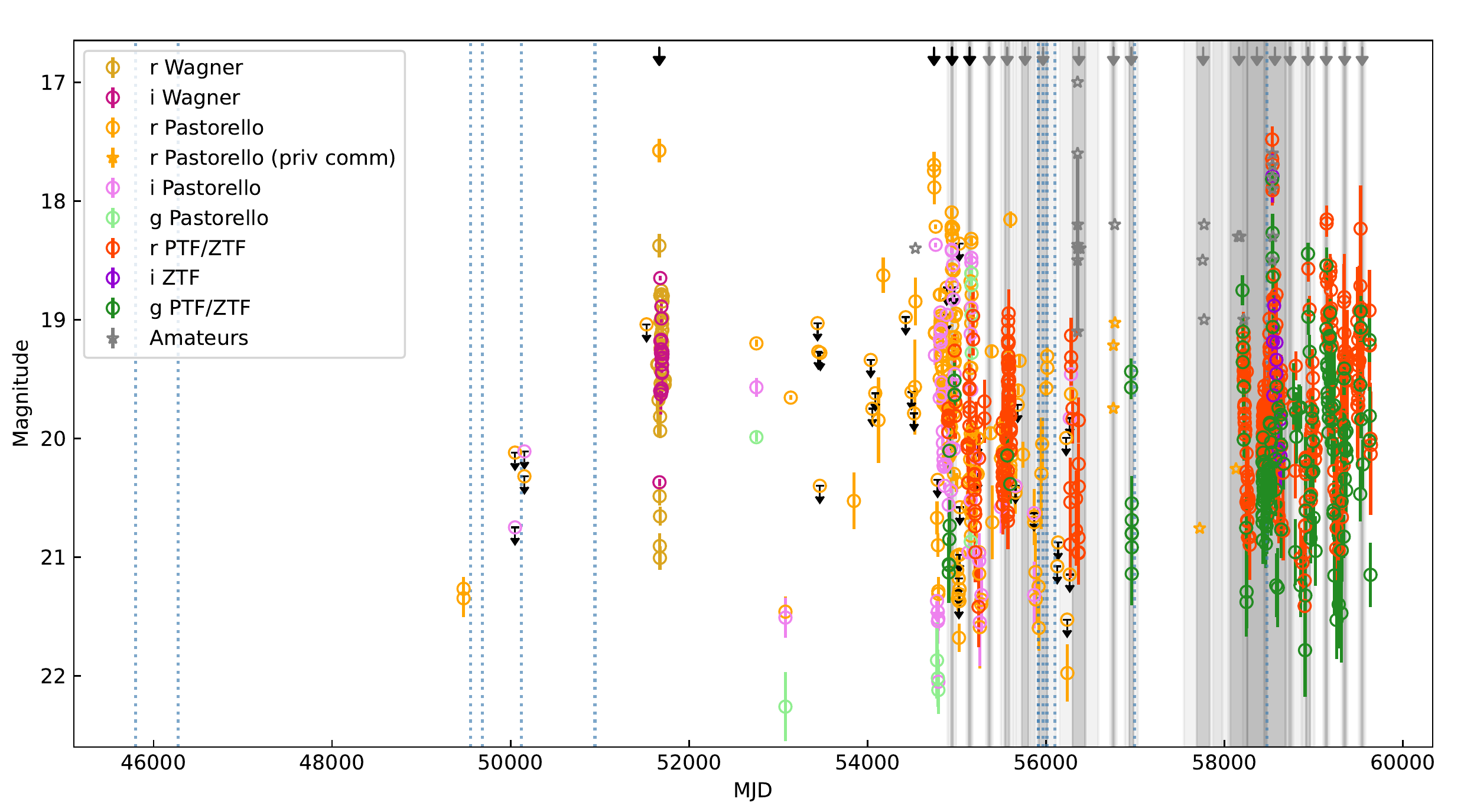}
        \caption{Optical light curve of AT~2000ch in Sloan r (orange), g (green), and i (purple) filters. Observations of amateur astronomers are included, shown as gray stars, {as are unpublished data points from A. Pastorello (priv. comm.), shown as orange stars}. Previously published photometry by \cite{Wagner2004}, \cite{Pastorello2010}, {and \citep{Pastorello2013}} {(converted to the Sloan photometric system; see Sec. \ref{sec:red})} is shown, as well as our photometric results from the PTF/ZTF survey observations. We only show the previously published data points in the same filter bands in which we added new observations. The vertical blue dotted lines indicate the dates of the observations that we collected in the radio range. Arrows at the top of the figure indicate the times of (black) previously known and newly detected or suspected (gray) eruptive episodes of re-brightening events based on Equation \ref{eq:mean_period} with their $1\sigma$ ($3\sigma$) uncertainty indicated by the gray (light gray) shaded areas.}
        \label{fig:lightcurve_full}
\end{figure*}

\section{History of AT~2000ch} \label{sec:hist}
    
    The transient denoted AT~2000ch resides in a luminous clump in the outskirts of its host galaxy NGC~3432, to which it is connected by an extended arm \citep{english1997}. NGC~3432 is a late-type edge-on dwarf barred spiral galaxy \citep[SBm, ][]{Bertola1966,Melisse1994} located at a distance of (9.42~\(\pm \)~0.66)\,Mpc~\citep{Irwin2012}, and is gravitationally connected to the dwarf galaxy UGC 5983. Together, they make up the galaxy pair Arp 206~\citep{Noreau1990}. For a more detailed description, we refer the reader to \citet{Pastorello2010} and references therein.
    
    The first known outburst of the SN imposter AT~2000ch, or 2000-OT following the naming convention of \citet{Pastorello2010}, was reported by \citet{discovery} as part of the Lick Observatory SN Search (LOSS). \citet{wagner2000} discuss the possibility of a superbright classical nova, while \citet{filippenko2000} on the other hand put AT~2000ch in connection with subluminous SN IIn and related those to a LBV nature.
    \citet{Wagner2004} found the R-band light curve of AT~2000ch to be erratic on short timescales (for reference see Fig. \ref{fig:lightcurve_full} and the first panel of Fig. \ref{fig:cutouts}); after its visual maximum at $m_R = 17.4\, \textrm{mag,}$ the transient faded to $m_R = 20.8\, \textrm{mag}$ within a week, and then re-brightened to $m_R = 18.6\, \textrm{mag}$ within four days, and from there it kept varying between $m_R = 18.6$ and $19.4\, \textrm{mag}$. The authors attributed the deep minimum to the evolution of a spherical, optically thick, ejected envelope, or a brief dust formation phase. \citet{Wagner2004} also examined optical spectra of the transient and detected prominent Balmer emission lines with full width at half maximum (FWHM) of $v_\textrm{FWHM}=1550\,\textrm{km\,s}^{-1}$ directly after the optical maximum and $1350\,\textrm{km\,s}^{-1}$ nine months later. The authors further detected weak He\,I and He\,II emission in some of the spectra, but noted a consistent absence of Fe\,II and [Fe\,II] emission lines usually present in spectra of LBVs, as well as any nebular emission lines. No spectral line showed a P Cygni profile.

    \citet{Pastorello2010} detected and studied three further outbursts  of the same transient in the years 2008 and 2009, namely
2008-OT, 2009-OT1, and 2009-OT2 (see Fig. \ref{fig:lightcurve_full} and panels two to four in Fig. \ref{fig:cutouts}). 
    The corresponding light curves continue to be rather complex and highly variable, each containing a deep minimum similar to the 2000-OT outburst, though at different times relative to the respective optical maximum. The second (2009-OT1) and third (2009-OT2) eruptions in this series both took place $\sim 200$\,days after their predecessors. Spectra obtained after the 2008-OT outburst differ in multiple ways from those following 2000-OT: the H$\alpha$ width had increased to $v_\textrm{FWHM} \approx 2300\,\textrm{km\,s}^{-1}$ suggesting a stronger outburst or less ejected mass; He\,I lines had grown stronger; and the integrated H$\alpha$ flux had doubled. Spectra taken two months after 2008-OT now displayed prominent P Cygni profiles and clearly contained Fe\,II and Ca\,II emission lines. Another month later, the P Cygni profiles in  He\,I and the Fe lines had vanished and He\,II $\lambda$4686 appeared as a prominent emission feature, indicating the presence of a higher temperature component. \citet{Pastorello2010} noted that the line appeared double peaked (inlet in their Fig. 9); however, this may at least be partly caused by contributions of Fe\,II, [Fe\,II], [Fe\,III], and He\,I in this wavelength range, which are all seen in various spectra of $\eta$ Carinae.
    Spectra of the 2009-OT1 event were similar to those taken one month after 2008-OT. However,
    \citet{Pastorello2010} do detect distinct blueshifted absorption features (identified as a, b, and edge) in H$\beta$ for 2009-OT1, which correspond to terminal wind velocities of up to v$_{\infty} = 7200\,$km s$^{-1}$ two days after maximum light, and slightly more moderate values of v$_{\infty} = 4800-5600\,$km s$^{-1}$ 20\,days later. Neither LBVs nor other stellar objects are known to have stellar wind velocity values as high as these, and furthermore, changes in terminal wind velocity of roughly 2000\,km/s within 20\,days are unknown. As already suggested by \citet{Humphreys2016}, these velocities could be attributed to a LBV giant eruption as shock velocities in the $\eta$ Carinae outer ejecta reach velocities almost as high \citep{Weis2012}.
    Over the subsequent month, the spectrum evolved so that the terminal velocity decreased to  $2400-2800\,\textrm{km\,s}^{-1}$, the P Cygni profiles grew stronger, and the continuum temperature dropped. 
    Pastorello et al. (2010) conclude their spectral analysis with two major observations: (1) During outburst, the transient resembles regular erupting LBVs and during quiescence it acquires spectral features commonly seen in young Wolf-Rayet stars (WRs). Stars with these features have been classified as Ofpe/WN9 by \citet{walborn1982} and \citet{bohannan1989}. LBVs often show this spectral type during their hot phase. (2) Spectra taken during and following the 2008-OT event closely resembled those taken of the imposter SN~2009ip. {\citet{Pastorello2013} investigated the similarity to SN~2009ip further, and published additional photometry of AT~2000ch spanning the years 2010 to 2012.}
    
    \citet{Pastorello2010} further analyzed the evolution of the SED for all four outbursts, which follows a consistent pattern: During the outbursts, the SED tends to be very blue ($U - B \approx -0.6, V - I \approx 0.5$), and becomes redder ($V - I \approx 1.0$) in coincidence with the post-outburst minimum of the light curve. During this phase, there also tends to be a large $R$-band excess resulting from H$\alpha$ emission. The maximum of the spectrum returns to its blue phase within $\sim 75-100$\,days. The authors attempted single black-body fits to the SED and obtained effective temperatures of $T = 8500-9500\,\textrm{K}$ during outburst ---though the $U$-band contains excess flux--- and temperatures down to $T = 5300\,\textrm{K}$ in quiescence. LBVs typically display the inverse behavior with lower {temperature ($T$)} and redder SED during outbursts compared to their quiescent phase.
    
    \citet{Kochanek2012} attempted to fit the optical and near-IR SED for the 2000-OT outburst with expanding shell models and reported poor fits. They concluded that the post-outburst minima in the light curve cannot be produced by obscuration from newly formed dust, because the change in brightness takes place too quickly for the evolution of the ejected shell. Instead, the authors proposed a scenario where the star produces a quasi-steady wind, so that the dust formation radius scales with the stellar luminosity $R_f \propto L_{*}^{1/2}$, resulting in weaker extinction and a bluer color during outbursts compared to quiescence. The authors do not address the varying presence of P Cygni profiles as presented by \citet{Pastorello2010}.
    
    Following these analyses, AT~2000ch was used for comparison with other irregular transients in the literature, often in the context of candidate stars for SN~2009ip analogs \citep{Smith2011, ofek2013b, Pastorello2013, Thoene2017, Elias-Rosa2018b, Pastorello2018, Reguitti2019, Pastorello2019, Fransson2022, Reguitti2022, Pessi2022}. Proposed candidates for this tentative class of transients include SN~2005gl, SN~2009ip, SN~2010mc, SN~2011ft, SN~Hunt~151, {SN~2013gc}, SN~2015bh, SN~2016bdu, SN~2018cnf, {SN~2019zrk,} and SN~2021foa.
    \citet{Pastorello2010} additionally compared AT~2000ch to the super-giant B[e] star S18 in the Small Magellanic Cloud, because both objects show a similar, double peaked profile for the He\,II\,$\lambda4686$ emission line. For S18, this was interpreted as an indication of accretion onto a hot companion star \citep{shore1987, Zickgraf}. \citet{Clark2013} expand on this comparison by showing that the light curve of S18 is similar in structure to that of AT~2000ch, but varies only over $\Delta m \sim 1\,\textrm{mag}$.

    \citet{Pastorello2010} compiled three possible sky positions for AT~2000ch within an angular distance of $\sim 1^{\prime\prime}$: (1) {Right Ascension\,$(\textrm{RA}) = 10^{\textrm{h}}52^{\textrm{m}}41.4^{\textrm{s}}, $\,Declination}\,$(\textrm{Dec}) = +36^{\circ}40^{\prime}08.5^{\prime\prime}$ from the original detection in LOSS \citep{discovery},  (2) $\textrm{RA} = 10^{\textrm{h}}52^{\textrm{m}}41.33^{\textrm{s}} \; \textrm{Dec} = +36^{\circ}40^{\prime}08.9^{\prime\prime}$ from the discovery of 2008-OT \citep{discovery3}, and (3) $\textrm{RA} = 10^{\textrm{h}}52^{\textrm{m}}41.256^{\textrm{s}} \; \textrm{Dec} = +36^{\circ}40^{\prime}08.95^{\prime\prime}$ from Sloan Digital Sky Survey (SDSS) images. In this work, we adopt the SDSS coordinates (3) as the most precise option (see Sec. \ref{sec:source} for reasoning).

\section{Observations} \label{sec:obs}

\subsection{Optical photometry}

We investigate a continuation of the optical light curve of AT~2000ch presented in \citet{Pastorello2010}. The resulting optical light curve of AT~2000ch shown in Figure \ref{fig:lightcurve_full} is made up of published observations from \citet{Wagner2004} and \cite{Pastorello2010} as well as re-measured archival data points from the Palomar Transient Factory (PTF) and Zwicky Transient Facility (ZTF). The archival data in the different filters of both surveys were already reduced and we only re-investigated the photometry to ensure the best possible detection rate for AT~2000ch (see below). We complemented these observations with reported detections of amateur astronomers. These are denoted by gray stars because (1) the photometric analysis and calibrations could be affected by larger uncertainties and (2) they used either a variety of $R$-filter systems that does not match our $r$ calibration or the observations are unfiltered or performed with a clear filter. Details for each data point can be taken from Table~\ref{tab:optical}; for 2000-OT and 2008-OT to 2013-OT, see also the corresponding tables in \citet{Wagner2004}, \citet{Pastorello2010}, and \citet{Pastorello2013}, respectively.

\subsection{Radio continuum}

Our analysis of this variable source started after its serendipitous detection within the data of the {Continuum HAlos in Nearby Galaxies -- an Expanded Very Large Array Survey (\textsc{chang-es})}. The \changes survey consists of 35 edge-on galaxies that have been observed with the Karl G. Jansky Very Large Array (JVLA) in two bands, 1.6\,GHz also denoted as L-band and 6\,GHz denoted as C-band, and three array configurations \citep[B, C, D, ][]{Irwin2012}. The \changes collaboration has released these data within three data releases. The first data release contains data from the D-Array \citep{wiegert2015changesDR1}. The second data release \citep{vargas2019changesDR2} adds H$\alpha$ imaging and  maps of the star-formation rate and the third data release contains the B-Array data \citep{irwin2019changesDR3}\footnote{The data is publicly available: \url{https://www.queensu.ca/changes}}. A fourth data release containing the C-Array data is currently in preparation (Walterbos et al., in prep.). 

We additionally searched for observations of the galaxy in several archives and chose five data sets of the historical Very Large Array (VLA) as well as one of the JVLA matching the following criteria:
either (1) the theoretical sensitivity was higher than $\approx 20\,\mu$Jy/beam, complementary to the low radio emission observed for SN~2009ip \citep{Hancock}, or (2) the source was not located farther then 
than 3$^\prime$  from the pointing center in order to reduce the effect of increasing noise at the primary beam edges, and (3) we ensured that any observation that passes (1) or (2) has an optimized (u,v)-coverage to reduce artifacts in the image plane, which can be misclassified as point sources (avoid snapshots).
We furthermore included one observation from the Westerbork Synthesis Radio Telescope (WSRT) and the LOw Frequency ARray (LOFAR) to achieve independent measurements from the VLA.
Based on these criteria, we selected the above-mentioned projects, which are summarized in Table~\ref{tab:radio_measurements} (first column).

\begin{figure*}
    \begin{minipage}{.33\textwidth}
        \centering
        \includegraphics[width=\textwidth]{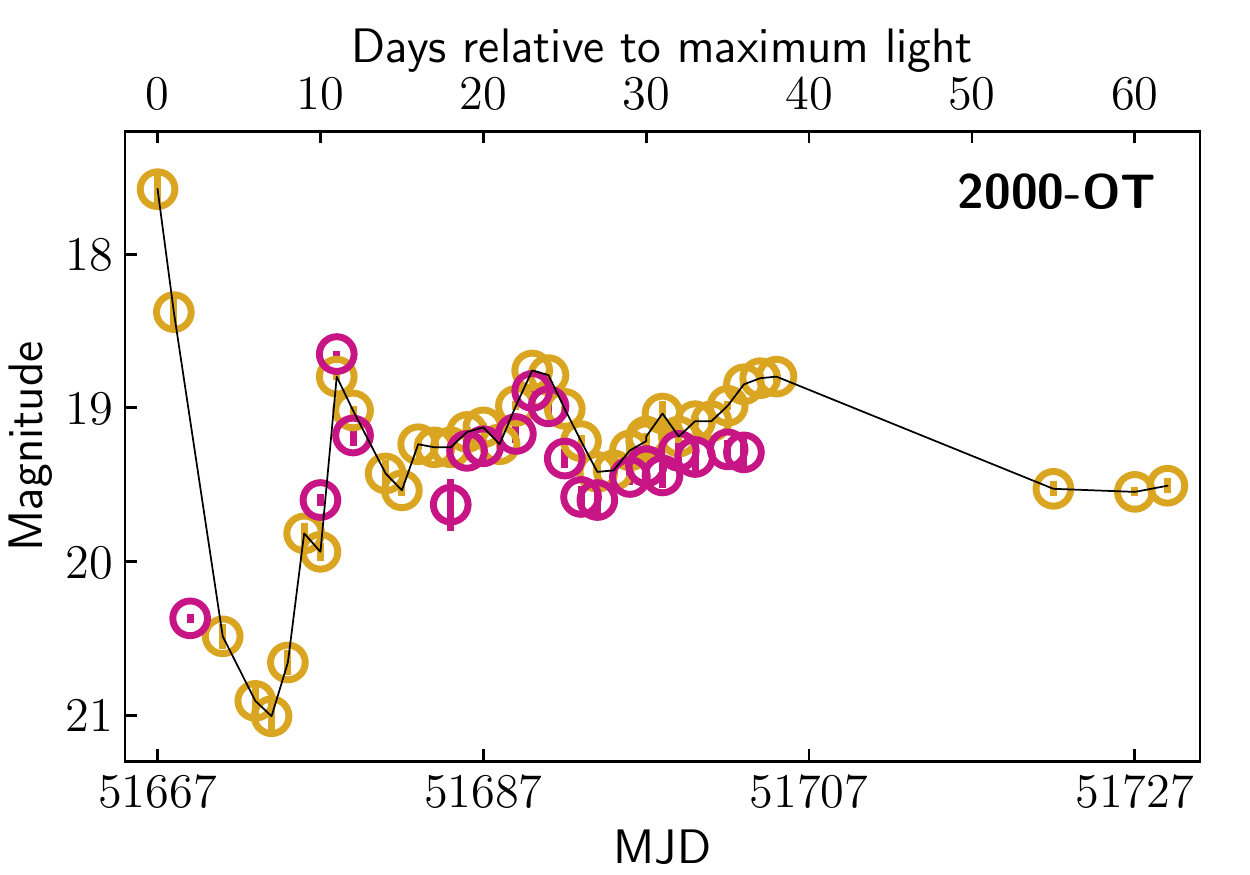}
    \end{minipage}  
    \begin{minipage}{.33\textwidth}
        \centering
        \includegraphics[width=\textwidth]{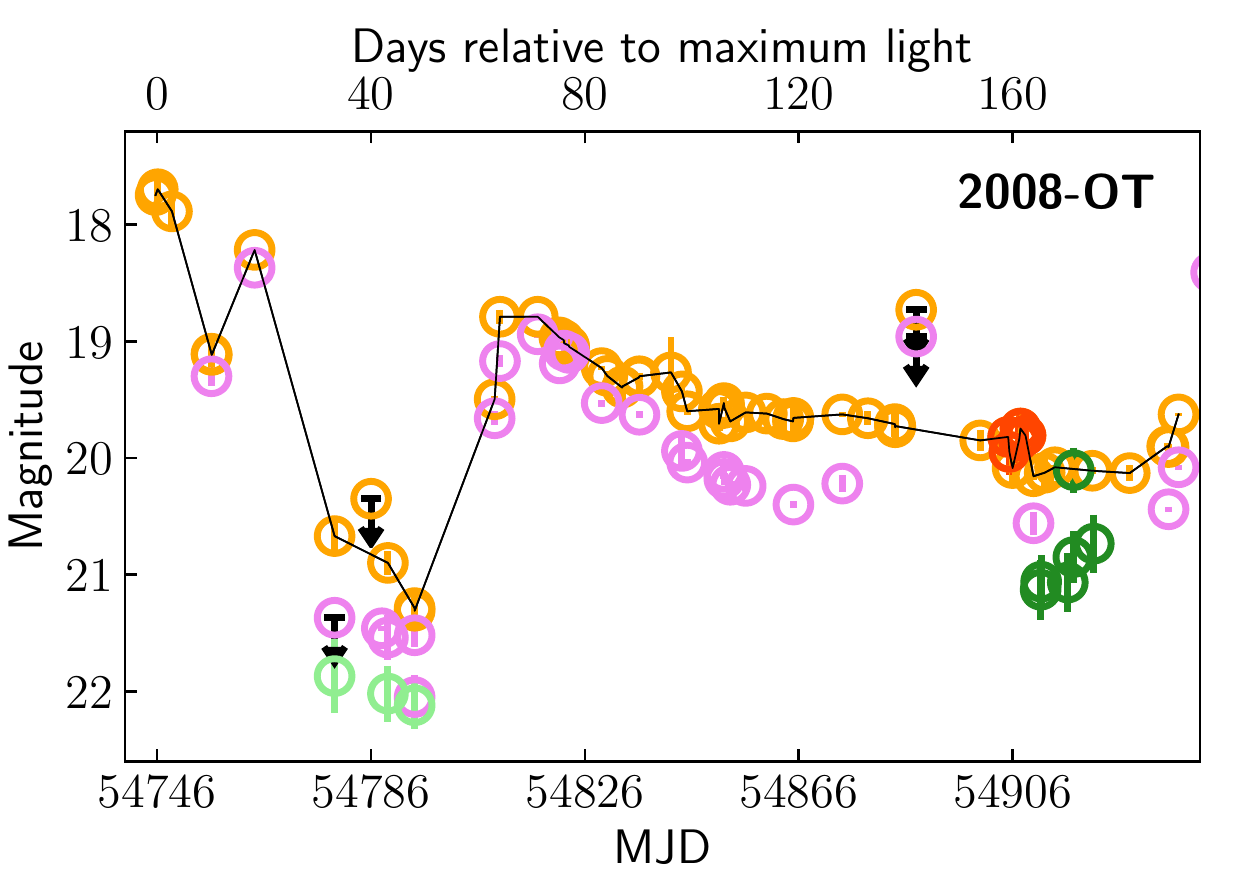}
    \end{minipage} 
     \begin{minipage}{.33\textwidth}
        \centering
        \includegraphics[width=\textwidth]{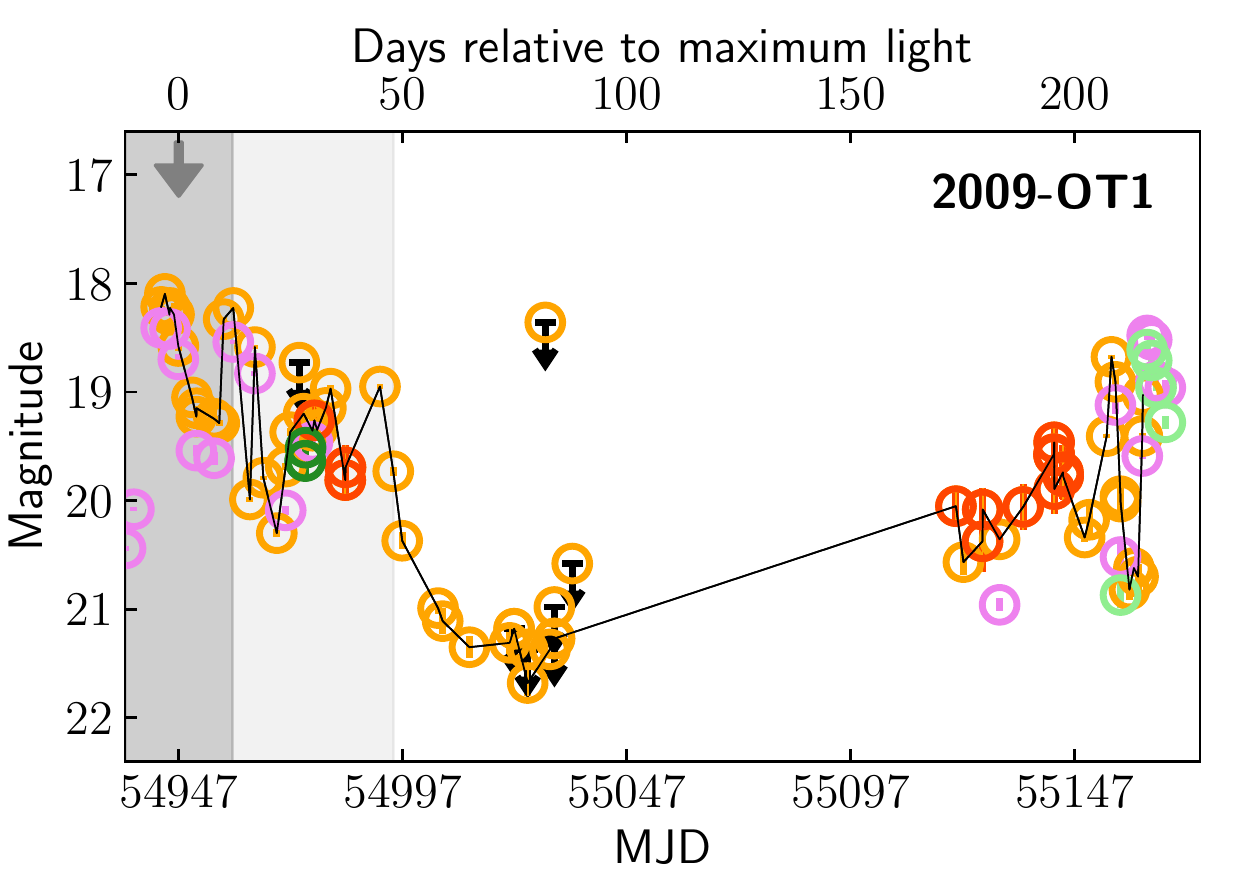}
    \end{minipage}  
       \begin{minipage}{.33\textwidth}
        \centering
        \includegraphics[width=\textwidth]{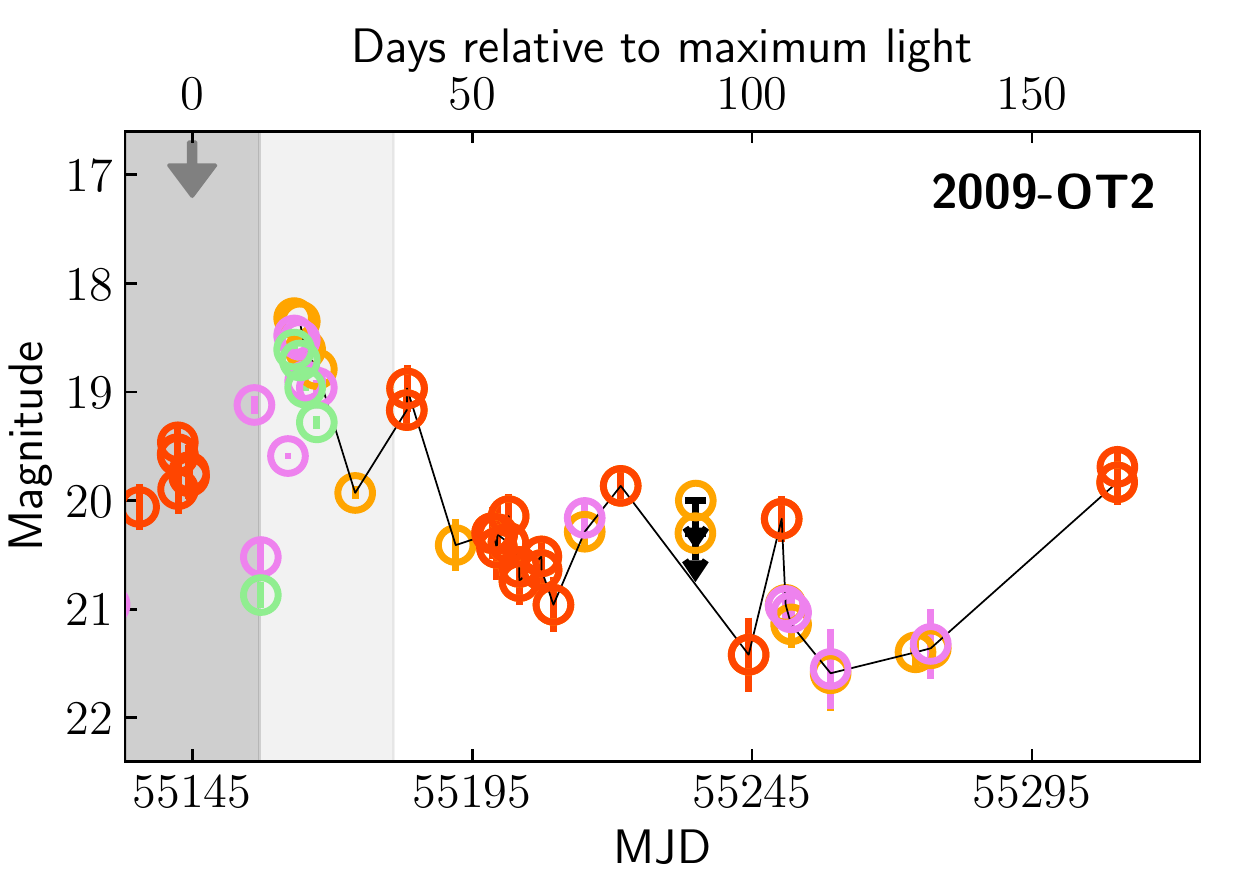}
    \end{minipage}  
    \begin{minipage}{.66\textwidth}
        \centering
        \includegraphics[width=\textwidth]{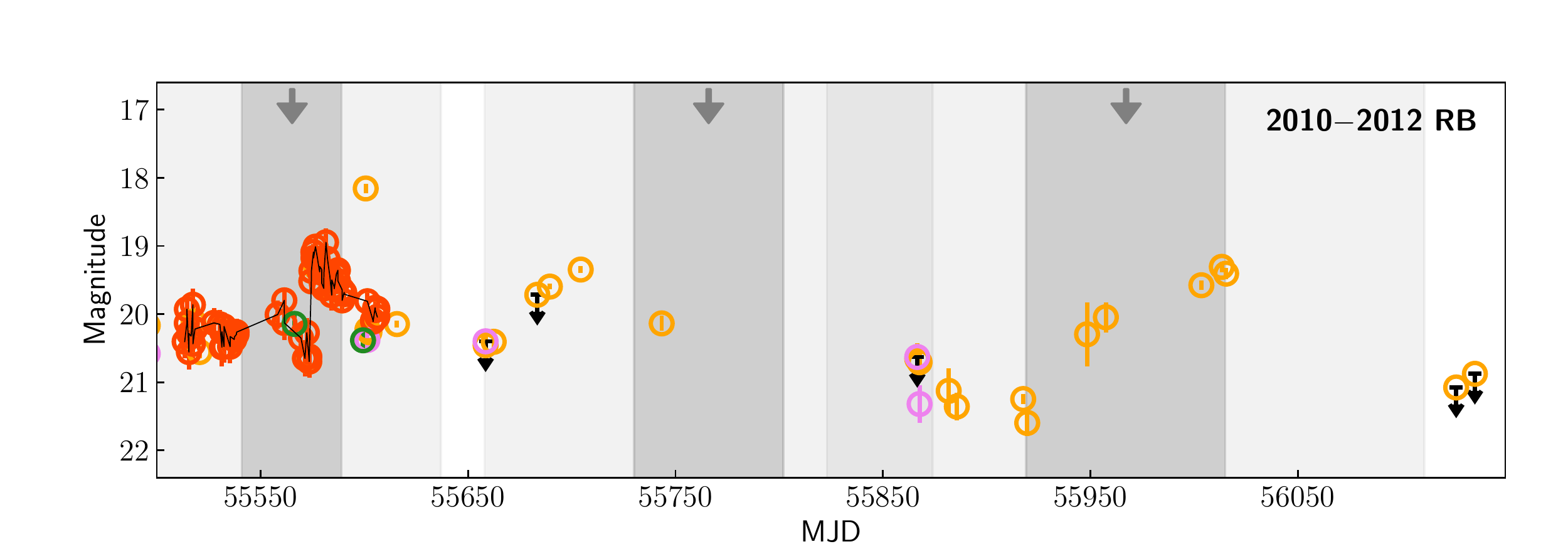}
    \end{minipage} 
     \begin{minipage}{.33\textwidth}
        \centering
        \includegraphics[width=\textwidth]{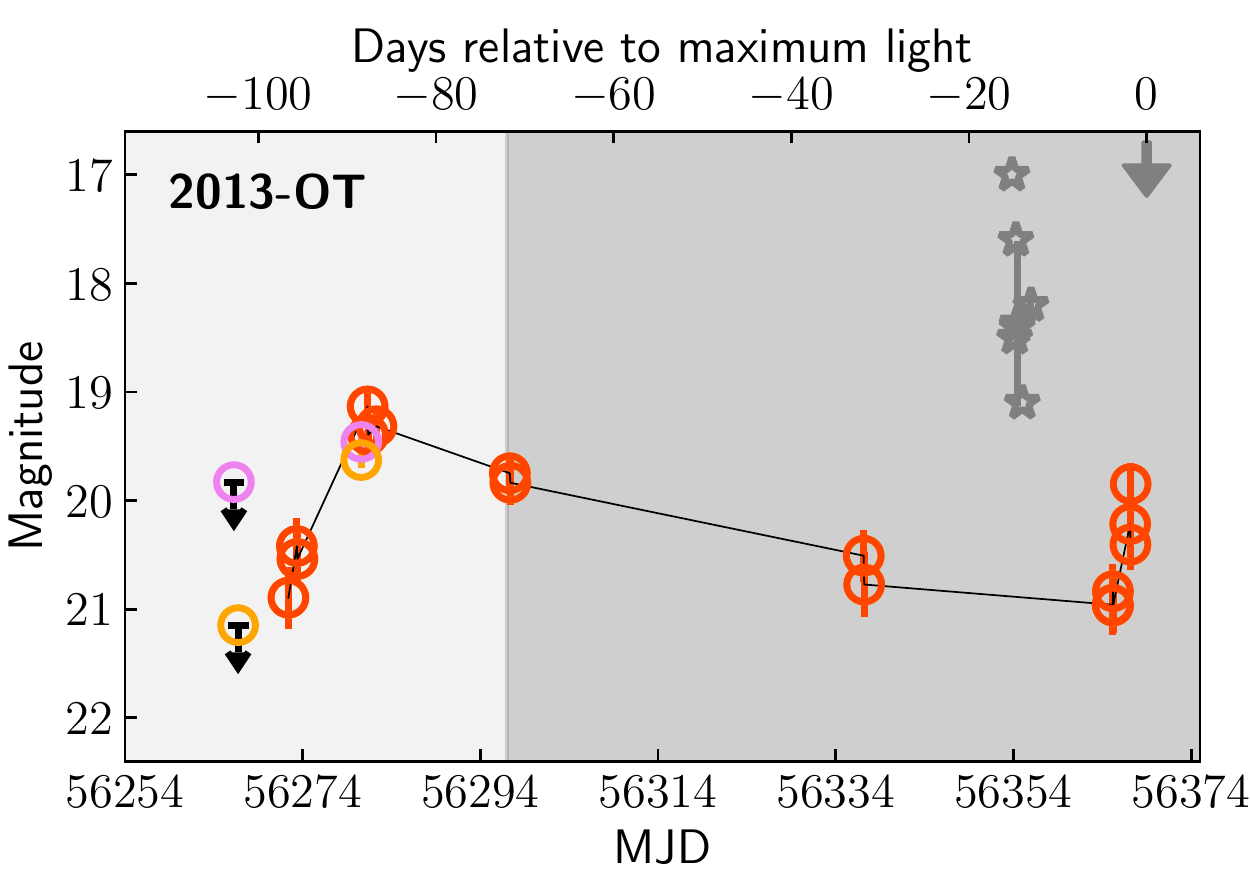}
    \end{minipage}
        \begin{minipage}{0.66\textwidth}
        \centering
        \includegraphics[width=\textwidth]{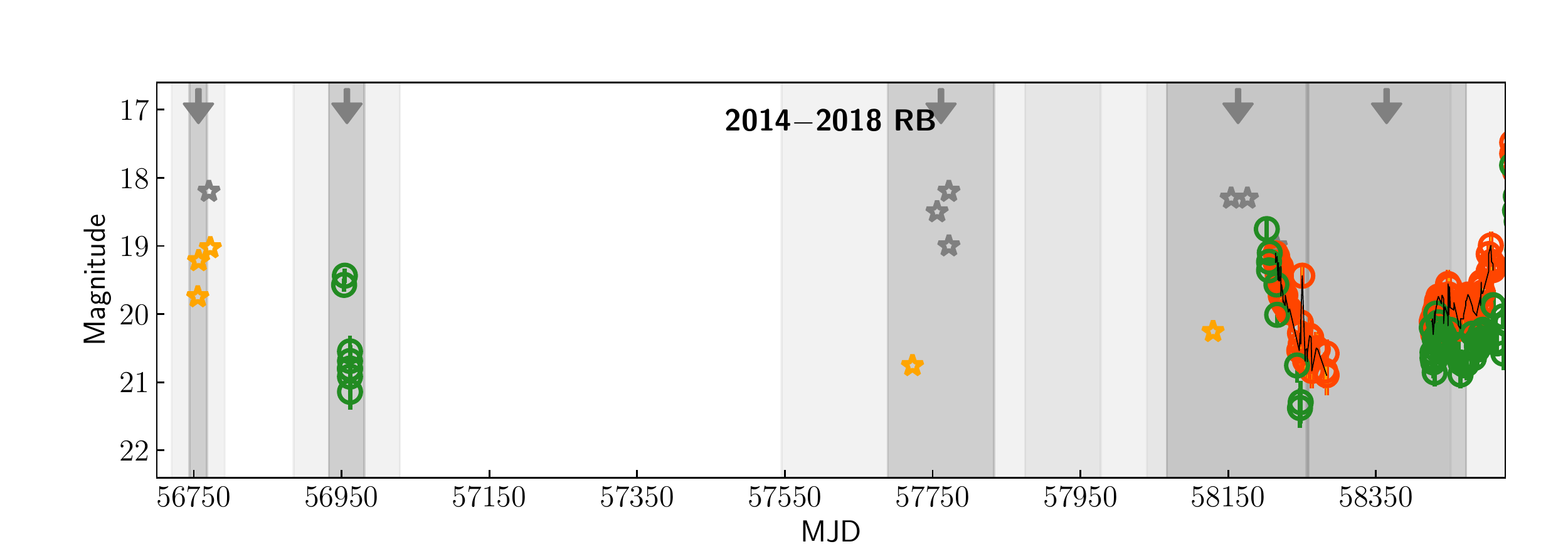}
    \end{minipage} 
        \begin{minipage}{.33\textwidth}
        \centering
        \includegraphics[width=\textwidth]{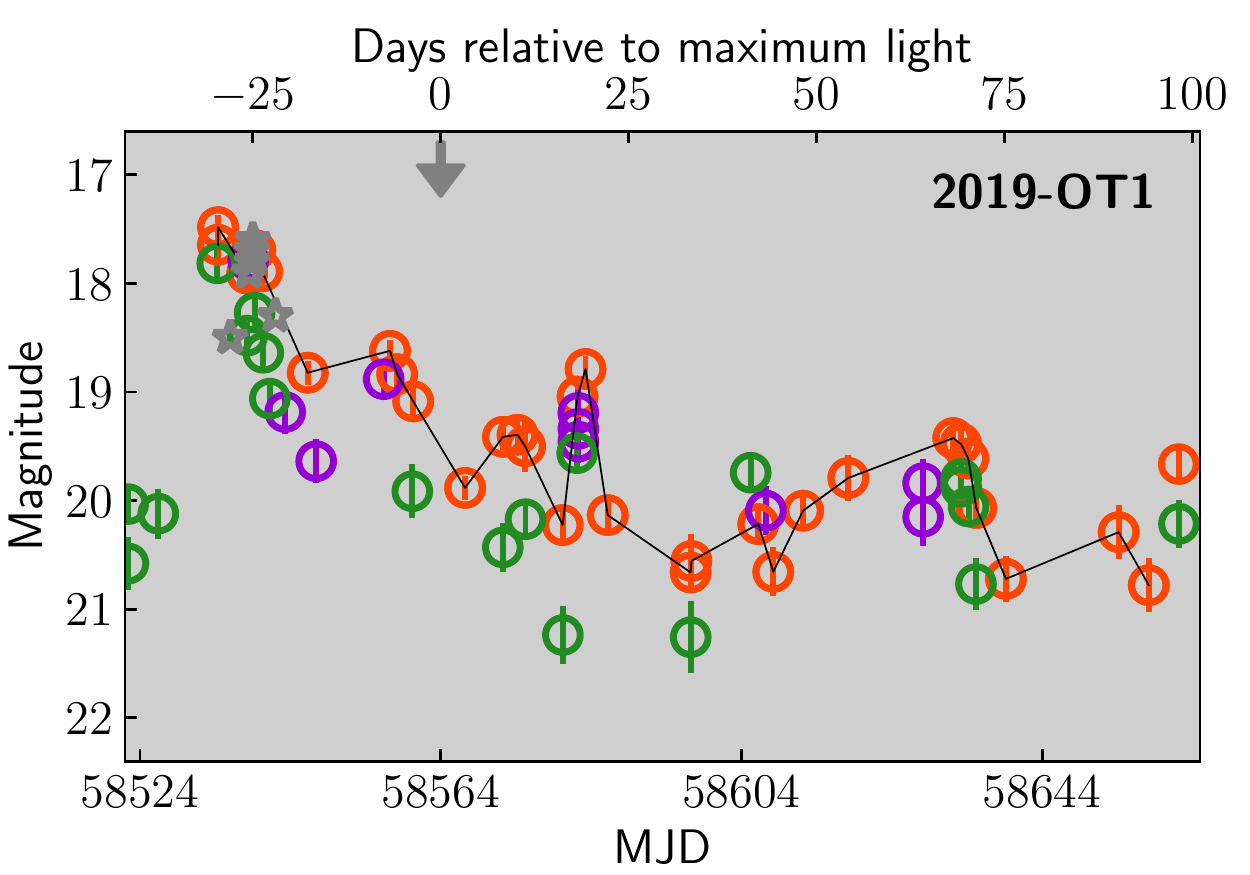}
    \end{minipage}  
    \begin{minipage}{.33\textwidth}
        \centering
        \includegraphics[width=\textwidth]{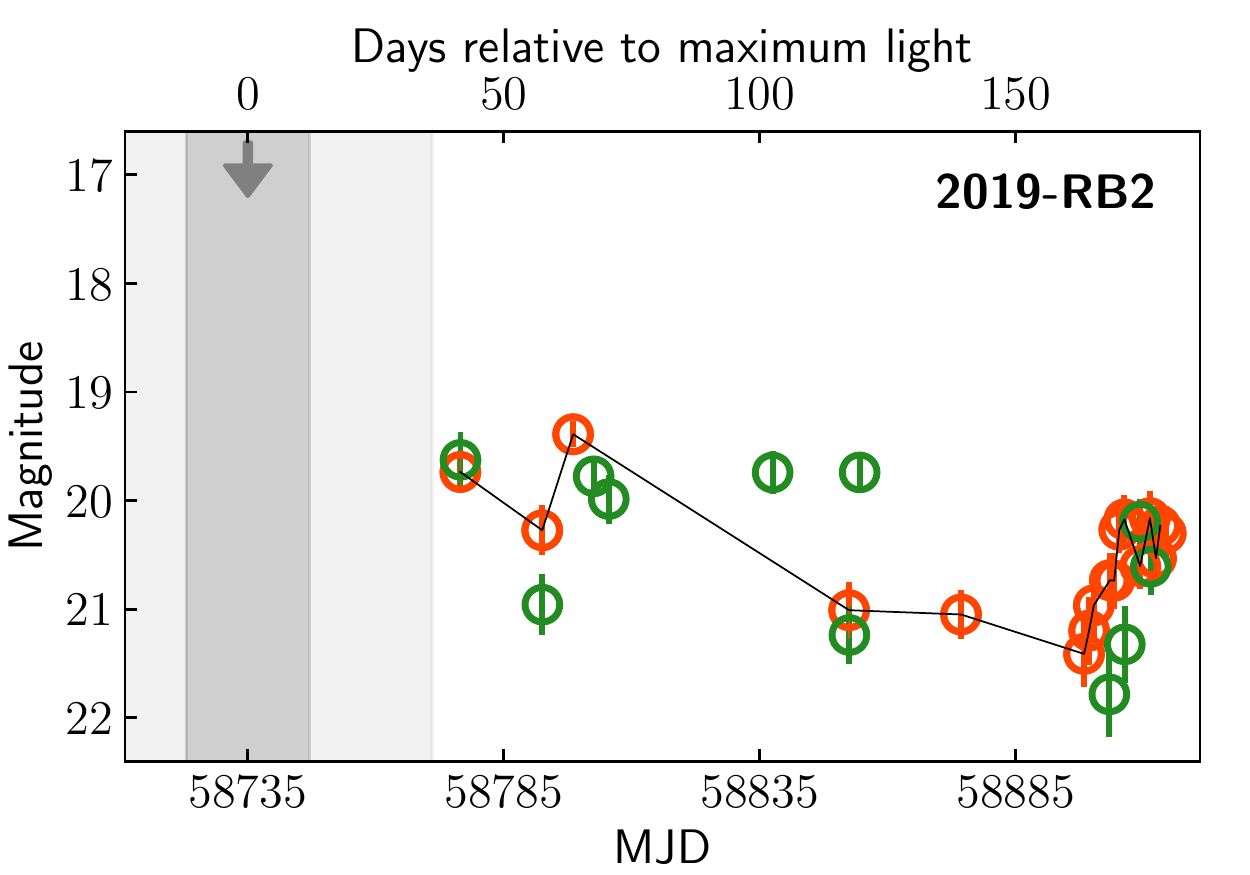}
    \end{minipage} 
     \begin{minipage}{.33\textwidth}
        \centering
        \includegraphics[width=\textwidth]{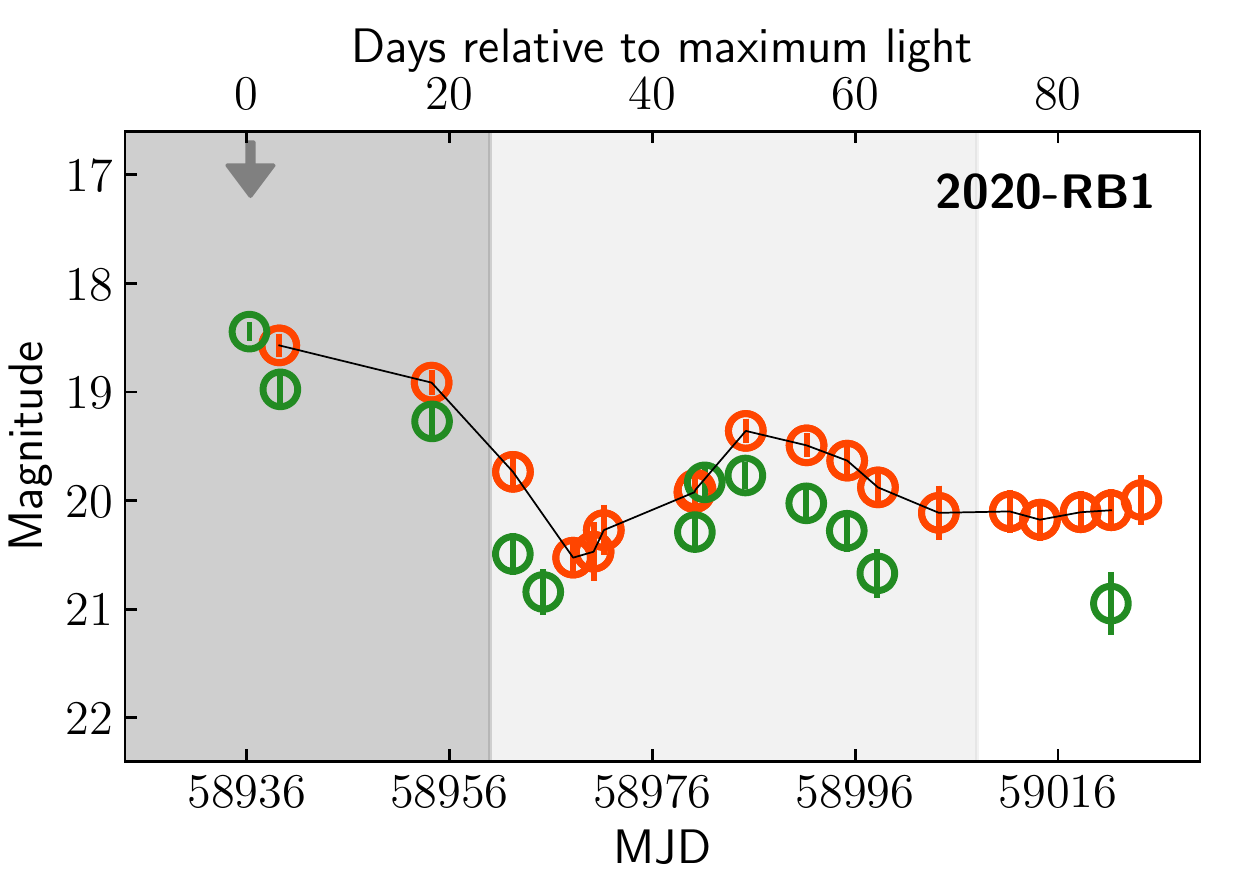}
    \end{minipage}
         \begin{minipage}{.33\textwidth}
        \centering
        \includegraphics[width=\textwidth]{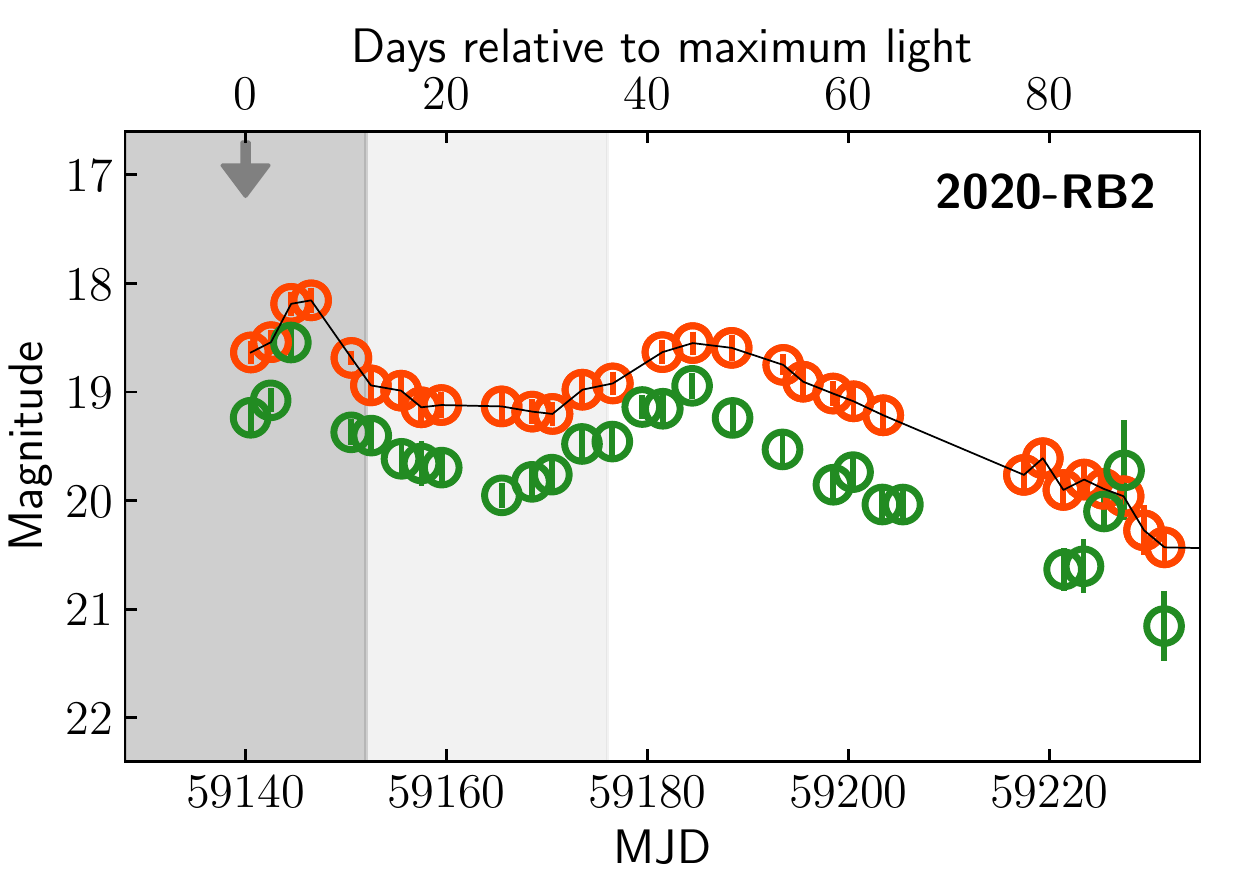}
    \end{minipage}
     \begin{minipage}{.33\textwidth}
        \centering
        \includegraphics[width=\textwidth]{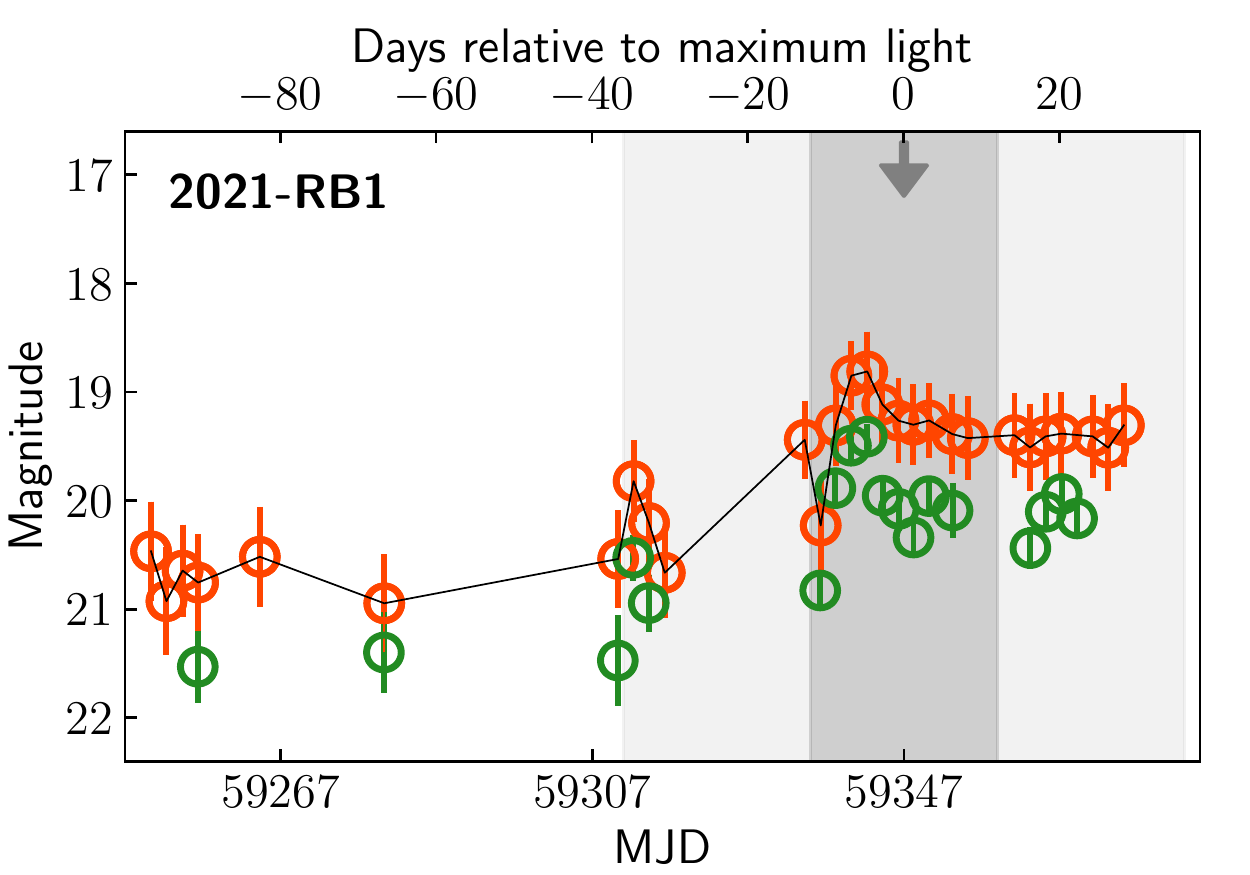}
    \end{minipage}
         \begin{minipage}{.33\textwidth}
        \centering
        \includegraphics[width=\textwidth]{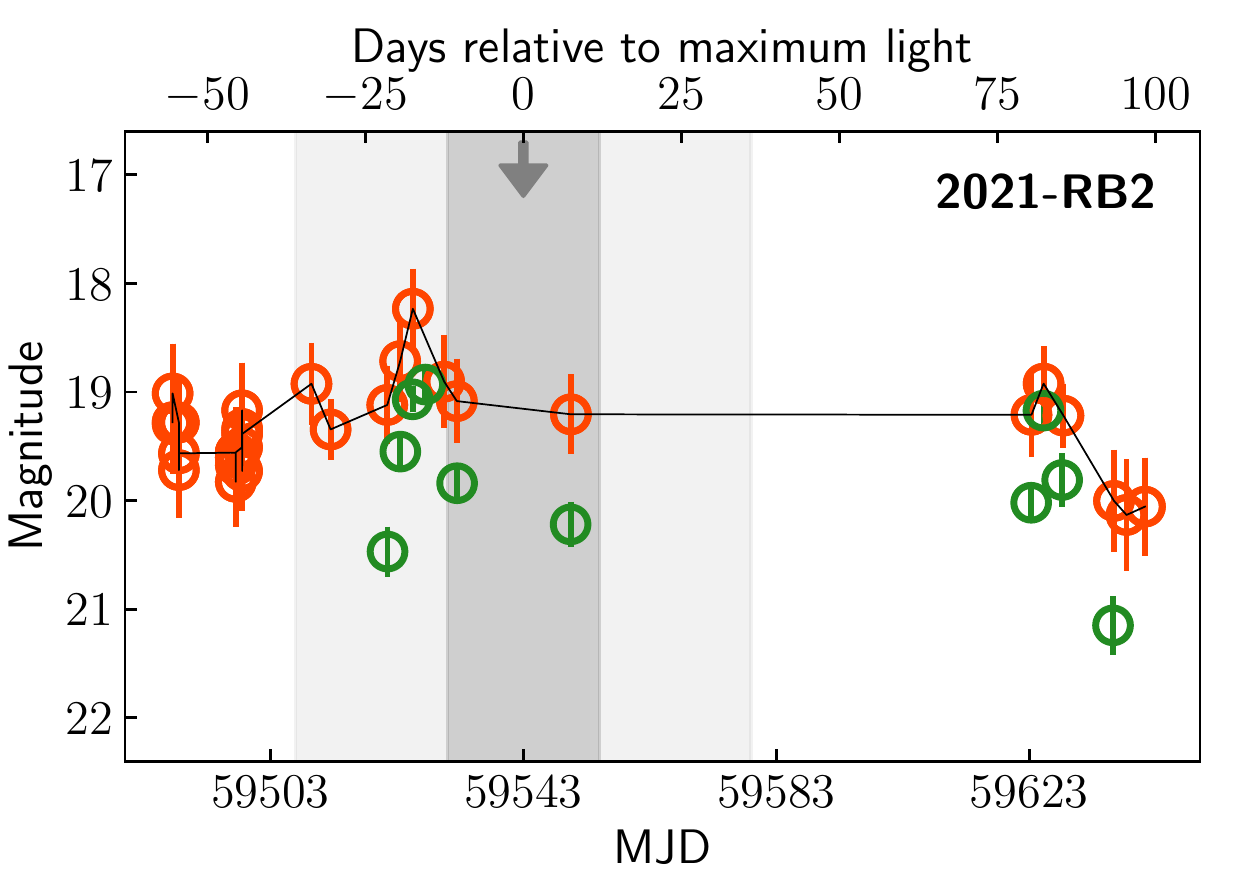}
    \end{minipage}
    \caption{Cutouts of the optical light curve presented in Figure \ref{fig:lightcurve_full}. Each subplot presents a set of consecutive measurements of an epoch of AT~2000ch, and the corresponding names of the outbursts described throughout the text are given in the top of each subplot. Each outburst is located at day 0 while newly detected or suspected (gray arrow) eruptive episodes or re-brightening events based on Equation \ref{eq:mean_period} with their $1\sigma$ ($3\sigma$) uncertainty indicated by the gray (light gray) shaded areas are additionally highlighted. The color code of each data point follows the legend shown in Fig. \ref{fig:lightcurve_full}{. We additionally show the less sampled brightness variation of the source between 2010 and 2012 as well as 2014 and 2018, which does not disagree with a {$\sim 201$\,day} period, even though the source seems to have undergone a more quiet phase (no reported outbursts).}} \label{fig:cutouts}
\end{figure*}

\section{Data reduction and analysis} \label{sec:red}

\subsection{Photometry}

To get an appropriate measure of the fluxes of our target in the data products provided by the PTF and ZTF surveys, we used DoPHOT \citep{dophot, dophot2} in an automated fashion. We estimated the necessary parameters (like Sky, Threshold, and Full Width Half Maximum) using Python 3.7 \citep{Python} and ran DoPHOT on a ($500 \times 500$)\,pixel grid with our target lying in the center. The photometric results were then calibrated with the SDSS ({data release} 16, \citet{SDSS}). The resulting data points in filter bands $g$, $r$, and $i$ therefore correspond to pseudo SDSS magnitudes. 
Data points taken from the literature were converted from the Johnson-Cousins system (Vega scale) to Sloan ABmags using the scaling relations given in \citet{Pastorello2010} {($r_\textrm{Sloan} = R_\textrm{Johnson}+0.18$ and $i_\textrm{Sloan} = I_\textrm{Johnson}+0.41$)} to ensure comparability to our newly added photometry.
We defined 30 standard stars in the field to compute the median offset and its standard deviation between the photometric catalog and SDSS for each observation, and corrected the magnitude of our target correspondingly. We estimated the uncertainty for each observation based on the error in photometry and the uncertainty resulting from the offset calibration. A summary of the derived photometric magnitudes and uncertainties is presented in Figure \ref{fig:lightcurve_full} and Table \ref{tab:optical}.

\begin{figure*}
    \begin{minipage}[t]{.49\textwidth}
        \centering
        \includegraphics[width=\textwidth]{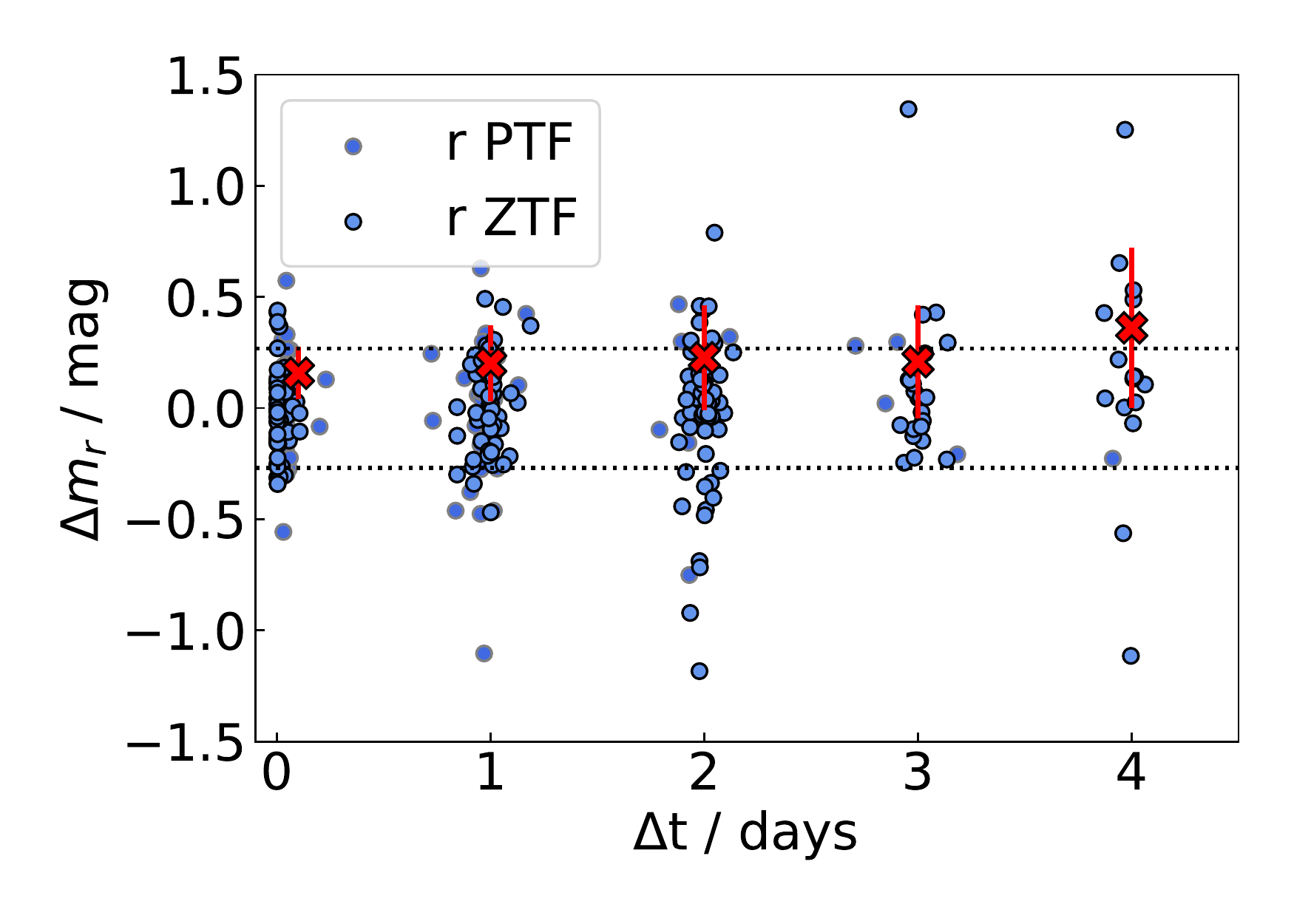}
    \end{minipage}  
    \hfill
    \begin{minipage}[t]{.5\textwidth}
        \centering
        \includegraphics[width=\textwidth]{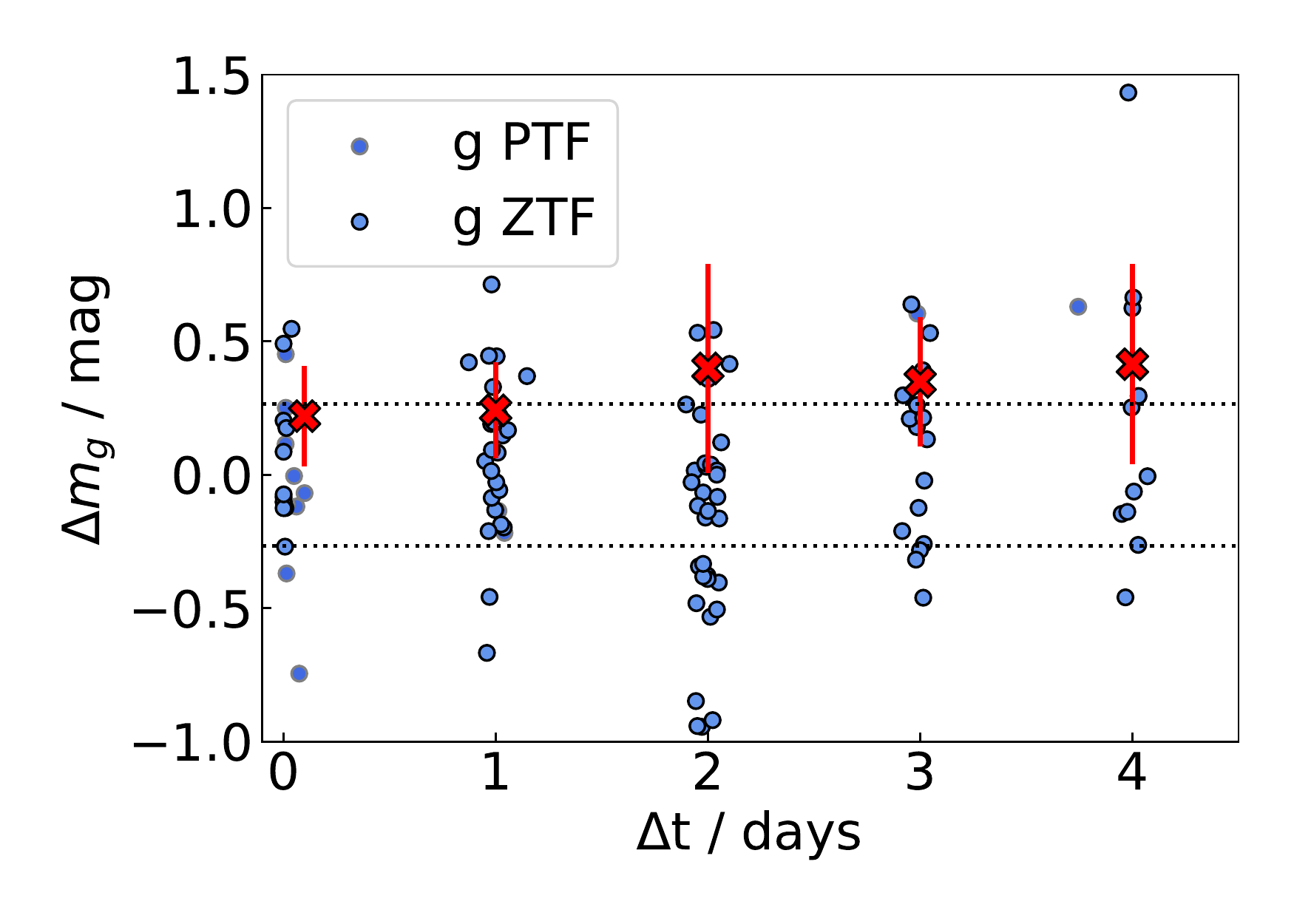}
    \end{minipage}
    \caption{Difference in brightness $\Delta m$ between consecutive PTF/ZTF measurements in the same filter as a function of the elapsed time between them $\Delta t$ for $r$ and $g$. The dotted lines mark $\pm\sqrt{2}\,\cdot\,$median photometric error in the corresponding filter. Red crosses mark the mean absolute deviation and the red error bar shows the standard deviation of absolute difference in brightness for each cluster of data points.} \label{fig:photvar_over_time}
\end{figure*}

Computing photometric color indices is generally problematic if the source varies in brightness on very short timescales, because multi-band observations are often taken in sequence and not simultaneously. To estimate the robustness of our data against this effect, we plot the change in brightness of consecutive measurements over the elapsed time between them for the PTF and ZTF data in $r$ and $g$ in Figure \ref{fig:photvar_over_time}. As can be seen for the $r$-band, both the mean absolute deviation ($MAD$, red mark in Fig. \ref{fig:photvar_over_time}) and standard deviation of absolute differences in brightness ($STD$, red interval in Fig. \ref{fig:photvar_over_time}) already increase significantly from $\Delta t \approx 0\,\textrm{days}$ {($MAD \approx 0.16\,\textrm{mag}$, $STD \approx 0.12\,\textrm{mag}$) to $\Delta t \approx 1\,\textrm{day}$ ($MAD \approx 0.20\,\textrm{mag}$, $STD \approx 0.17\,\textrm{mag}$)}, which we interpret as stellar variability between consecutive measurements. {The increase in the $MAD$ from $\Delta t \approx 0\,\textrm{days}$ ($MAD \approx 0.22\,\textrm{mag}$) to $\Delta t \approx 1\,\textrm{day}$ ($MAD \approx 0.24\,\textrm{mag}$) is marginal in the $g$-band and the $STD \approx 0.18$ remains constant, but the general trend that both parameters increase toward more elapsed time remains in the $g$-band as well.} Therefore, we compute color indices only where multi-band observations are available from the same night. If multiple measurements in the same filter band exist on the same day, only the data point with the smallest temporal offset is used for this computation rather than stacked photometry in order to mitigate an additional uncertainty due to variations on timescales of hours as much as possible. The nominal measurement errors for color indices include only the photometric error and do not scale with the time difference between the individual measurements.

\subsection{Radio continuum}
\label{subsect_radiocont}

We performed the data reduction based on the Common Astronomy Software Application package \citep[CASA,][]{2007ASPC..376..127M}. Interactive flagging was performed to mitigate the impact of radio frequency interference (RFI) for the data sets older than 1998. For the more recent data sets, the automated flagging routines TFCrop and RFlag were applied. Any corrupted data found during calibration was flagged manually, as were channels affected by the prominent hydrogen line. The data of project AW393 and AN0025  show the typical horn profile of such line emission and were therefore flagged. A standard cross-calibration was performed thereafter.
Self-calibration and imaging is done setting interactive masks around all visible sources to mitigate coarse artefacts and scattering of source fluxes. Phase-only self-calibration with a solution interval of 10 minutes was performed on the first four data sets (see Tab. \ref{tab:radio_measurements}). As there was marginal improvement, no further self-calibration was performed. In the case of project AI0073, a second phase and amplitude self-calibration was applied spanning the duration of the observation. The final imaging was done using tclean with a robust weighting of -1.0, a cleaning threshold of 1\,mJy, and a cell size of one-fifth of the synthesized beam. 

In case of the WSRT and LOFAR data, we used the publicly available data products and refer the reader to \citet{2001ASPC..240..451V} and Shimwell et al. in prep, respectively, for  further details.
All maps from the \changes project used in this work were cleaned using a Briggs robust weighting of zero and no uv-taper. While no self-calibration was applied to the C-Array C-band data, the C-array L-band and B-array L-band maps were self-calibrated. Multi-scale cleaning was applied to the C-Array maps \citep[][Walterbos et al. in prep.]{irwin2019changesDR3}.

\subsubsection{Flux}

Each radio image was analyzed using the source-finding software PyBDSF \citep{pybdsf}. An adaptive threshold of 3$\sigma$ was used as an island and pixel threshold for each analysis. We were only able to detect a source at the same position as the optical source of AT~2000ch for the \changes data sets. In cases of nondetection, we derived an upper limit for the radio flux of AT~2000ch using the local noise map generated automatically during each run of PyBDSF. The value at the position of AT~2000ch was extracted and multiplied by three to give a $3\sigma$ limit. In cases of a detection, we used the peak flux of the island, which we could associate with the position of the source. Flux errors were derived by accounting for a 5\% uncertainty in the absolute flux and the estimated local noise level given by PyBDSF.

\subsubsection{Spectral index} \label{sec:SIdata}

With AT~2000ch being highly variable in the optical range, we expect such a behavior to also be possible in the radio range. Significant changes can be exhibited on timescales of hours or days, which is why we can only investigate the spectral index for data taken within a few days at most. Only the two D-array \changes observations from December 2011 are separated by one day; however, we were not able to detect any radio emission on the 12 December 2011 at 1.5\,GHz. To estimate the steepest possible spectral index, we take our estimated upper limit into account.
No further consecutive observation has been performed within a few days because they are separated by at least 40\,days.

Nevertheless, the spectral index in radio data can also be estimated in-band. To this end, an observation with a significant fractional bandwidth is needed. Such an analysis can only be made for data where AT~2000ch is clearly detected ($>5\,\sigma$) because the data need to be split into two bands, reducing the signal-to-noise ratio (S/N) by about the square-root of two.
Two individual observations fulfill both criteria, a large fractional bandwidth and proper S/N, \changes project D-array and C-array with a central frequency of 5.998\,GHz and 2\,GHz bandwidth in December 2011 and February 2012, respectively. We use both data sets to estimate the spectral index between the upper and lower part of the band.

In case of the observation from the 17 December 2011, we split the data into two parts, centered at 6.027\,GHz and 5.322\,GHz, in order to optimize the S/N in the high-frequency regime; however, we were not able to detect the source in PyBDSF and can therefore only provide limits. We took the full uv-range because the exclusion of shorter baselines ---which could mitigate the effect of diffuse galaxy emission at the position of AT~2000ch--- would result in no detection of the source in either band. In the case of the observation from 13 February 2012, we split the data set into two parts, each covering 1\,GHz of bandwidth, resulting in central frequencies of 6.508\,GHz and 5.490\,GHz. For each data set, several images were generated where we sequentially excluded more of the shorter baselines ($\geq500$\,m to $\geq1500$\,m) to mitigate the contribution of the diffuse emission from NGC~3432. Source finding was then performed as described above. The source could be detected in all our generated images. To find the optimal balance between sensitivity and contribution of diffuse emission, we selected the images where the peak flux of the source and its integrated flux were lying within their respective errors. This was the case for a (u,v)-range of $\geq1000\,$m.

\section{Results} \label{sec:res}

\subsection{Source identification}\label{sec:source}
\begin{figure*}
        \centering
        \begin{subfigure}{.33\linewidth}
        \centering
        \includegraphics[width=\hsize]{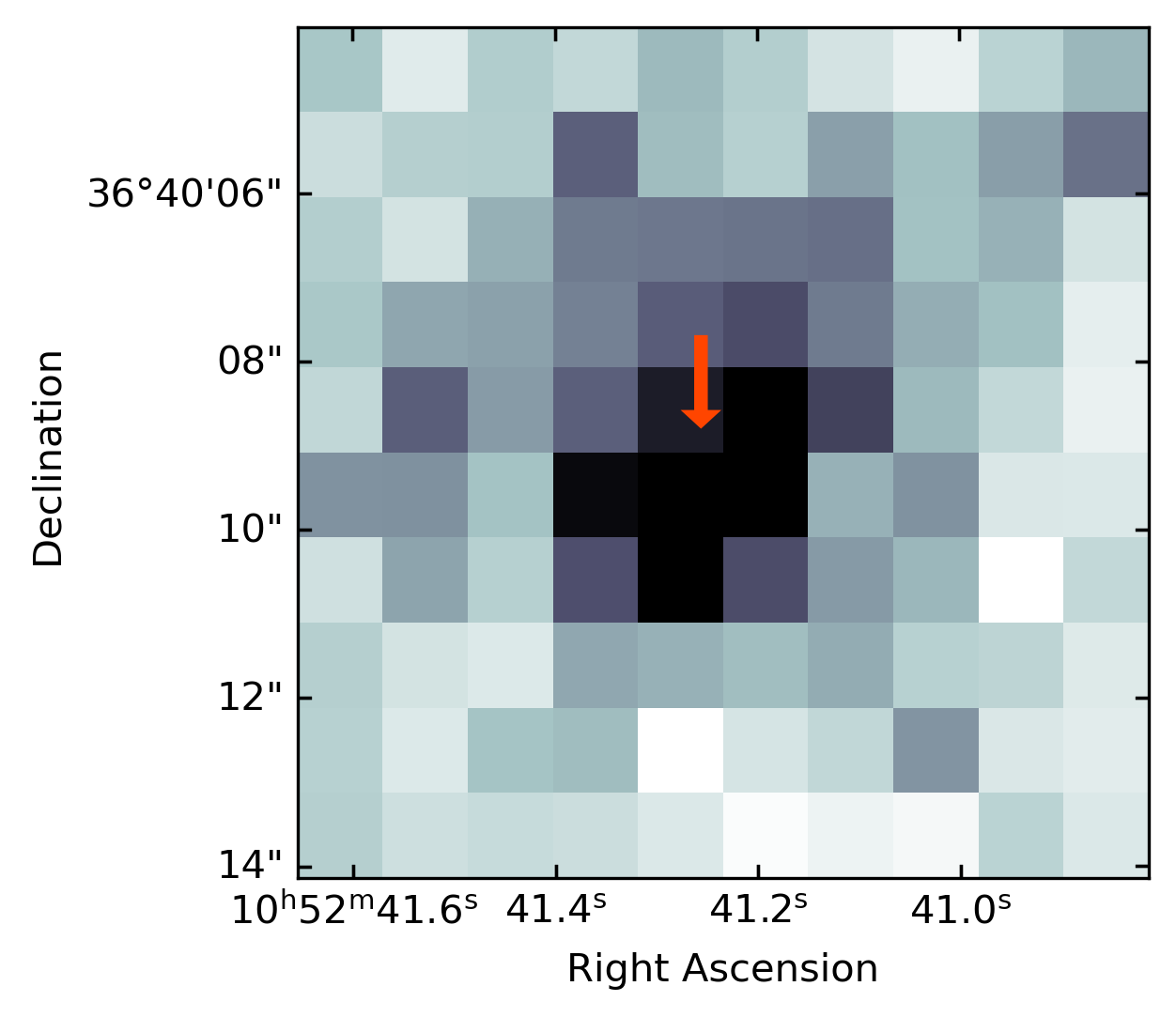}
        \end{subfigure}
        \begin{subfigure}{.33\linewidth}
        \centering
        \includegraphics[width=\hsize]{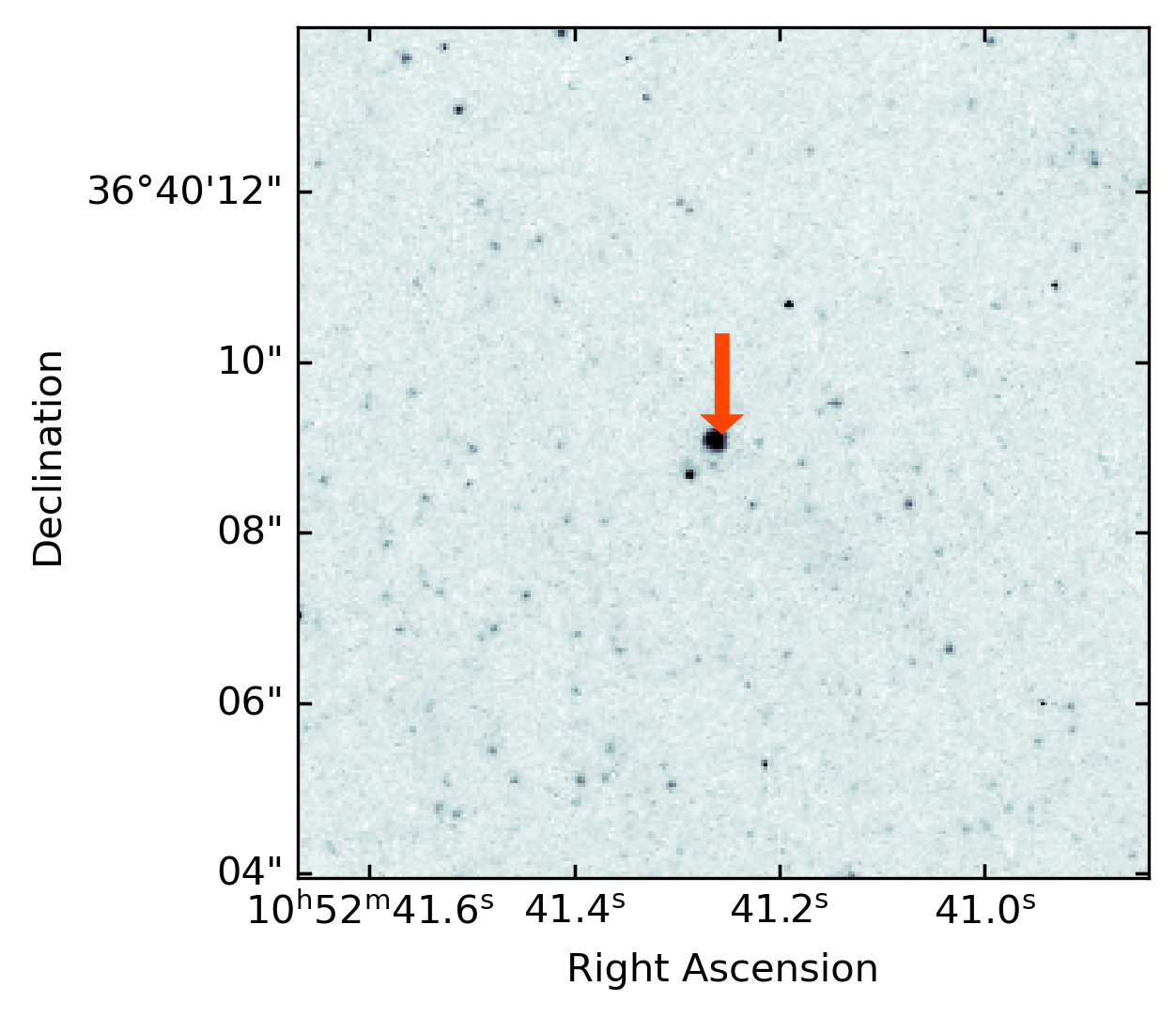}
        \end{subfigure}
        \begin{subfigure}{.33\linewidth}
        \centering
        \includegraphics[width=\hsize]{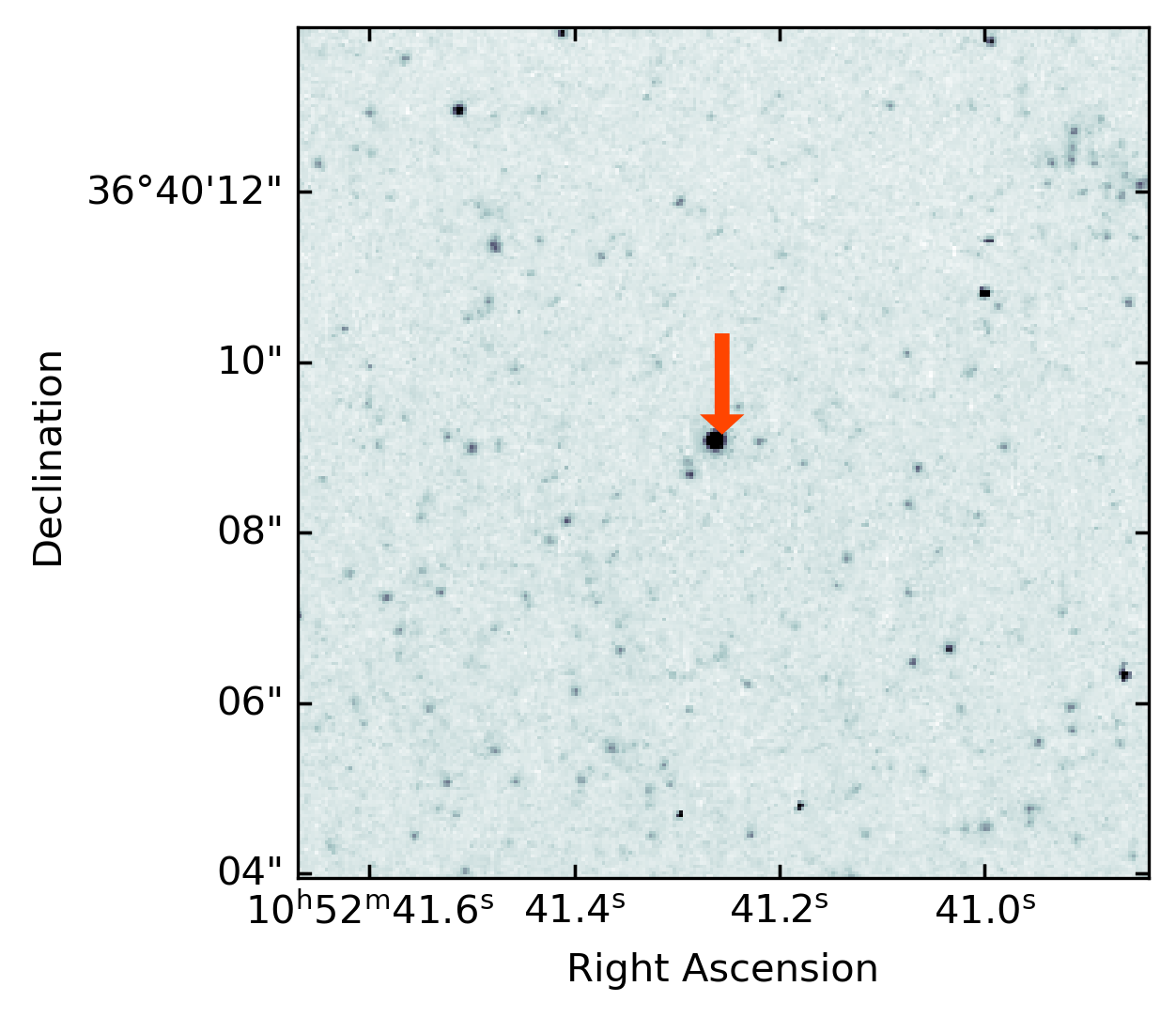}
    \end{subfigure}
    \\
        \begin{subfigure}{.33\linewidth}
        \centering
        \includegraphics[width=\hsize]{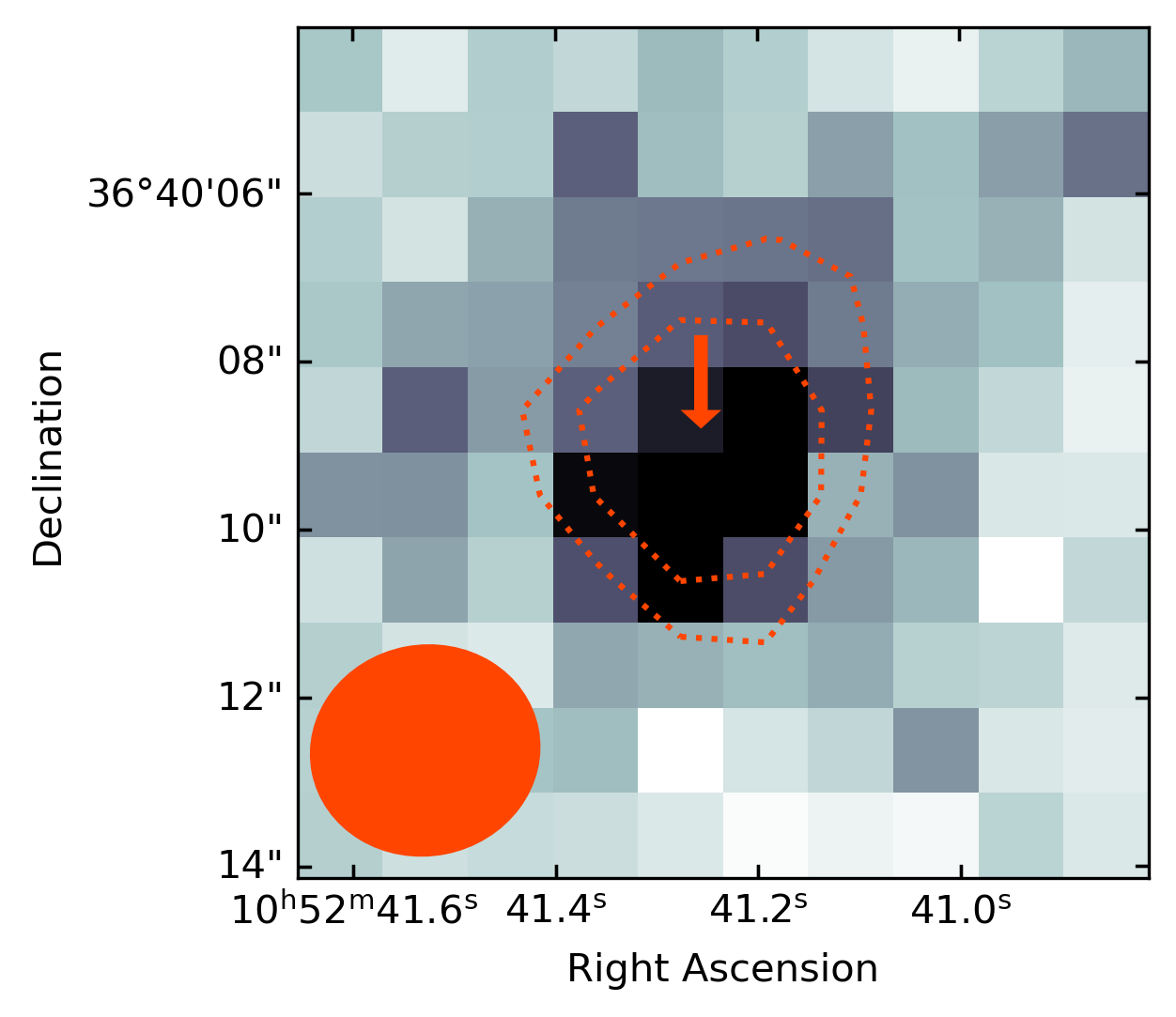}
        \end{subfigure}
        \begin{subfigure}{.33\linewidth}
        \centering
        \includegraphics[width=\hsize]{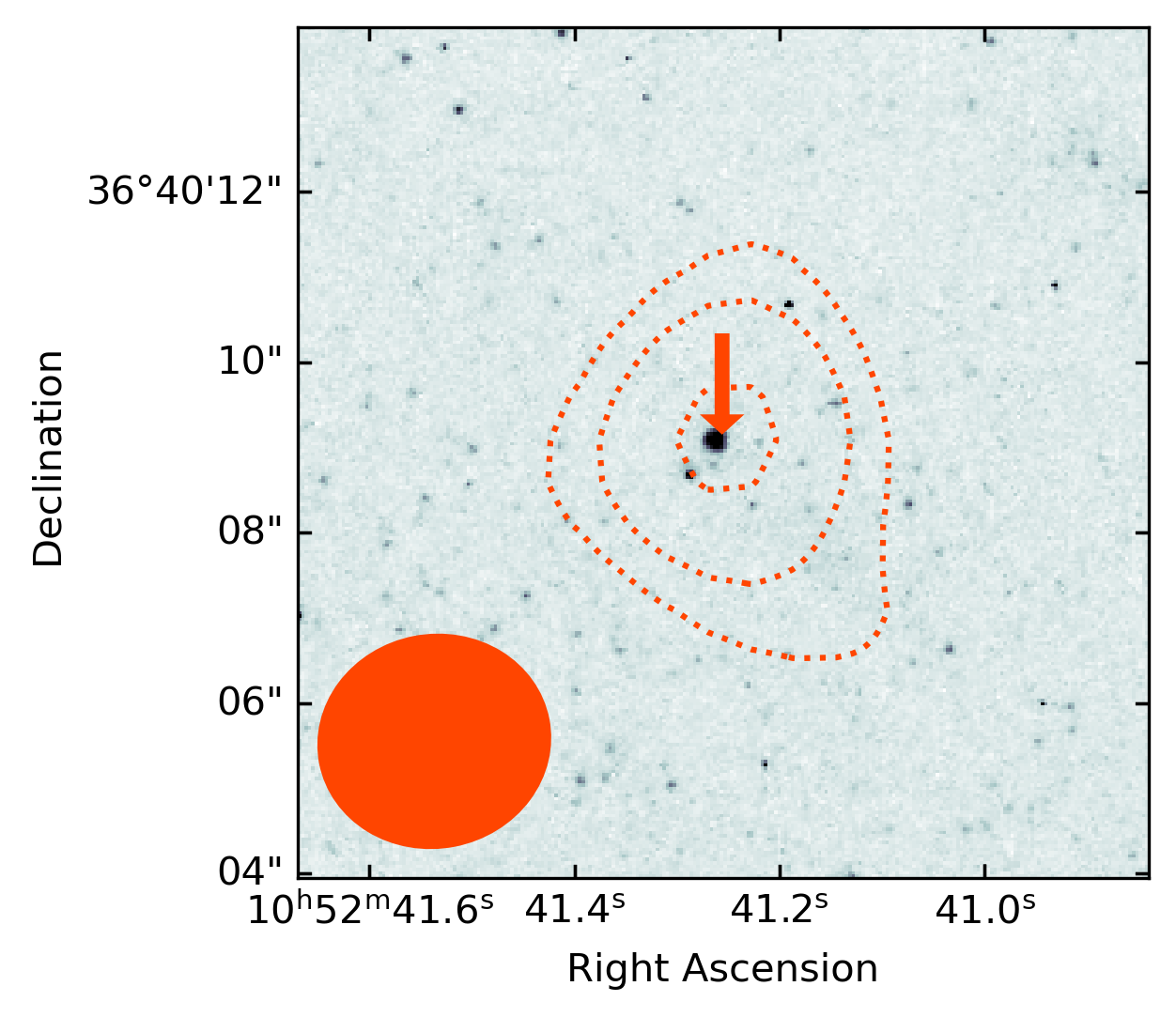}
        \end{subfigure}
\caption{Source identification in the optical and radio regimes. Close-up images (10''$\times$10'' boxes) of AT~2000ch from ZTF and the HST. The red arrow points to the position of AT~2000ch: \sdsspos. \textbf{Top panels:} ZTF r-Band image (left), HST F814W image (middle), HST F555W image (right). 
\textbf{Bottom panels:} Radio contours from the \changes C-Array C-Band map are overlaid in red. The resolution is indicated by the red beam in the left corner. ZTF r-Band image plus 3$\sigma$ and 6$\sigma$ contours (left), F814W image plus 3$\sigma$, 6$\sigma$, and 9$\sigma$ contours (right). 
}
\label{fig:hubble_ztf_pos}
\end{figure*} 
As mentioned above, three different positions for AT~2000ch have been reported. However, given they are all within 1$^{\prime\prime}$, these are most likely only astrometric offsets.
Here, we show $10^{\prime\prime} \times 10^{\prime\prime}$ close-up images of AT~2000ch\footnote{Based on observations made with the NASA/ESA Hubble Space Telescope, and obtained from the Hubble Legacy Archive, which is a collaboration between the Space Telescope Science Institute (STScI/NASA), the Space Telescope European Coordinating Facility (ST-ECF/ESA) and the Canadian Astronomy Data Centre (CADC/NRC/CSA).} in Figure \ref{fig:hubble_ztf_pos} to clarify the position of AT~2000ch to be most likely \sdsspos. The top left panel shows an example image of a detection of AT~2000ch and representative of PTF and ZTF data. Comparing the central brightness with space -based observations of the Hubble Space T (HST; top middle and right panels), we find a clear position match (as indicated by the red arrow in Fig. \ref{fig:hubble_ztf_pos}). Interestingly, the HST images reveal a second source with a very small projected distance ($\sim0.5\arcsec$). The resolution of the optical data sets from PTF and ZTF used in this work does not allow us to distinguish these two sources. One pixel of these surveys resembles a $1^{\prime\prime}\times1^{\prime\prime}$ grid in which both sources are enclosed. Taking the ground-based resolution (seeing influence) and the projected distance between the two sources into account, we expect the photometric and spectroscopic observations by \cite{Wagner2004} and \cite{Pastorello2010} to also be affected by the second source.
The potential influence of the second source is discussed in Section \ref{sec:disc}.
Based on the peak emission, we attribute the detected variability to the brighter source.

To ensure that the observed radio continuum emission can be associated with the optical emission of AT~2000ch, we overlaid the contours on the optical images in the bottom panels of Figure \ref{fig:hubble_ztf_pos}. We find the emission to be in good agreement with the location of AT~2000ch. Moreover, the peak emission in radio continuum (inner red contour) perfectly matches the bright source that we identified as AT~2000ch (red arrow) in the bottom right panel (Fig. \ref{fig:hubble_ztf_pos}). The extension of the radio emission corresponds to the resolution of the data (indicated by the red beam in the left corner in the bottom panels). The radio emission of AT~2000ch is further studied in Section \ref{sec:radio_var}.

\subsection{Optical light curve} \label{sec:lightcurve}
\begin{figure*}[ht]
        \centering
        \includegraphics[width=1.0\textwidth]{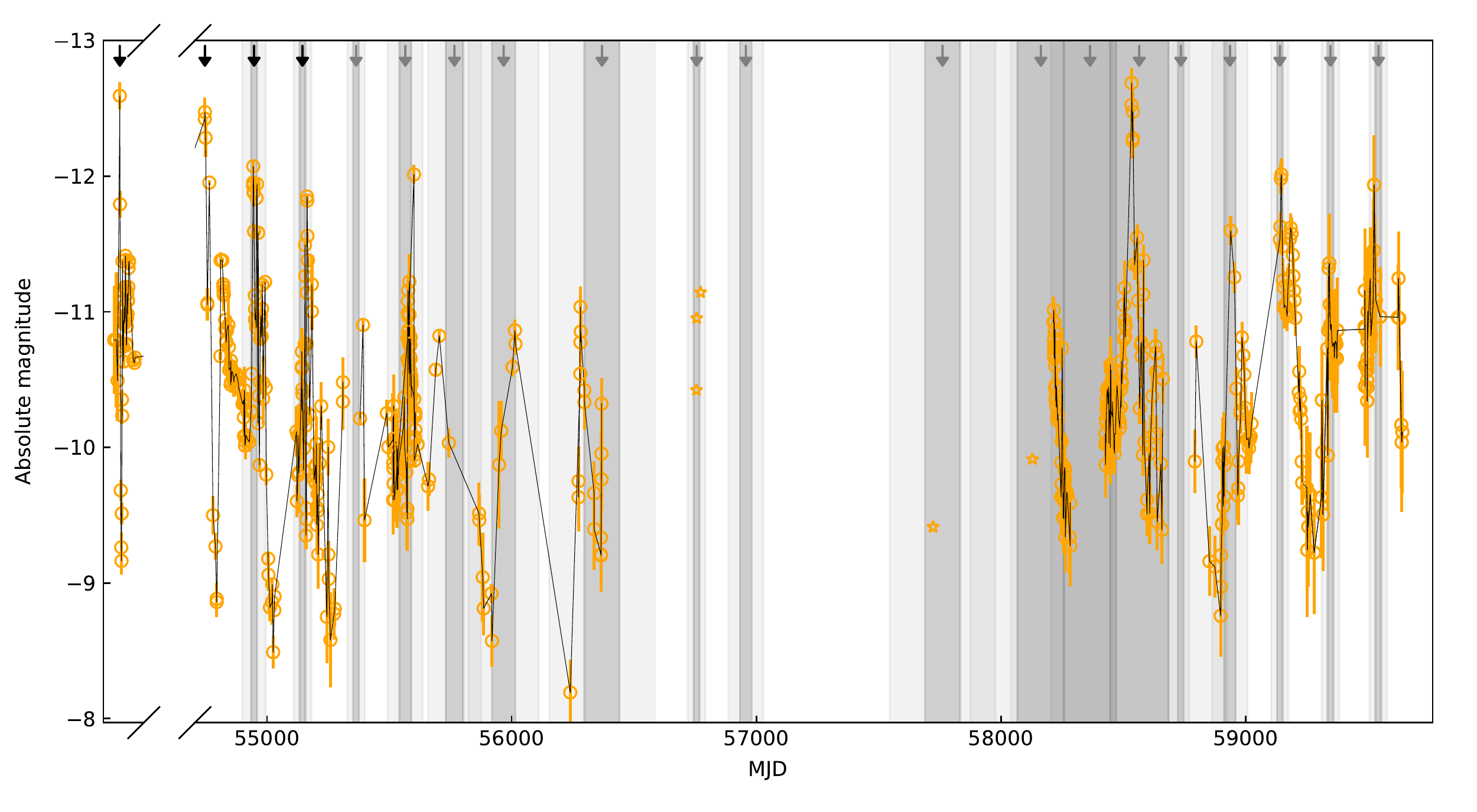}
        \caption{Optical light curve of AT~2000ch with all available $r/R$-band magnitudes transformed to Sloan $r$-band absolute magnitudes.}
        \label{fig:lightcurve_r}
\end{figure*}
\renewcommand{\arraystretch}{1.2}
\begin{table}[ht]
        \caption{Elapsed time between peaks of consecutive, well-sampled outbursts or re-brightenings in the r-band light curve.}
        \centering
        \begin{tabular}{M{5.0cm} M{2.8cm}}
                \hline\hline
                Interval between & Elapsed time (days)\\
                \hline
                2008-OT \& 2009-OT1& 197.79\\
                2009-OT1 \& 2009-OT2& 219.30\\
                2020-RB1 \& 2020-RB2& 207.32\\
                2020-RB2 \& 2021-RB1& 195.76\\
                2021-RB1 \& 2022-RB2& 183.24\\
        \end{tabular}
        \label{tab:outburst_timing}
\end{table}
\renewcommand{\arraystretch}{1}
The optical light curve of AT~2000ch is presented in Figure \ref{fig:lightcurve_full}. Based on the new data points from PTF and ZTF, we show the corresponding SLOAN filter magnitudes also from \citet{Wagner2004}, \citet{Pastorello2010}, {and \citet{Pastorello2013}} using the corrections already given in \citet{Pastorello2010}. {We made cutouts of each outburst and re-brightening phase (Fig. \ref{fig:cutouts}).} 
We find that consecutive outbursts or re-brightenings occur $\sim$ 201\,days apart, which is consistent with the elapsed time between 2008-OT and 2009-OT1, as well as between 2009-OT1 and 2009-OT2. Table \ref{tab:outburst_timing} lists the elapsed time between maximum light in the $r$-band for consecutive outbursts and re-brightenings, where the light curve is sampled sufficiently well. We arrive at a mean period $\pm$ standard deviation $\sigma$ of ($201 \pm 12$)\,days. This indicates, that the mode of instability of AT~2000ch may have been consistent starting from 2008-OT, or possibly earlier, up until now. This average periodicity allows us to model the timing of the $n$th outburst $T_n$ after an outburst occurred at the known time $T_0$ with
\begin{align}\label{eq:mean_period}
    T_n = T_0 + n \cdot (201 \pm 12)\,\textrm{days}.
\end{align}
Equation \ref{eq:mean_period} is only meaningful on shorter timescales than nine iterations, or $\sim$ 5\,years, at which point the uncertainty starts to cover the time-axis completely. This periodicity model therefore does not allow any inferences on the timing of the 2000-OT outburst, because no other re-brightening was reported within a time-span of 8 years.
We indicate the position of the predicted outbursts from this periodicity since 2008-OT with gray shading in the plots of the light curve using the nearest reliably detected outburst or re-brightening to determine the error on this prediction. We view an outburst to be reliably detected in the context of this periodicity analysis if there is a clear peak in the light curve within the $3\sigma$ interval of the respective predicted time of maximum light. We further accept 2013-OT as reliably detected based on the data collected by amateur astronomers. Whenever such a reliable detection occurs, we ``update'' the $T_0$ parameter in Eq. \ref{eq:mean_period} to the time of the measured maximum light for subsequent predictions.

We aligned this analysis to the shape of the light curve in Sloan $r$ where possible. Therefore, the evolution of AT~2000ch in Sloan $r$ is shown in Figure \ref{fig:lightcurve_r}. We marked the previously identified outbursts with a black arrow, and again the position of possible outbursts {or measured or expected maximum brightness of the transient in the re-brightening} with a gray arrow. {We assume an outburst to be detected if its maximum brightness is $>18.5$\,mag and it shows a sufficient sampling of a rapid magnitude increase and decrease before and after the measured maximum light over a timescale of $\sim$ 1 week, respectively. We further find a re-brightening event to be reliable when the sampling is again sufficient enough to identify a comparable shape of the light curve but smaller magnitude variations and on longer timescales.} 

The evolution spans almost 20 years with the stellar object still being variable. AT~2000ch reaches an absolute magnitude of between $r=-12$\,mag and $r=-13$\,mag during outburst and $r\approx-9.5$\,mag during quiescence, spanning about 4\,mag within 200\,days. The evolution within the individual outburst is found to be diverse but also highly affected by the observational coverage. Outbursts marked with the black arrow (Fig. \ref{fig:lightcurve_r}) are already discussed in detail in \citet{Pastorello2010}; here, we discuss the evolution based on the added data following the naming convention of the outbursts introduced by \cite{Pastorello2010}. {A re-brightening of the source and its expected maximum (based on the {assumed} periodicity of the events) where we do not have enough data coverage for a detection are denoted RB hereafter. Overall, we report 2 new outbursts and 13 significant re-brightening events, three of which are only sparsely covered but} are shown to complement the discussion of the periodicity in Section \ref{sec:disc}.

\textbf{2010--2012 RB}: {Based on the {$\sim 201$\,days} of re-brightening periodicity of the transient (see Eq. \ref{eq:mean_period}) we mark the expected maximum brightness of the transient during 2010 and 2012 with gray arrows {({Modified Julian Date} $(MJD)\approx55565,\;55766,\;55967$) and show the corresponding uncertainties.} With the given observations, we do not find the transient to be in outburst; it probably experienced a phase of re-occurring re-brightening{s}. 
Unfortunately, the cadence of the light curve during each re-brightening period is too low to make conclusive remarks about the offset or a different re-brightening cadence in that time. Nevertheless, the next detected outburst agrees with a {$\sim 201$\,days} cadence measured between 2008 and 2009 and interpolated to 2013.}

\textbf{2013-OT:} Including the observations of amateur astronomers (V. Nevski, E. Romas and I. Molotov), we report that AT~2000ch reached its maximum on March 2, 2013 ($MJD=56353$) with -13.17\,mag (unfiltered){, which agree{s} with the findings of the unpublished observation of A. Pastorello (priv. comm.)}. Unfortunately, the coverage of the light curve is very low; nevertheless, the magnitude is found to be fading by 4\,mag within 12\,days. Based on Equation \ref{eq:mean_period} we were able to predict the outburst to occur at $MJD=56369\pm72$ (gray arrow and darker gray shaded region in Fig. \ref{fig:cutouts}), which is 17\,days off from the detected maximum and therefore lies within the uncertainty.

{\textbf{2014--2018 RB:} During 2014 and 2018, the source was not
frequently observed; however, we identified five re-brightening phases with maximum brightnesses at least 1\,mag lower than in outburst. The transient has not reached a brightness higher than $r\approx 18$\,mag. The variations in the light curve (increase and decrease over a few days) agree with the behavior of the transient in outburst and also with the expected {201-day} period marked with gray arrows {($MJD\approx56756,\;56957,\;57761,\;58163,\;58364$) and the corresponding uncertainties of the predictions marked by gray shaded regions} in the middle panel of Figure \ref{fig:cutouts}. Due to the low cadence, we cannot conclude whether the transient was undergoing a more quiet phase of re-occurring brightening or continued its eruptive phase where the outburst has not been monitored. We note that the detections of Masakatsu Aoki in 2014 are in disagreement with the three detections of A. Pastorello (yellow stars) at the same time and we therefore do not show the measurement in Figure \ref{fig:cutouts}.}

\textbf{2019-OT1:} AT~2000ch reaches a maximum absolute brightness of $r=-12.7$\,mag on 20 February 2019 ($MJD\approx 58534$). We sample 100\,days of this epoch detecting several phases of re-brightening, which is in agreement with the findings shown in the top four panels of Figure \ref{fig:cutouts}. Here, the brightness is again changing by about 1\,mag within only a few days. AT~2000ch reaches its minimum in June 2019 ($MJD=58658$) with $r=-9.4$\,mag. {With respect to the last detected outburst in 2013, we find the predicted outburst of $MJD=58565 \pm 120$ to agree with the predicted period. The prediction deviates by 31\,days from the maximum brightness detected during 2019-OT1.}

{\textbf{2019-RB:} Based on the measured re-brightening of the transient,} we predict AT~2000ch has reached its maximum brightness at the beginning of September 2019 {($MJD\approx 58735$)}. Unfortunately, there exist no further observations before and within the next 50\,days, and also the subsequent 100\,days are only sparsely covered but coincide with our findings of previous epochs. We identify changes in brightness of about 1\,mag within a few days, 70\,days after outburst, and AT~2000ch reaches its minimum on 19 February 2020 with $r=-8.8$\,mag. We detect a rise of about 1.5\,mag within a few days thereafter, introducing the transition to the next {significant re-brightening.} 

{\textbf{2020-RB1:} We predict that the transient has reached its maximum brightness} at the end of March 2020 {($MJD\approx 58936$)}. We measure a maximum brightness of $r=-11.6$\,mag on 31 March, 2020 ($MJD=58939$), which is significantly lower than during the previous detected outburst but {consistent with the uncertainty of the prediction. We assume that the transient has reached a lower brightness at its maximum.} About 50\,days after AT~2000ch reached its maximum, we detect a minimum of $r=-9.6$\,mag. The shape of the light curve can generally be compared with the past outbursts, but the first 50\,days are only sparsely covered.

\textbf{2020-RB2:} We find a maximum brightness of $r=-12.0$\,mag on 24 October, 2020 ($MJD\approx 59146$), which is slightly lower than the maximum brightness found during previous outbursts. {The detected brightness differs from the prediction by 6\,days, which still agrees with the uncertainty of the prediction.} We find the trend of the light curve thereafter to be much smoother than in the previous cases. {It is therefore likely that the transient has undergone a calmer phase of re-brightening.} Within the 60\,days after {maximum brightness}, AT~2000ch is re-brightening only twice ($\Delta r \approx 0.5\,$mag) and rather slowly instead of showing a peak. This difference in behavior during the evolution of the light curve might indicate a change of physics concerning the stellar object, the surrounding wind, or the circumstellar envelope. 

{\textbf{2021-RB1:} We find the maximum brightness ($MJD\approx 59342$) to coincide with the expected {$201\pm12$}\,day period {($MJD\approx 59347$)}, but this is much fainter than during outburst. The monitored brightness variation only spans 2\,mag, which agrees with the previously stated transition to a calmer phase. The re-brightening occurred over 25\,days and then the source was found to fall in brightness  by about 0.5\,mag thereafter, also over 25\,days. The timescales of both phenomena are significantly longer than the eruption event of this transient.} 

{\textbf{2021-RB2:} The next brightness increase also coincides with a rough {$201$-day period detected at $MJD\approx 59525$ and predicted at $MJD\approx 59543$. The measurement agrees with a 3$\sigma$ uncertainty of the prediction.} The event is found to be 1\,mag brighter than the previous event but we measure only minor variations of 1\,mag in brightness. From the few observations thereafter, we can argue that the transient was possibly calm over 75 days and became fainter by 1\,mag over the 25\,days thereafter. However, for a conclusive analysis, a high cadence light curve should at least be investigated over $\sim 201$\,days to evaluate whether the detected calmness is a sampling effect or a change in the intrinsic properties of the source. Follow-up observations over the next years could provide information on whether the transient has evolved or this calmer phase belongs to the eruptive behavior of the transient.}

\subsection{Color evolution}
\begin{figure*}[ht]
        \centering
        \includegraphics[width=1.0\textwidth]{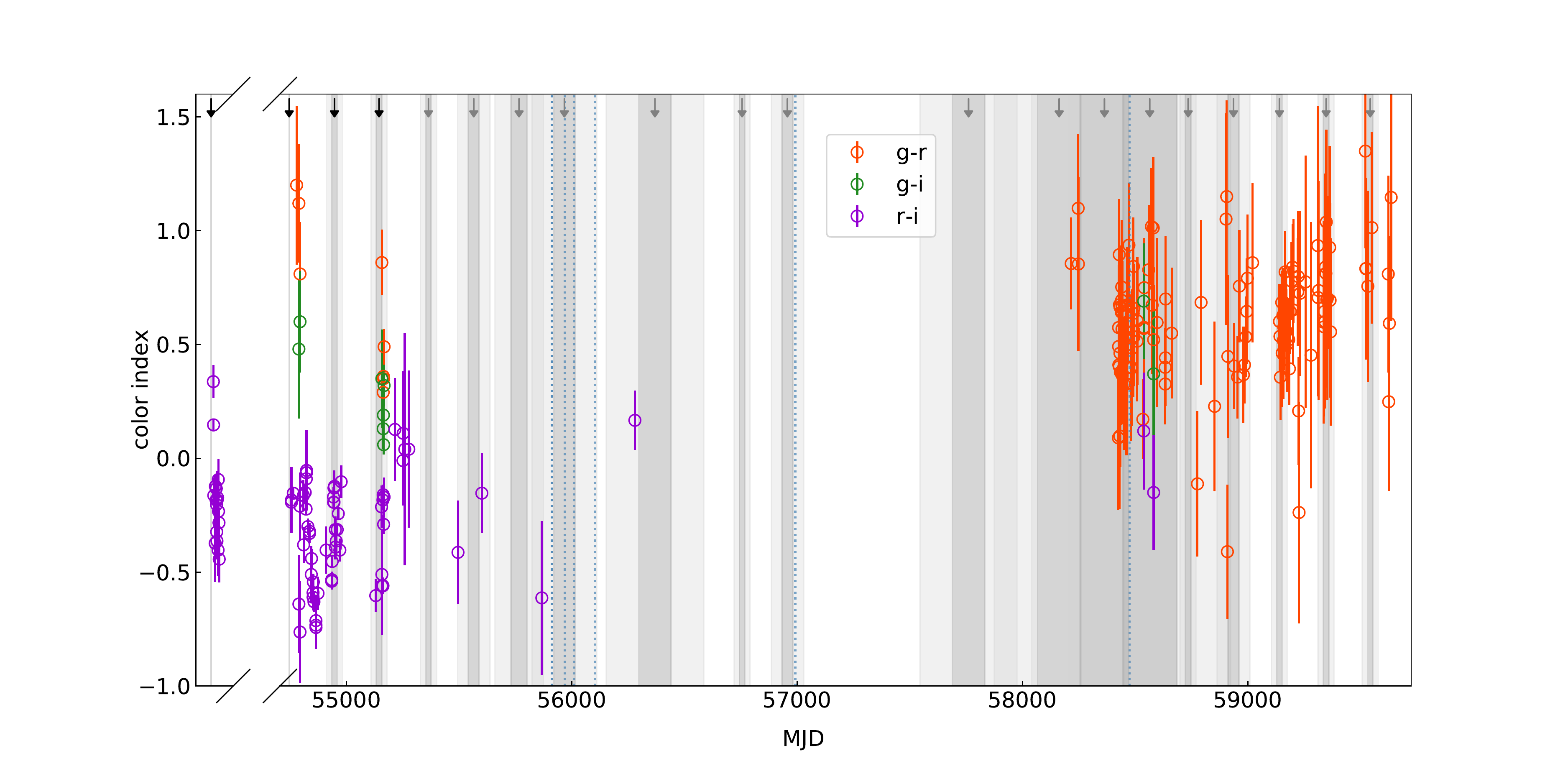}
        \caption{Full color evolution of AT~2000ch in $g-r$ (orange), $g-i$ (green), and $r-i$ (purple). Previously published color values by \citet{Wagner2004}, {\citet{Pastorello2010}, and \citet{Pastorello2013} (converted to the Sloan photometric system)} are shown as well as our photometric results from the PTF/ZTF survey observations. We only show the previously published data points for those color indices where we have added new observations. The vertical blue dotted lines indicate the dates of the observations that we collected in the radio range and the arrows mark the times of known (black) {and newly detected or suspected (gray) eruptive episodes based on Equation \ref{eq:mean_period} with their $1\sigma$ ($3\sigma$) uncertainty indicated by the gray (light gray) shaded areas}.}
        \label{fig:colorcurve_full}
\end{figure*}

The photometric color information based on the Sloan filters $g$, $r$, and $i$ from PTF/ZTF is shown in Figure \ref{fig:colorcurve_full}. We note that we only show those colors from \cite{Wagner2004} and \cite{Pastorello2010} that contain the same filter bands. A more detailed SED discussion containing more color indices can be found within these latter studies, but we provide a summary in Section \ref{sec:hist}.
We aim to study the color evolution by comparing the individual epochs, but, as mentioned above, some epochs are sparsely covered and do not provide two independent filter measurements within one day. Therefore, as can be seen in Figure \ref{fig:colorcurve_full}, for some epochs (outbursts marked by the colored arrows) we have no color information. While the color evolution based on the PTF/ZTF data is focused on $g-r$ (due to the fact that we only have very few measurements in Sloan $i$ from ZTF), during the epoch of 2000-OT no $g$-filter observations were performed and only a few between 2008-OT and 2009-OT2.

\begin{figure*}
    \begin{minipage}[t]{.33\textwidth}
        \centering
        \includegraphics[width=\textwidth]{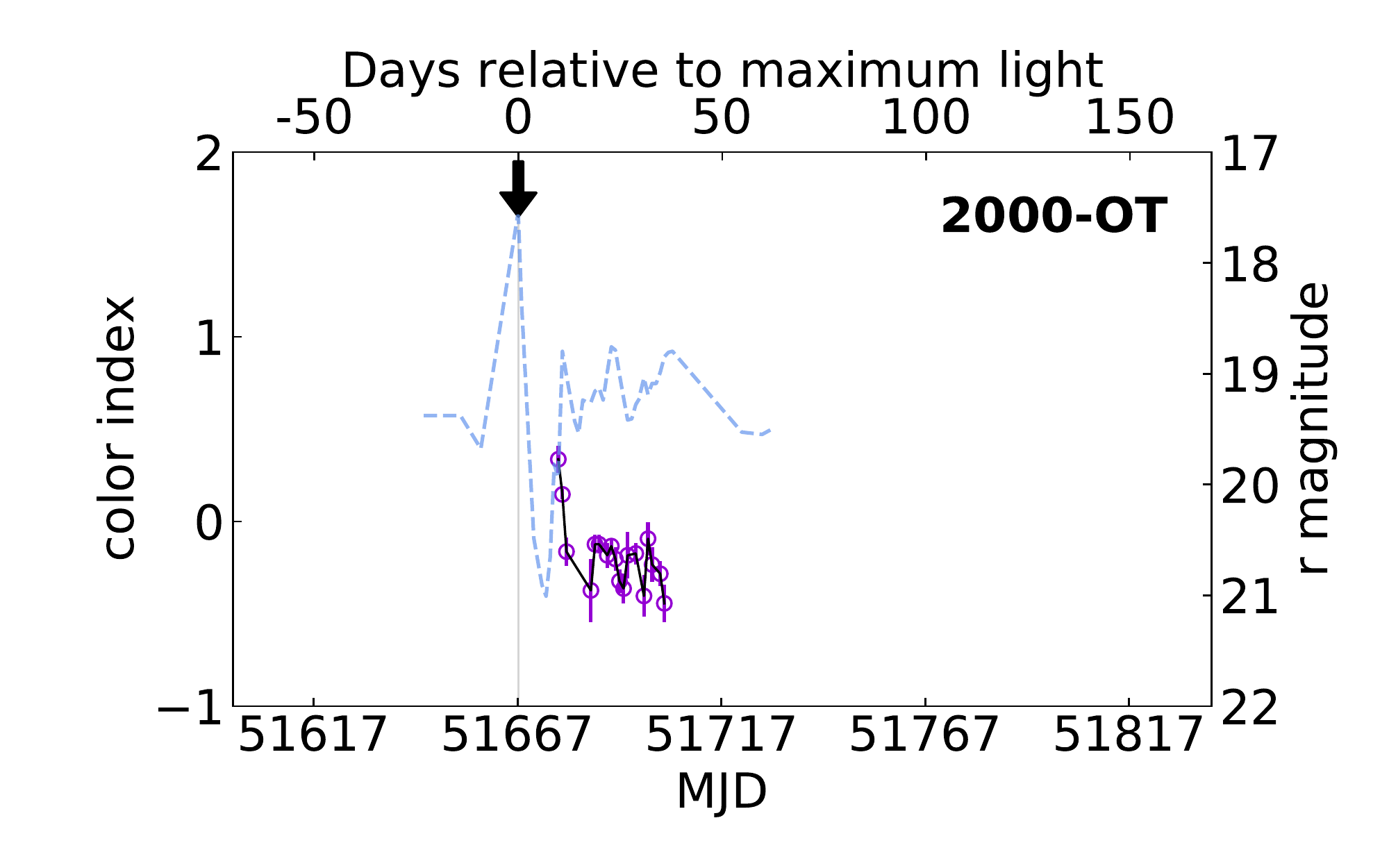}
    \end{minipage}  
    \hfill
    \begin{minipage}[t]{.33\textwidth}
        \centering
        \includegraphics[width=\textwidth]{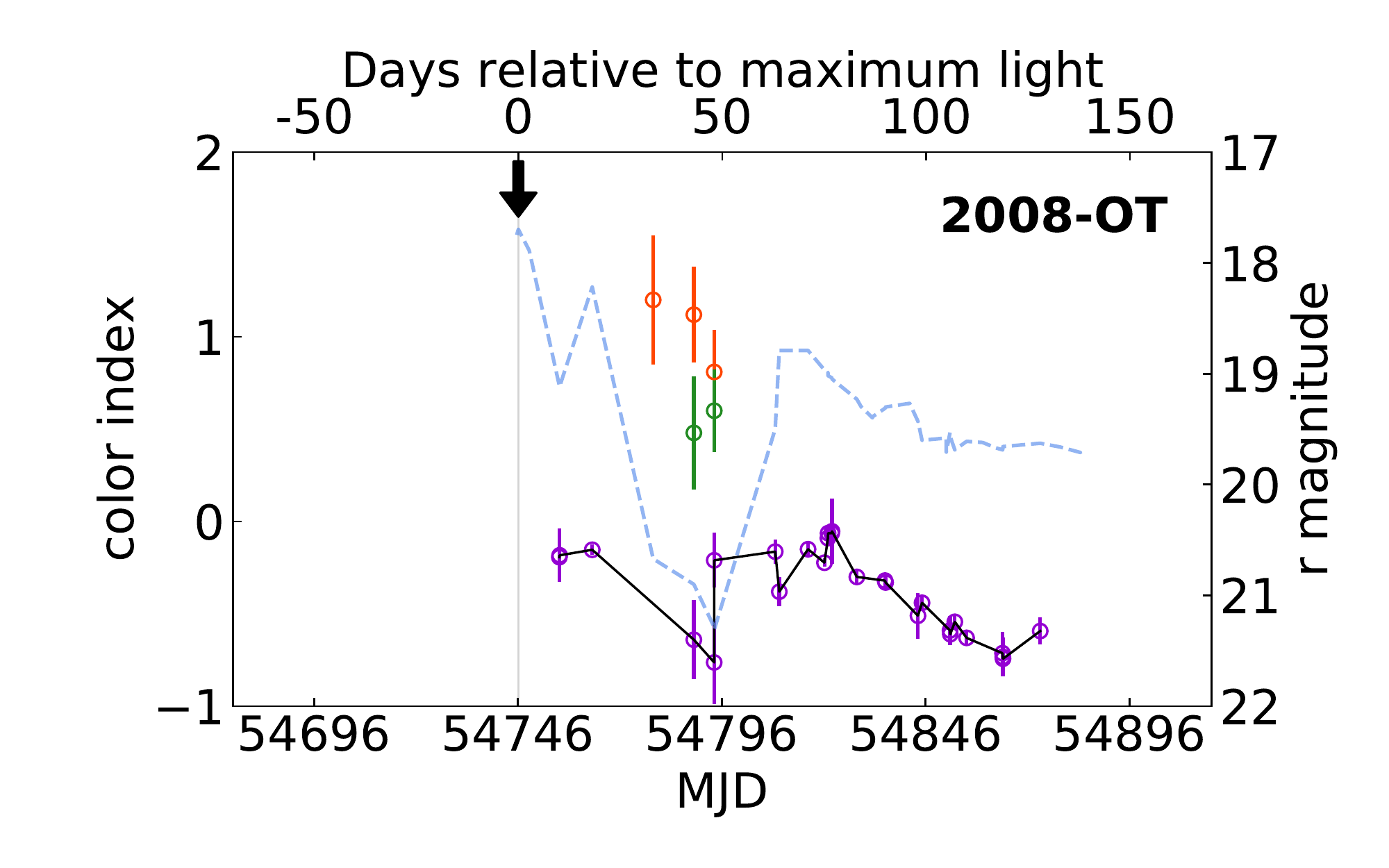}
    \end{minipage} 
     \begin{minipage}[t]{.33\textwidth}
        \centering
        \includegraphics[width=\textwidth]{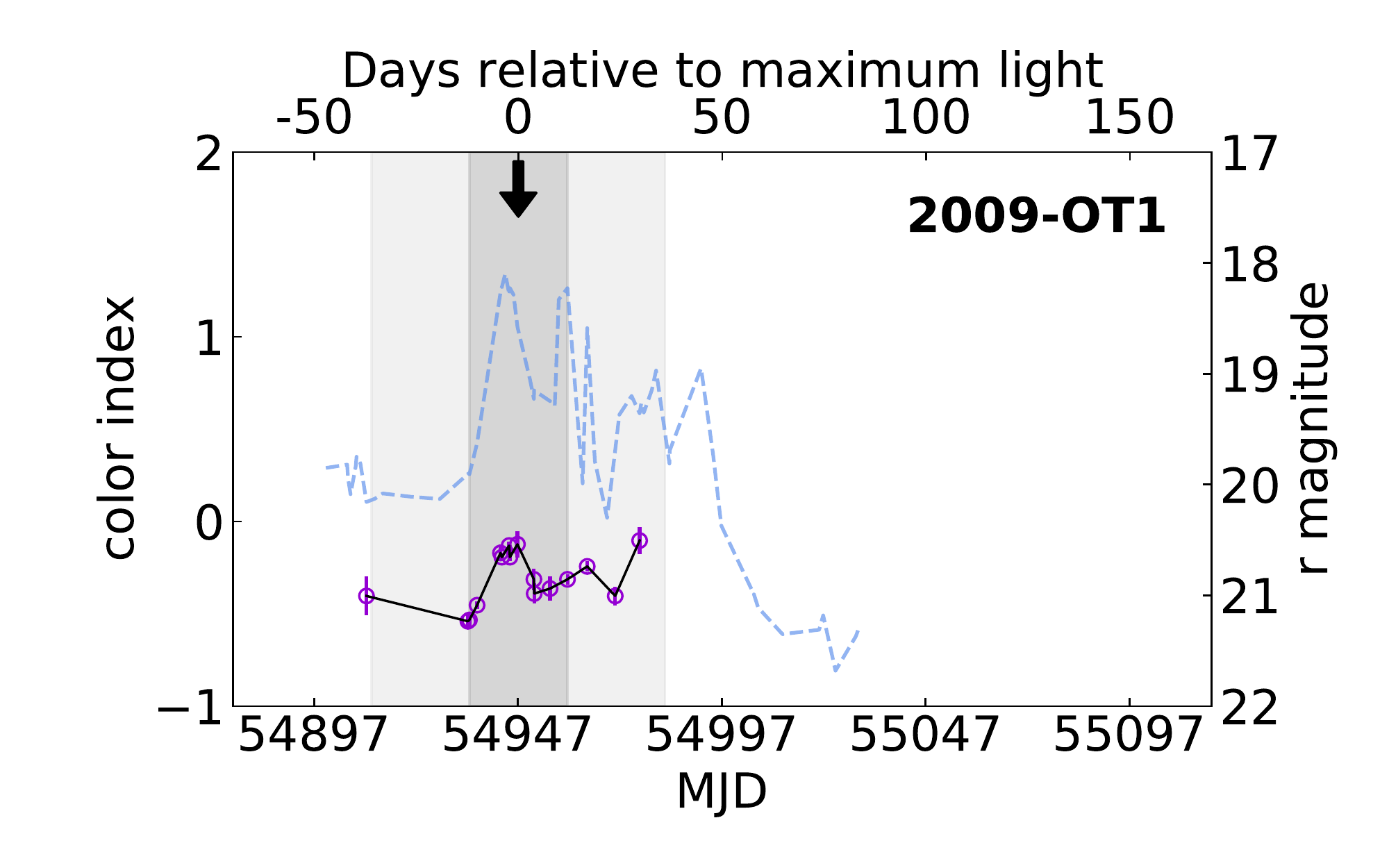}
    \end{minipage}  
     \begin{minipage}[t]{.33\textwidth}
        \centering
        \includegraphics[width=\textwidth]{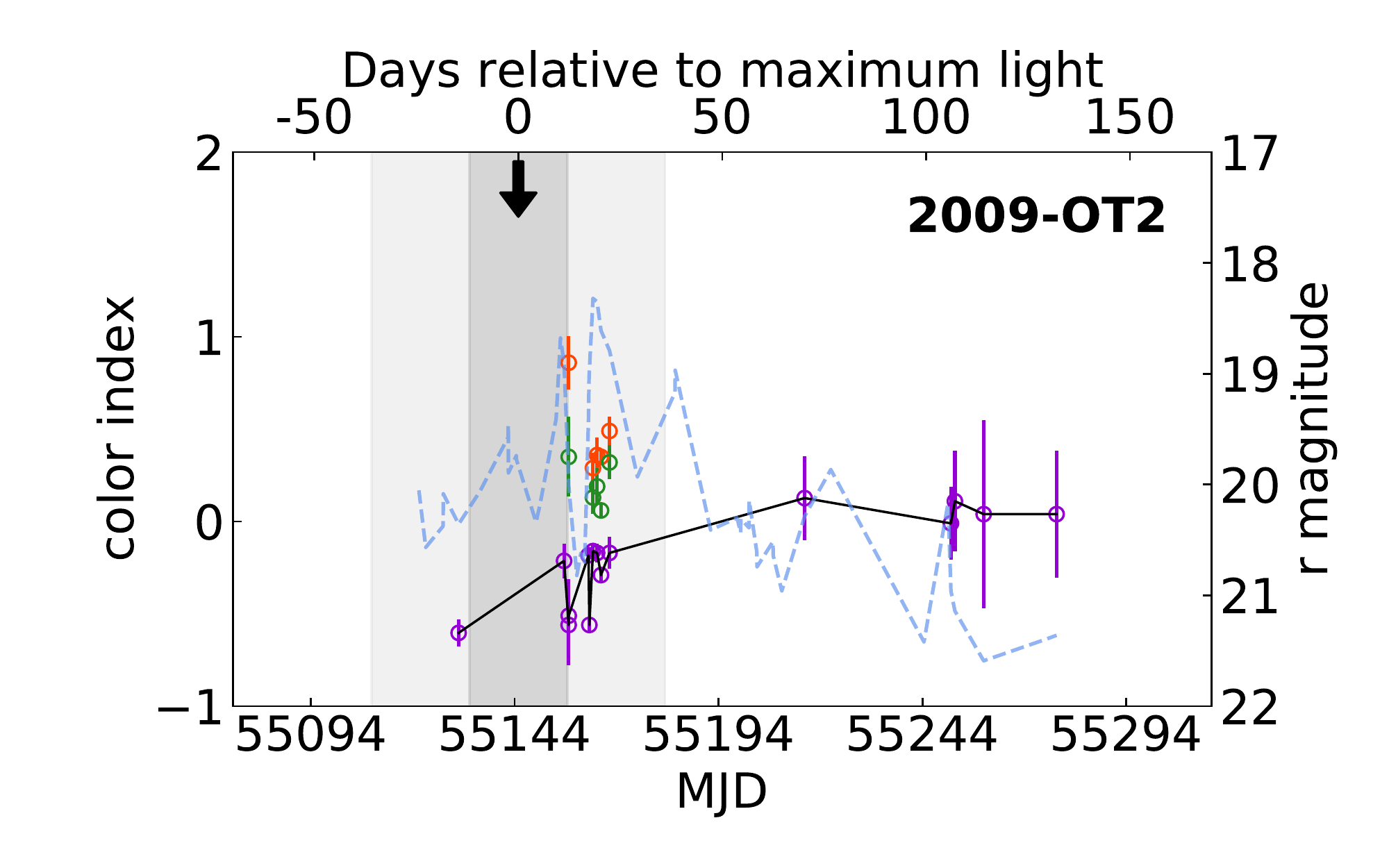}
    \end{minipage}  
    \hfill
    \begin{minipage}[t]{.33\textwidth}
        \centering
        \includegraphics[width=\textwidth]{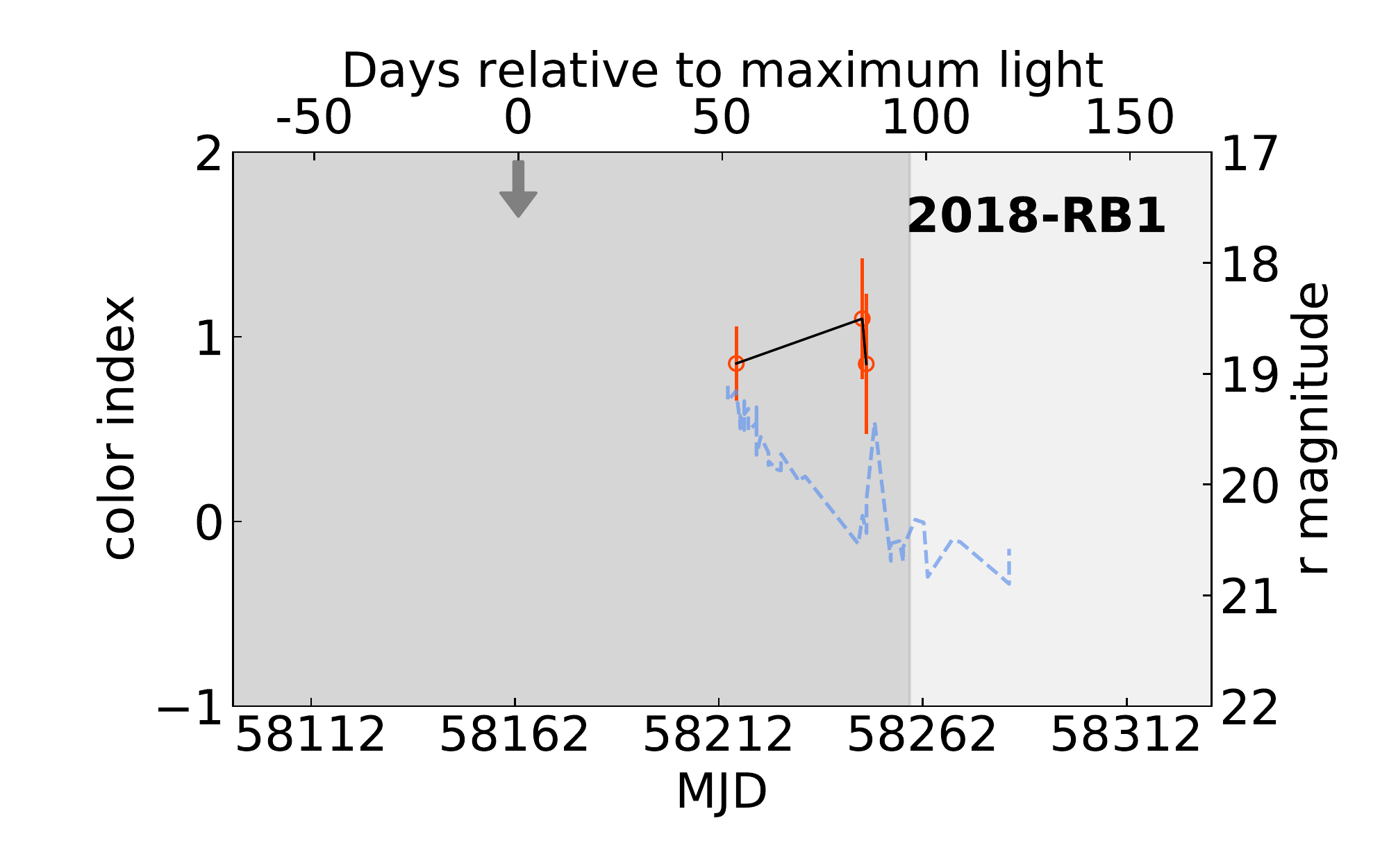}
    \end{minipage} 
     \begin{minipage}[t]{.33\textwidth}
        \centering
        \includegraphics[width=\textwidth]{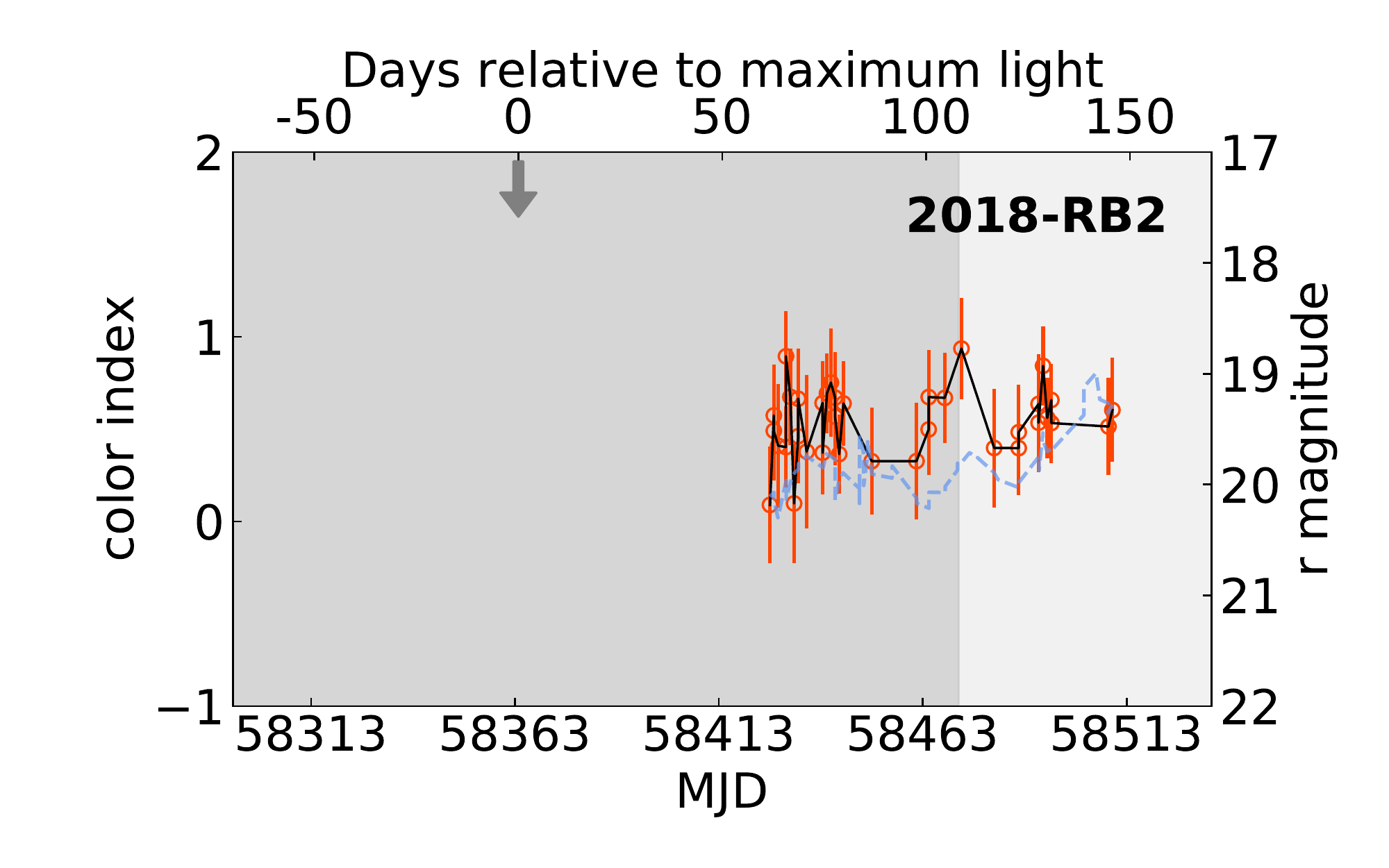}
    \end{minipage}
    \begin{minipage}[t]{.33\textwidth}
        \centering
        \includegraphics[width=\textwidth]{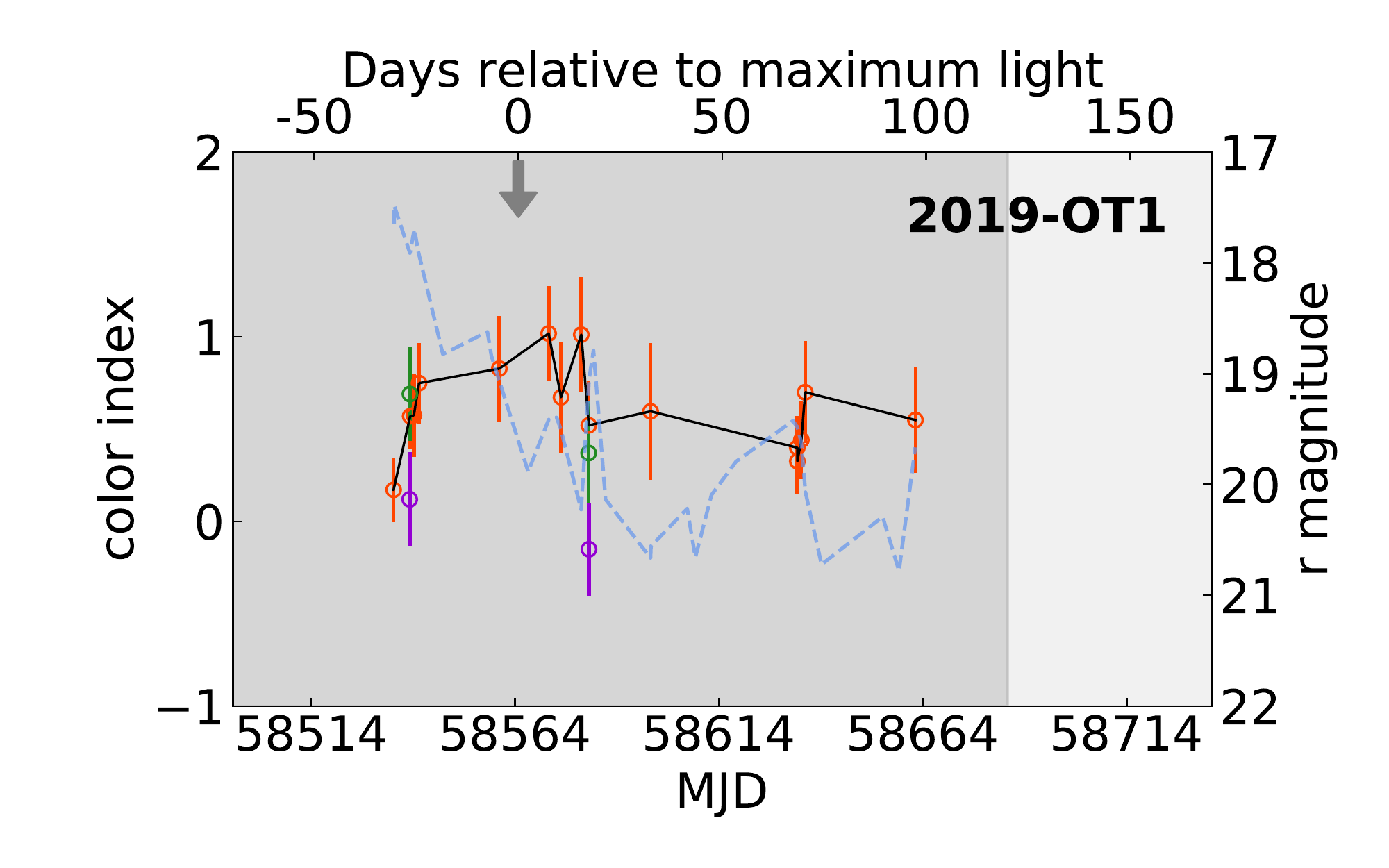}
    \end{minipage}  
    \hfill
    \begin{minipage}[t]{.33\textwidth}
        \centering
        \includegraphics[width=\textwidth]{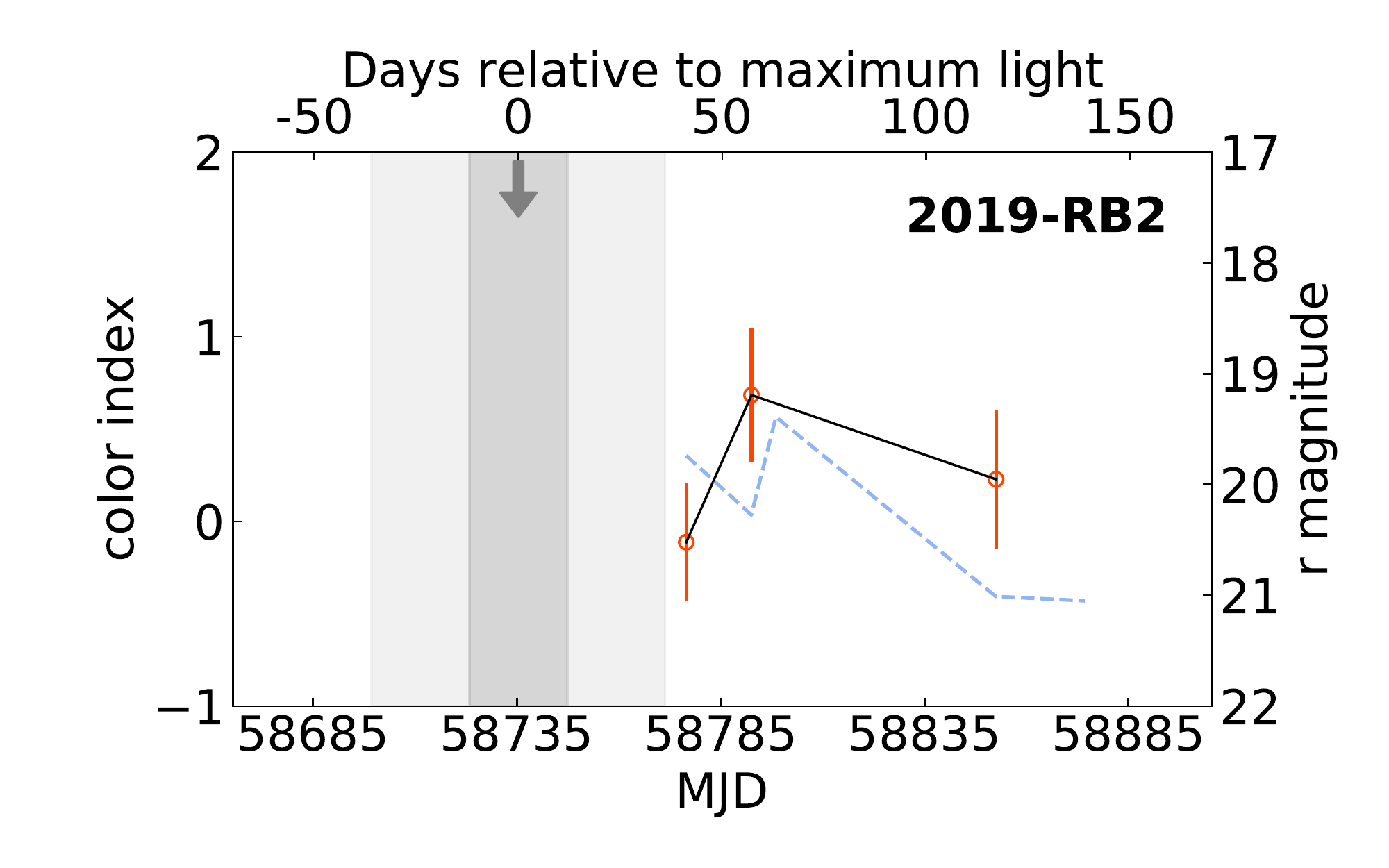}
    \end{minipage} 
     \begin{minipage}[t]{.33\textwidth}
        \centering
        \includegraphics[width=\textwidth]{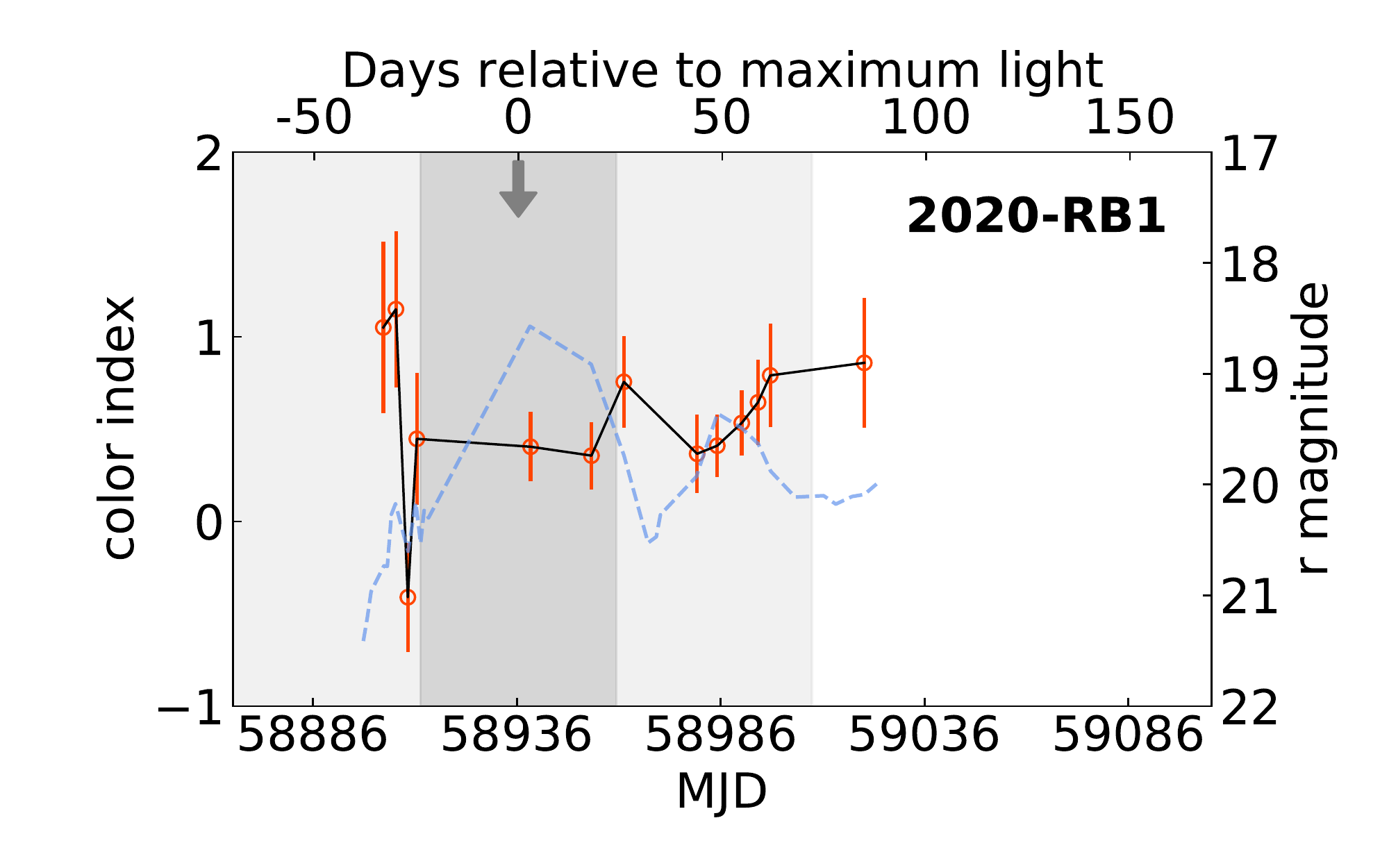}
    \end{minipage}   
     \begin{minipage}[t]{.33\textwidth}
        \centering
        \includegraphics[width=\textwidth]{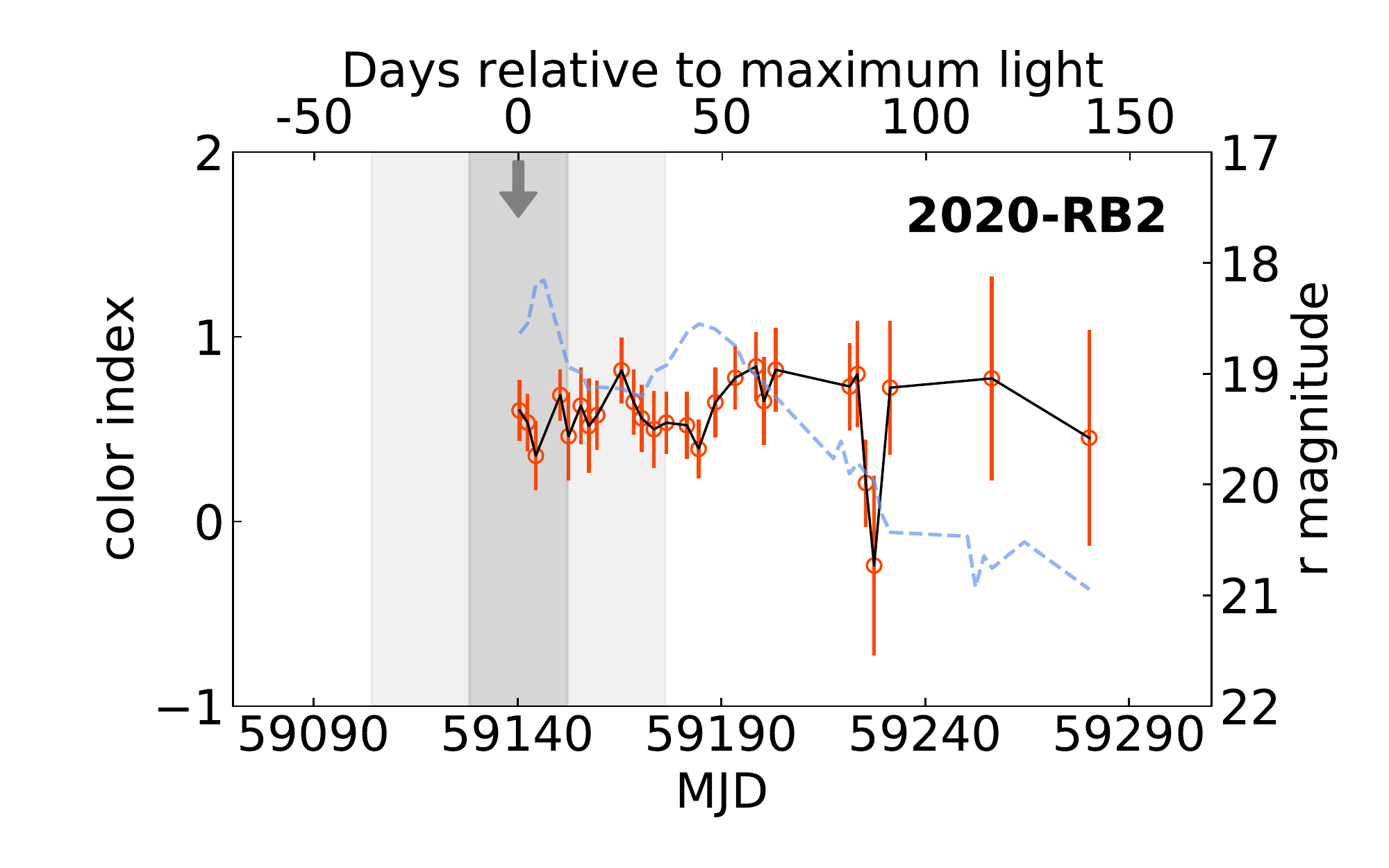}
    \end{minipage} 
     \begin{minipage}[t]{.33\textwidth}
        \centering
        \includegraphics[width=\textwidth]{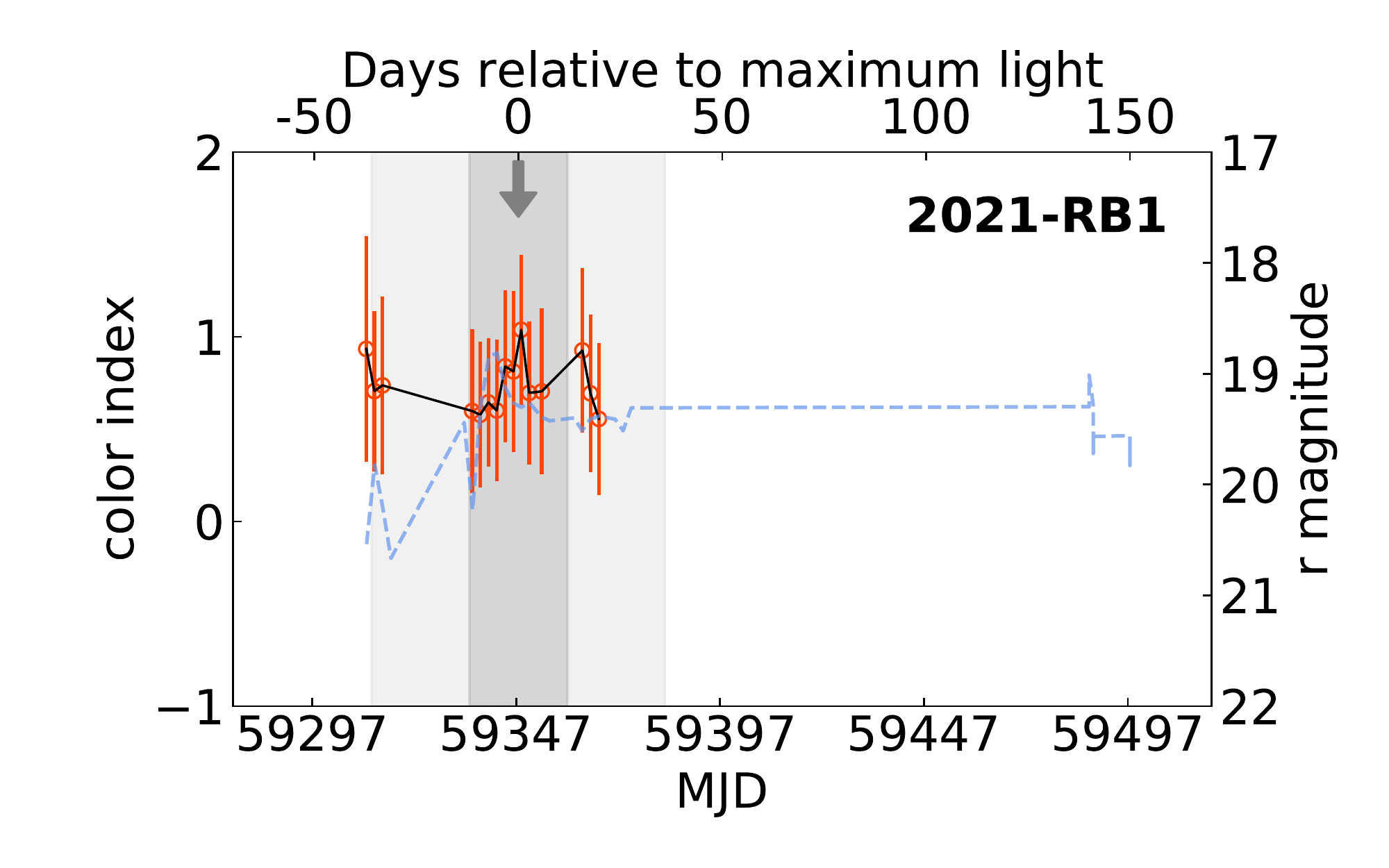}
    \end{minipage} 
     \begin{minipage}[t]{.33\textwidth}
        \centering
        \includegraphics[width=\textwidth]{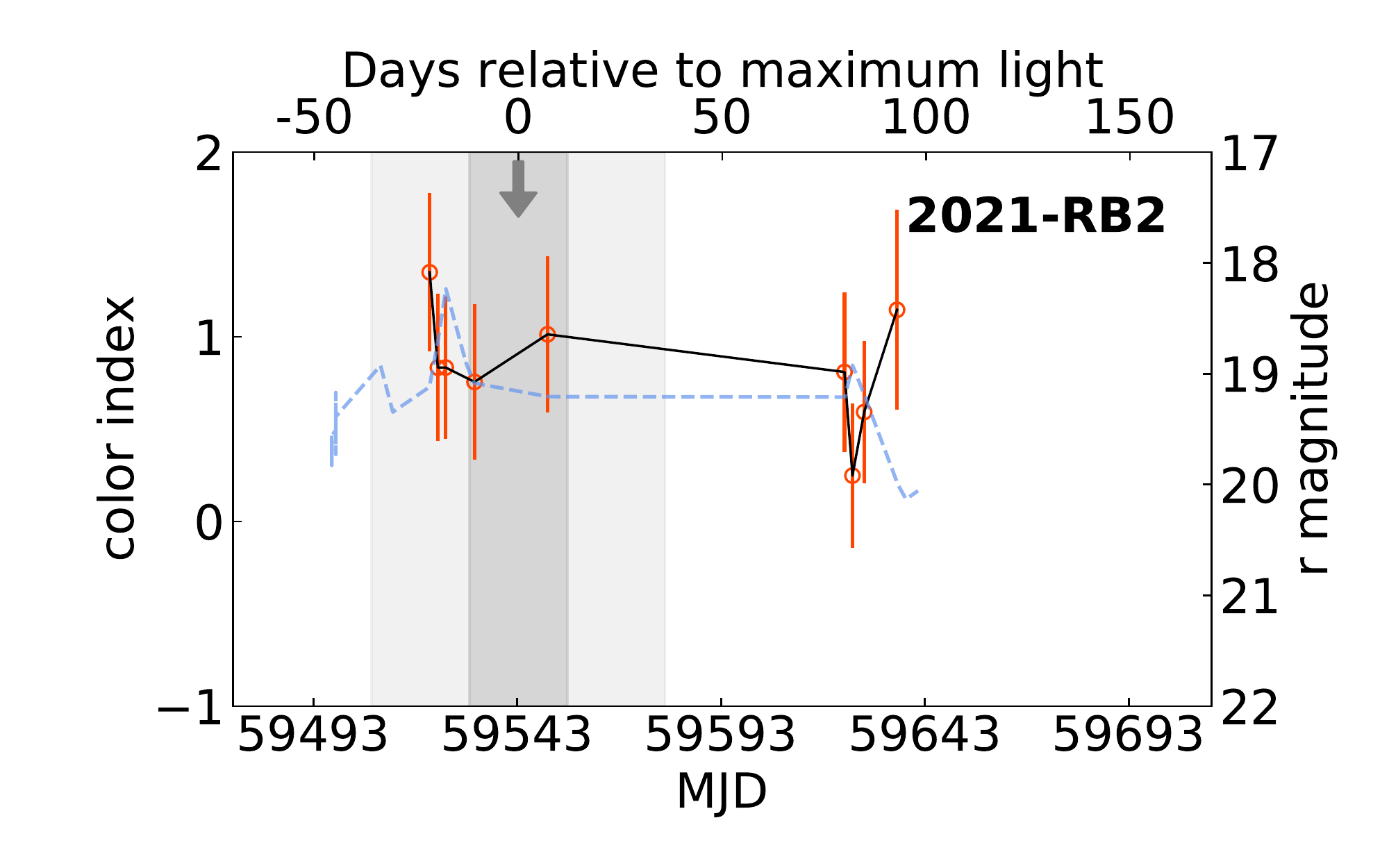}
    \end{minipage} 
    \caption{Cutouts of the color evolution shown in Figure \ref{fig:colorcurve_full} for each outburst with available data. {Each cutout shows the time interval from 50\,days prior to the estimated time of maximum light based on Equation \ref{eq:mean_period} to 150\,days afterward. The respective time of outburst is marked by an arrow (black for known outbursts and gray for suspected and newly detected ones) and the gray shaded area corresponds to its uncertainty in the same way as in Fig. \ref{fig:colorcurve_full}. Orange data points correspond to $g-r$, green to $g-i,$ and purple to $r-i$}. The color with the best sampling is connected by a solid black line. The blue dashed line shows the r-band light curve for comparison.} \label{fig:outbursts_colorcuts_with_r}
\end{figure*}

Figure \ref{fig:outbursts_colorcuts_with_r} shows the color evolution (solid line) of AT~2000ch with respect to the corresponding measured or suspected visual maximum (marked by arrows) compared to the $r$-band light curve (dashed line). Each cutout shows the time interval from 50\,days prior to the assumed peak luminosity to 150\,days afterwards. 
We mention two cautionary remarks. First, any brightness variations of the transient on the order of hours will introduce an additional random error for the computed color indices not reflected by the error bars in Figures \ref{fig:colorcurve_full} and \ref{fig:outbursts_colorcuts_with_r}. Second, \citet{Pastorello2010} showed that both the continuum flux in $r$-band and the H$\alpha$ flux can vary considerably. This degeneracy makes any inference from a color involving the $r$-band somewhat problematic. The $g$-band contains only minor contamination from the weaker H$\beta$ line, and the $i$-band does not cover any strong stellar emission lines.

\textbf{\textit{r - i}} (purple data points in Fig. \ref{fig:outbursts_colorcuts_with_r}): We detect a qualitative tendency for the $r-i$ color curve to be correlated with the light curve, meaning that brighter phases of the transient tend to coincide with larger (redder) $r-i$ values. This can be seen, for example, during 2008-OT, 2009-OT1, and for 2009-OT2 up to the visual maximum around $MJD = 55155$. However, there are exceptions to this trend. For example, after the visual maximum of 2009-OT2 at $MJD > 55205,$ the color remains on a flat plateau of $r-i \approx 0.0$ while the light curve is dimming on average, modulated by variations of $\Delta r \leq 0.5\,\textrm{mag}$, and during the minimum of the light curve following 2008-OT around $MJD = 54794$ we observe a sharp reddening in color from $r-i = -0.8$ to $-0.2$ within a few hours without a corresponding change in the light curve.
$r-i$ mostly varies between values of $0.0$ and $-0.7$, though larger values of up to $0.3$ have been observed a few days after maximum light of 2000-OT and 2019-OT1, as well as during the dimming $\geq 50\,\textrm{days}$ after 2009-OT2.

\textbf{\textit{g - r}} (orange/yellow data points in Fig. \ref{fig:outbursts_colorcuts_with_r}): Conversely to the behavior in $r-i$, there are multiple instances where the $g-r$ color is qualitatively anti-correlated with the light curve; for example {2018-RB1}, 2019-OT1, {2020-RB1, 2020-RB2,} {and 2021-RB2.} However, counterexamples can also be found, such as the sparse available data for 2009-OT2 and {2019-RB2}, and the two very brief and deep minima in the color curve shortly before {2020-RB1} at $MJD=58909$ and during the dimming $93\,\textrm{days}$ after {2020-RB2} at $MJD=59227$. In these latter two cases, $g-r$ falls by $1.5$ $(1.0)\,\textrm{mag}$ for a brief interval of $5-10\,\textrm{days}$, while the $r$-band light curve evolves relatively smoothly. This rapid change in color is therefore caused by briefly enhanced fluxes in the $g$-band (plausible explanations for this are discussed in Sect. \ref{sec:disc}). These minima at $g-r \approx -0.4$ and $-0.2$ are also unusually deep considering that the transient mostly varies within the interval $0.0 \leq g-r \leq 1.0$. Also, we rarely observe values above that range. Between $30$ and $50\,\textrm{days}$ after 2008-OT, for example, $g-r$ reaches $\approx1.2,$ but in this instance the data points are consistent within the nominal errors with the slightly more moderate behavior $30-50\,\textrm{days}$ after 2019-OT1, and so there is no clear evidence for any extraordinary behavior.
\\
In much the same way as the light curve is somewhat different for each eruptive event, we do not find any regular behavior of the color curve from one outburst to another, although the average sampling rate of color space in our data is likely too sparse to make a strong assertion on this point. We have a sufficient sampling rate for a convincing argument only for 2018-OT$2^*$ and {2020-RB2}, but there the data cover different phases relative to the assumed optical maximum.

\textbf{\textit{g - i}} (green data points in Fig. \ref{fig:outbursts_colorcuts_with_r}): The $g-i$ coverage in our data is too sparse for a detailed analysis; there are a total of 11 data points spread out over three eruptive events. However, in general, $g-i$ seems to behave similarly to $r-i$. The main difference is a roughly constant offset of $\sim 1\,\textrm{mag}$ most likely caused by the excess H$\alpha$ emission contained in the $r$-band.

\subsection{Variability in the radio range}\label{sec:radio_var}
\renewcommand{\arraystretch}{1.2}
\begin{table*}[ht]
        \caption[Detection properties]{Radio data sets of AT~2000ch.}
        \centering
        \begin{tabular}{M{2.9cm} M{0.7cm} M{2.7cm} M{2.7cm} M{2.9cm} M{2.7cm}}
                \hline\hline
                Project ID & Array      & Date (dd.mm.yyyy) &   Frequency \( \nu \) (GHz)         & Radio flux \(S_{ \mathrm{\nu} } \) ($\mu$Jy)  & 1$\sigma$-noise ($\mu$Jy)\\
                \hline
                AN0025  & C & 09.04.1984 & 1.417 &  $<849$ & 283 \\
                AI0023  & C & 27.07.1985 & 1.46-1.51 & $<140$ & 47 \\
                AB628 & B & 17.07.1994 & 1.36-1.43 & $<15$ & 5 \\
        AW393   & C & 28.11.1994 & 1.417-1.418 & $<651$ & 217 \\
                WSRT & - & 09.02.1996 & 1.418-1.421 & $<5400$ & 1800 \\
                AI0073  & A     & 13.05.1998 & 1.39-1.46 & $<207$       & 69 \\
                \multirow{5}{*}{\parbox{1.6cm}{10C-119 (\textsc{chang-es})}} & D & 17.12.2011 & 4.98-6.89 & $53.3 \pm 9.8$ & 8.9 \\
                 & D & 18.12.2011 & 1.25-1.89 & $<189.9$ & 63.3 \\
                 & C & 13.02.2012 & 4.98-6.89 & $50.2 \pm 4.5$ & 3.7 \\
                 & C & 25.03.2012 & 1.25-1.89 & $163 \pm 34$ & 32       \\
                 & B & 24.06.2012 & 1.25-1.89 & $111 \pm 24$ & 22.6 \\
                \multirow{2}{*}{14B-150} & \multirow{2}{*}{C} & \multirow{2}{*}{01.12.2014} & 1.01-1.97 &  $<82.7$ & 27.6 \\
                & & & 4.49-6.38 & $<5.5$ & 1.8 \\
                LOFAR - DR2  & HBA & 23.12.2018 & 0.145 & $<390$ & 130 \\
                \hline
        \end{tabular}
        \label{tab:radio_measurements}
        \tablefoot{Columns from left to right are: project ID, telescope array configurations, frequency range in GHz, the flux of AT~2000ch or a 3$\sigma$ upper limit in $\mu$Jy, and the 1$\sigma$-noise level in $\mu$Jy.}
\end{table*}
\renewcommand{\arraystretch}{1}
\begin{figure*}
        \centering
        \includegraphics[width=1.0\textwidth]{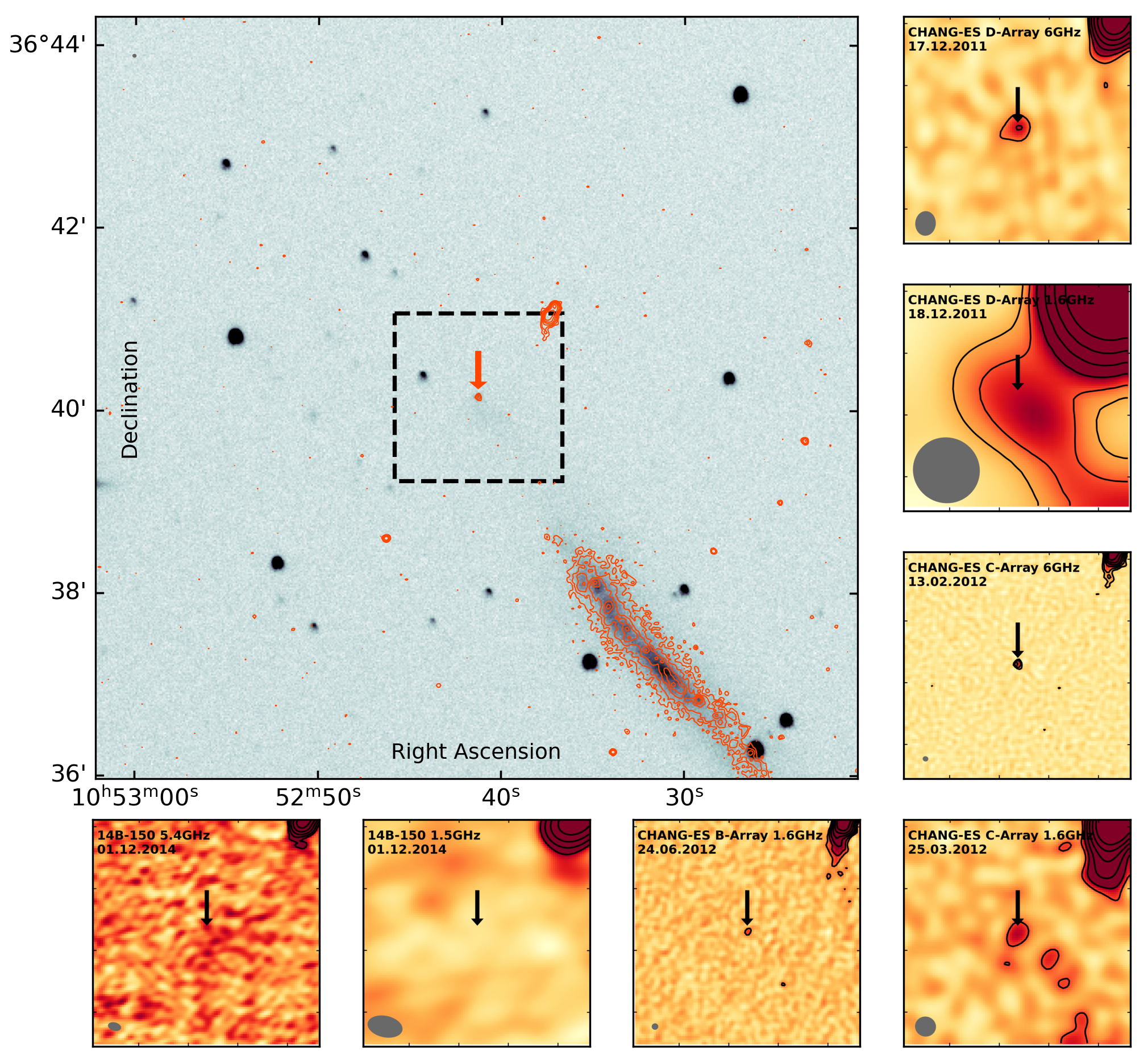}
        \caption{Detections and examples of nondetections of AT~2000ch in radio continuum data. The large panel shows the complete ZTF r-Band frame ($500^{\prime\prime} \times500^{\prime\prime}$). The red and black arrows point to the position of AT~2000ch: \sdsspos. Small panels show $110^{\prime\prime} \times110^{\prime\prime}$ cutouts (indicated as black box in the large panel) of radio maps listed in Table \ref{tab:radio_measurements} (\changes and 14B-150). As listed in Table \ref{tab:radio_measurements}, we report detections in the \changes D-Array C-Band maps, C-Array C-Band  and L-Band maps, and B-Array L-Band maps.}
        \label{fig:radio_detail}
\end{figure*}

Figure \ref{fig:radio_detail} presents the radio continuum detections of AT~2000ch in different frequency bands, telescope configurations, and epochs. For reference, we also show the nondetections during two different epochs, which we discuss in more detail in Section \ref{sec:disc} for the analysis of the variability of AT~2000ch in the radio range.  
We find the continuum emission to be point-like in the case of each detection (compare the black contour next to the black arrow with the resolution of the data shown by the gray beam in the bottom left corner in Fig. \ref{fig:radio_detail}).  The extent is generally found to be slightly larger than one beam which is an indication of a more extended radio emission surrounding AT~2000ch.
Moreover, we find the radio emission to be mostly isolated from the emission of the galaxy and other possible radio sources in the galaxy outskirts where AT~2000ch lies (see e.g., the large panel in Fig. \ref{fig:radio_detail} showing the radio contours for the whole ZTF $r$-Band frame) except for the array configuration with the lowest resolution (D-Array 1.6\,GHz).

\begin{figure*}[ht]
        \centering
        \includegraphics[width=1.0\textwidth]{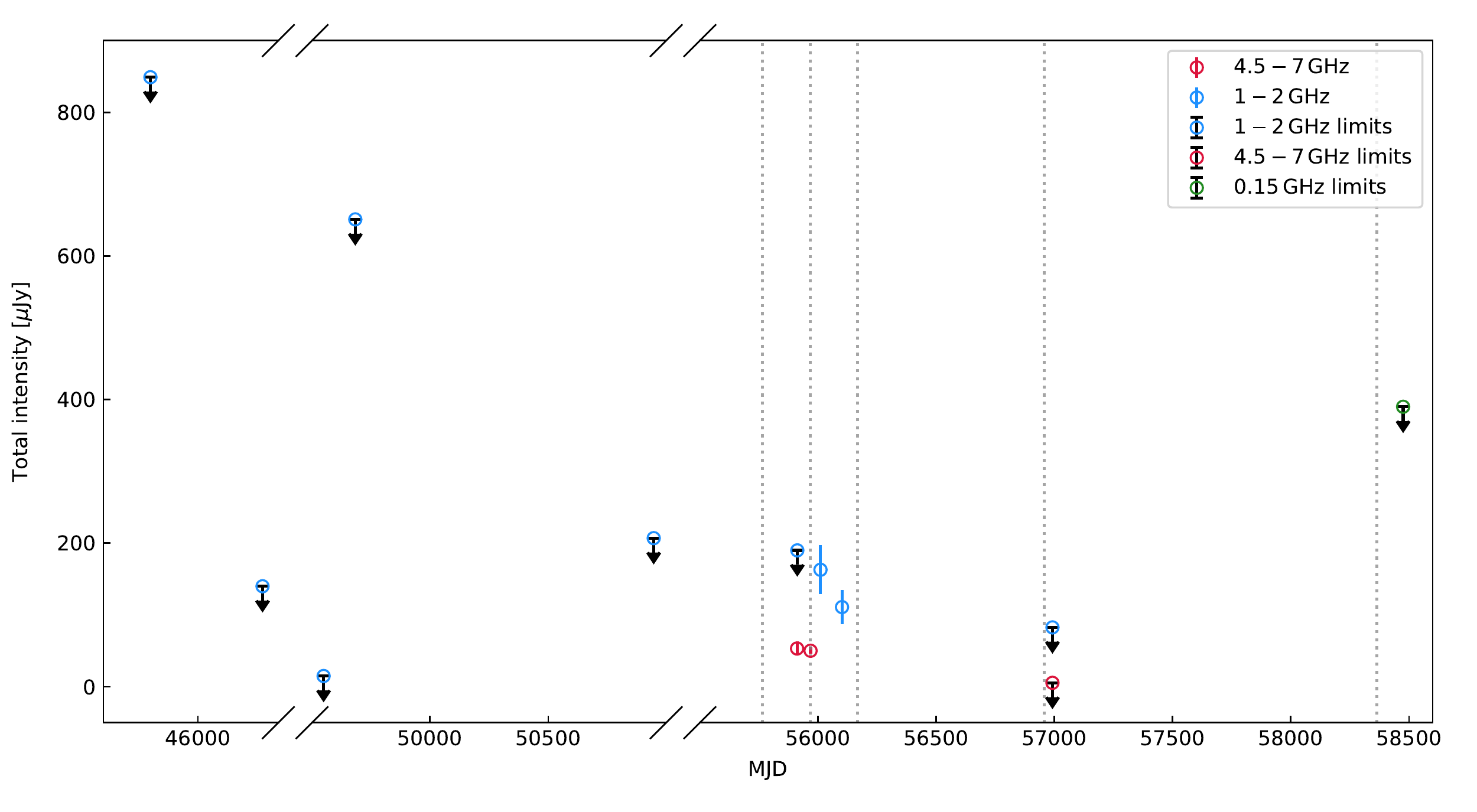}
        \caption{Radio light curve of AT~2000ch.  Observations at different frequencies are merged in broader bands. Predicted optical outbursts are marked by the gray dotted lines for orientation. The observation from the WSRT is not shown for visual purposes.}
        \label{fig:lightcurve_radio}
\end{figure*}

We analyzed 14 different measurements to investigate the radio emission of AT~2000ch between 1984 and 2018 (see Table \ref{tab:radio_measurements} and Fig. \ref{fig:lightcurve_radio}). We did not detect AT~2000ch between 1984 and 1998; based on the measurement from 1994 (AB628) in particular, we find the radio emission to be very low ($<15\,\mu$Jy). The next observations were performed between 2011 and 2012 as part of the \changes survey. Here, we detect AT~2000ch four times in two different frequency bands and three different array configurations. {The resulting resolution of the sources and the S/N ratio varies with the frequency and array configuration and therefore affect the detection of our source.} We find a total intensity of about 50\,$\mu$Jy at 6\,GHz {in} December 2011 and February 2012 (for exact values, see Table \ref{tab:radio_measurements}). Furthermore, we find the total intensity to reach $(163\pm34)\,\mu$Jy and $(111\pm24)\,\mu$Jy at 1.6\,GHz at the end of March 2012 and at the end of June 2012, respectively. While the absolute values indicate a variability within these 90\,days, this variable is within the uncertainties. Both estimates are nevertheless significantly higher than the upper flux limit found before 2000-OT, especially compared to 1994. A possible variation between the detections can be evaluated by estimating the spectral index between the different frequencies, which we discuss in the following section. Nevertheless, a key result demonstrating the variability in the radio range after 2000-OT is the observation from 1 December, 2014. The two measurements at 1.5\,GHz and 5.5\,GHz provide upper limits of 83\,$\mu$Jy and 5.5\,$\mu$Jy, respectively, which are significantly lower than the detected flux values stated above. We find AT~2000ch to produce less radio continuum emission in the {re-brightening} event of 2014 than in the epochs of 2011 and 2012.
\subsection{Spectral index}
The origin of the radio emission can be further investigated using the spectral index ($S\propto\nu^\alpha$ with $S$ as radio brightness, $\nu$ the corresponding frequency, and $\alpha$ the spectral index). Emission from thermal plasma exhibits an intrinsic spectral index of $\alpha=-0.1$ while relativistic electrons gyrating in a magnetic field show values of $\alpha\leq-0.7$. Steeper values are generally associated with an energy loss of electrons during propagation, which is also referred to as aging. Additional effects leading to flatter spectral indices can be caused by absorption effects such as free-free absorption by thermal electrons, which are usually expected in the MHz and low GHz regime.
As described in Section \ref{sec:SIdata}, we performed three measurements to study the spectral index of AT~2000ch during different phases, one determined from two different frequency observations performed on consecutive dates and two in-band: The first measurements were performed on {17 December, 2011.} While splitting the C-band we found the noise ratio to be too low in the higher frequencies for a direct detection. Nevertheless, we find an upper limit of $S_\textrm{high}<72\,\mu$Jy and $S_\textrm{low}=89.7\pm5.8\,\mu$Jy at 6.027\,GHz and 5.322\,GHz, respectively, restricting the spectral index to $\alpha < -1.76$. {Measurements were also performed on 17 and 18 December, 2011.} Based on the upper limit found for AT~2000ch on the 18 December, we are able to restrict the spectral index to flatter than $-0.9$. Taking the diffuse emission of the host galaxy into account, which contributes to the upper limit measurement (see shape of the radio contours in Fig. \ref{fig:radio_detail}), we expect the spectral index to be reasonably flatter. To reach a purely thermal value for the spectral index of -0.1, AT~2000ch should only reach a continuum emission of about 60\,$\mu$Jy, which might be the case. {The last measurement of the three was made on 13 February, 2012.} By splitting the broadband \changes C-array data at 6\,GHz into two, we measured peak fluxes of $S_\textrm{high}=33.6\pm5.4\,\mu$Jy and $S_\textrm{low}=67.4\pm5.8\,\mu$Jy for frequencies of 6.508\,GHz and 5.490\,GHz, respectively. This results in a value of $\alpha=-2.79$. 

Our findings indicate strong energy losses due to synchrotron emission; however, optically thick ejecta material can also cause such steep spectra (see Sect. \ref{sec:disc}).
Several factors can influence the spectral index measurements. Even though we excluded the short baselines for our measurements of the spectral index where possible (see Sect. \ref{subsect_radiocont}), a contribution from the diffuse emission of the galaxy cannot be completely excluded. In addition, a variability within the time-frame of the observations cannot be ruled out, which would lead to unpredictable spectral indices. Our optical light curve (Fig. \ref{fig:colorcurve_full}) shows changes over an order of more than a magnitude within several hours. To investigate a possible change of the spectral index within the timescale of several hours, we tried to image the \changes data sets with time-frames of two hours. Unfortunately, the (u,v)-coverage became too sparse to reliably generate an image where the flux of the source could be measured.

\section{Discussion} \label{sec:disc}

\subsection{Evolution of AT~2000ch}
We report {2} additional outbursts of AT~2000ch and {13 significant re-brightening events} resulting in a total of {6 detected outbursts of which 4 were reported in the} literature by \cite{Wagner2004} and \cite{Pastorello2010}.
We find a predicted period of roughly {201\,days (see Eq. \ref{eq:mean_period}), taking consecutive outbursts and re-brightening events into account (see Table \ref{tab:outburst_timing}), which is stable over the measured timescale of the light curve. The period agrees with the prediction of \cite{Pastorello2010}.}
This latter authors were not able to observe such a period over two decades in the previously mentioned candidates for 2009ip analogs (compare, e.g., \citet{Pastorello2013, Thoene2017, Pastorello2018, Pastorello2019}), {but the cadence of their light curve is significantly lower than in the case of AT~2000ch.} 
Based on the periodicity, we assume that AT~2000ch has undergone several epochs of eruptive outburst or {significant re-brightening events} of the length of {roughly 201}\,days for each epoch within the past 20 years. We additionally assume a comparable brightness of AT~2000ch during outburst, which we find to be reasonable based on the findings from 2000-OT, 2008-OT, 2009-OT1, 2009-OT2, 2013-OT, and 2019-OT1 (see Sect. \ref{sec:lightcurve}).

Looking at the sufficiently sampled light curves from 2000-OT; 2008-OT to {2009-OT2; 2010-RB; and 2019-OT1 to 2021-RB2,} we find common characteristics, even though their overall shape is not easy to compare, mostly because of the sampling but also because AT~2000ch is reaching its minimum at a different point in time after the outburst {or significant re-brightening} (e.g., only a few days passed in the case of 2000-OT, but some weeks passed in the case of the latest reported re-brightening events). 
We generally find a change of brightness of $1-2$\,mag (and sometimes even more) within a few days, arising multiple times within $\sim 50$\,days after the outburst. Thereafter, the light curve shows a smoother trend, but a change of 1\,mag is still frequent. The coverage of the light curve before the {maximum brightness} is low in most of the epochs, but from 2009-OT1, 2019-OT1, and 2020-RB1, we argue that a few days before AT~2000ch reaches its maximum the brightness increases rapidly, including strong fluctuations ($\Delta r \approx 2$\,mag, see 2009-OT1 in Fig. \ref{fig:cutouts}) within several hours or a few days. The minimum is reached in 2019-OT1 and more recent outbursts are found to be less faint, which could be the result of the combination of the telescope sensitivity and the sampling of the light curve.

Remarkably, we find the latest {re-brightenings}, {2020-RB2 to 2021-RB2,} to be clear exceptions when compared to the other {events, and the differences} cannot be traced back to the sampling of the light curve. The light curve is found to evolve much more smoothly within the first few weeks after the {assumed} outburst. We do not detect large-scale fluctuation on short timescales. The brightness decreases slowly by only $\approx 0.5\,$mag, re-brightens by an even smaller amount, and then decreases steadily thereafter. We did not detect such a calm evolution in other {events} so shortly after the {transient reached its maximum brightness.} 
{We note that the transient might already have experienced such a phase before, where we did not detect any eruptive outburst ($2010-2012$ and $2014-2018$), but this could also be an effect of the low cadence of the light curve and is therefore neglected in the following discussion. Additional observations from that time are needed to shed light on this particular scenario.}

We find qualitative evidence that the $r-i$ color tends to be correlated to the $r$-band light curve, meaning that phases of higher luminosity coincide with redder colors. This is consistent with the finding of \citet{Pastorello2010} that the transient displays lower continuum temperatures ---and therefore also cooler temperatures--- during outburst than during quiescence. 
However, in $g-r$, the transient shows mostly the inverse: low (blue) $g-r$ values during maxima in the light curve and large (red) values during minima. In both cases, the respective color approaches a value of $\approx 0$ at maximum light. This behavior can be explained if the H$\alpha$ flux causes a smaller relative excess in the $r$-band during outbursts than during quiescence. Alternatively, the transient would need to have changed in structure between the years 2009 (the end of regular $r-i$ coverage in the data) and 2018 (the beginning of regular $g-r$ sampling), in some way that resulted in the inversion of its color evolution over the cycle length.

Other noteworthy features in the color evolution were the extremely brief, deep minima in $g-r$ prior to {2020-RB1} and following {2020-RB2}. The former took place during a short episode of fast fluctuation in the light curve of about $\Delta r \approx 0.5\,\textrm{mag}$, and so it is possible that this anomalous color value is in part caused by an intrinsic change in brightness between the times of measurement in the $g$- and $r$-filter. However, the latter is clearly caused by a brief period of enhanced flux in $g$ (see bottom left panel in Fig. \ref{fig:cutouts}).
One plausible scenario explaining this could be that the line of sight to the star incidentally passes through a cavity in the circumstellar envelope. In that case, we would expect to measure larger amounts of blue continuum flux, while the $r$-band would remain dominated by the surface of the circumstellar envelope. Alternatively, the $g$-band may have been briefly enhanced by emission line fluxes. \citet{Pastorello2010}, for example, noted the variable presence of the He\,II\,$\lambda4686$ emission line. 
We can estimate the necessary excess flux if we assume the color remains largely constant during the steady dimming in $r$ at $MJD \approx 59227$ following {2020-RB2}. In that case, the excess enhanced the source brightness from $m_{g,0} = 20.95\,\textrm{mag}$ to $m_{g} = 19.72\,\textrm{mag}$, corresponding to a mean flux density of $\tilde{f}_g \approx 32 \times 10^{-29}\,$erg\,s$^{-1}\,$Hz$^{-1}\,$cm$^{-2}$ over the width of the band. Using an effective width of {$W_\textrm{eff} = 1064.68\,$\r{A} and central wavelength $\lambda_\textrm{eff} = 4671.78\,$\r{A} taken} from the SVO Filter Profile Service \citep{rodrigo2012, rodrigo2020}, we can estimate the integrated excess flux via
\begin{equation}
    F \approx \tilde{f}_g \frac{c\, W_\textrm{eff}}{\left(\lambda_\textrm{eff}-\frac{W_\textrm{eff}}{2}\right) \left(\lambda_\textrm{eff}+\frac{W_\textrm{eff}}{2}\right)} \approx 4.7 \times 10^{-14}\,\frac{\textrm{erg}}{\textrm{s\,cm}^2},
\end{equation}
where $c$ is the speed of light in a vacuum. For comparison, \citet{Pastorello2010} measured a similar value of $4 \times 10^{-14}\,$erg$\,$s$^{-1}\,$cm$^{-2}$ for the average H$\alpha$ flux during the 2008-OT event, and \citet{Wagner2004} measured half that value for H$\alpha$ during 2000-OT. 
We find it unlikely that a sudden increase in emission line fluxes in the $g$-band can explain this excess{, as the strongest line} in this spectral range is typically H$\beta$, which is always weaker than H$\alpha$, and especially weaker than the H$\alpha$ flux measured during the optical maximum. In fact, the $r$-band brightness is steadily decreasing during this time, suggesting rather a decrease in the Balmer series fluxes than a sudden increase. The [O\,III]$\,\lambda\lambda\,4959,5007$ doublet should be absent, because the circumstellar material will be too dense for its production. This leaves mostly He\,II\,$\lambda4686$ as a major contributor to the excess, which is far too weak to cause the necessary amount of flux.
If this excess is instead caused by a serendipitous crossing of a cavity in the circumstellar envelope over the line of sight, this might suggest the star experienced lower mass-loss rates in the most recent eruptions than in previous ones. Together with the much smoother light curve evolution of {2020-RB2}, a case can be made that the transient may be currently transitioning into a period of relative calm.

\subsection{AT~2000ch, a double source?}

While identifying our measurements with the reported positions of AT~2000ch in the literature by taking ground- and space-based images, in two independent  measurements in different filters of the HST  we find a second source located southeast of the bright spot that we classified as AT~2000ch. The sources are separated by a projected angular distance of $0.5^{\prime\prime}$, which translates into a physical distance of $\textrm{d} \approx 23$\,pc assuming that the second source is located at the same distance as AT~2000ch (not a foreground or background source). At such large physical distances, we can assume no physical interaction between the two sources; for example, strong stellar winds would build an envelope and expand further on only subparsec scales. However, with ground-based photometry and spectroscopy being limited by seeing, they typically reach a resolution of $\approx 1^{\prime\prime}$ or even worse. With the reported angular distance, we find that the measurements made so far as part of this study, but also those of \cite{Wagner2004} and \cite{Pastorello2010}, favor the superposition of these two unrelated sources.

To analyze the influence of the companion source on the photometric measurements, we estimate the flux ratios of AT~2000ch and the second source using aperture photometry on the Hubble images. We report the following flux ratios (R=$f_{\textrm{SN2000ch}}/f_{\textrm{2ndSource}}$) on 22 October, 2016: R\textsubscript{F555W}$\approx$9;  R\textsubscript{F814W}$\approx$5. Based on our photometric analysis, we assume that AT~2000ch reached its maximum in the middle of June, 2016, and therefore we expect to measure such ratios between the two objects in the more quiescent phase of AT~2000ch. During the different epochs, we find a mean brightness of $r\approx20.0$\,mag at this phase, which resembles the combined brightness of the two objects. Under the assumption that the flux relation is only about 5 (worse case) between the two objects in $r$-band, we find a brightness of $r\approx20.2$\,mag and $21.9$\,mag for AT~2000ch and the second object, respectively. AT~2000ch is therefore found to also be the dominant source in quiescence and is only marginally less bright than the measured combined magnitude. The contribution of the second source increases the total brightness only by 0.2\,mag, which still lies within the uncertainties of most data points. Moreover, this effect becomes even smaller during outburst and re-brightening. The contribution of the second source can therefore be neglected in the photometric study of AT~2000ch. 

In radio continuum, we find no sign of a double source. Based on the clearly identified peak emission in radio continuum with the position of AT~2000ch in the HST image, radio continuum emission of the second source would be expected to produce a southeastern elongation in the radio contours. We identify no such shape or relocation of the continuum peak towards the second source and therefore conclude that our radio continuum findings are not affected by it.

In their spectral analysis, \citet{Pastorello2010} discern two distinct types of spectral behavior from AT~2000ch: near the outburst the transient displays a bluer continuum and emission lines of H, He\,I, and Fe\,II with prominent P Cygni profiles and a slower $v_\textrm{FWHM}$ of $1600-2200\,$km\,s$^{-1}$, while during quiescence the continuum is shifted redwards, P Cygni profiles are absent, He\,II\,$\lambda4686$ appears in emission and line widths are generally broader ($v_\textrm{FWHM} = 2400-2800\,$km\,s$^{-1}$). From the fact that a spectrum taken at 6 May, 2000, during the deep minimum following 2000-OT is quantitatively similar to one taken on 24 April, 2009, near maximum light of 2009-OT1, it becomes unlikely that the companion in the HST image dominates the total flux of the composite source at any phase of the transient. Still, at broadband flux ratios of $R \leq 10$ some of the spectral features of AT~2000ch and its SED may have been modulated by the companion source to a measurable degree. High-resolution spectra at different phases are needed to study the contribution.

\subsection{Origin of the radio variability}

We analyzed six individual measurements before the first discovered and subsequently verified outburst in the optical regime in 2000 and detect no radio continuum emission at the location of AT~2000ch. For instance, in 1994 we found the continuum emission to be lower than 15\,$\mu$Jy at 1.4\,GHz. Only a few months before, AT~2000ch was detected in $r$-band at a brightness comparable to its minimum brightness reached during the several eruptive phases after 2000 ($r\approx21$\,mag at $MJD=49474$) that are accompanied by strong stellar winds. We therefore find it likely that AT~2000ch (1) was not accompanied by a strong stellar wind during this time and (2) had no intrinsically strong magnetic field; otherwise we would have observed it with the sensitivity reached in 1994.
Based on the four detections between 2011 and 2012, we find that AT~2000ch reaches radio continuum emission of about several tens to a few hundred $\mu$Jy depending on the observed frequency. We find the continuum emission of AT~2000ch at 6\,GHz to be within the limits found for SN~2009ip \citep[($<66\,\mu$Jy at 18\,GHz)][]{Hancock}. We expect that with a higher S\textbackslash N,\ SN~2000ip would have been detected in radio continuum taking the low emission values of AT~2000ch into account and the larger distance of SN~2009ip. 

Does the radio continuum variation coincide with the optical activity and might it therefore be caused by the thermal contribution due to stellar winds? Those winds would be expected to increase during sufficient brightness variations, heating the surrounding gas, and also radiating in radio continuum.
To answer the first part of the question, we can compare the radio continuum detections to the optical light curve. Unfortunately, even though we investigated several new epochs of data points complementing the analysis done by \cite{Wagner2004}, \cite{Pastorello2010}{, and \cite{Pastorello2013},} {there were only a few observations performed between the end of 2011 and the middle of 2012} (see orange circles around $MJD=56000$ in Fig. \ref{fig:lightcurve_full}, where radio observations are indicated by the blue dotted lines), where we detected AT~2000ch in the radio range. {Also taking into account} the {tentatively} stable periodicity of AT~2000ch of about {201}\,days, we can assume that AT~2000ch has undergone {at least a re-brightening event at $MJD\approx 55766,55967,56168,56957,58364$} (indicated by the gray dotted lines in Fig. \ref{fig:lightcurve_radio}). 
We therefore assume the emission from mid-December, 2011, at 6\,GHz was measured at a time where AT~2000ch is expected to have had a comparably {lower brightness with respect to its maximum} ($\approx 150$\,days after {the predicted maximum brightness}) while the emission from February, 2012, belongs to a phase where AT~2000ch is expected to show extreme {or moderate (in case of the re-brightening events)} variations (only a few days after {the predicted} {maximum brightness}). We find the continuum emission in both cases to be about $50\,\mu$Jy. Therefore, we cannot verify a clear correlation of the intensity of the radio continuum emission with the optical variation on comparable timescales while only taking its emission into account (see spectral index discussion below and the determination of the optical depth of the ejecta in Sect. \ref{sec:massloss} for a clearer picture).
Looking at both detections at 1.6\,GHz, one at the end of March, 2012, and the other at the end of June, 2012, we find them to be in agreement with a {less bright and also possible calmer phase of the optical transient} expected from the sufficiently sampled light curves at different epochs and the few detections during that time. Even though the total intensity at the end of June coincides with the uncertainties measured in March, the decreasing trend is in agreement with the transient becoming less active and therefore coincides with an envelope of gas becoming cooler. Remarkably, we did not detect AT~2000ch in 2014 at $MJD=56992$, even though only 40\,days earlier we detected a significant re-brightening of the transient in $g$-filter. AT~2000ch is therefore assumed to have gone through a {phase of significant re-brightening} in the optical regime but producing no significant continuum emission ($<5.5\,\mu$Jy and $<82.7\,\mu$Jy at 5.5\,GHz and 1.5\,GHz); we also found the optical depth of the ejecta to decrease (Sec. \ref{sec:massloss}) and we therefore conclude AT~2000ch to be less bright in 2014 with respect to the radio continuum detections from 2011 and 2012. {The latest radio continuum detections agree with the reported phase of moderate re-brightening events between 2014 and 2018.}

To generally evaluate whether the origin of the continuum emission is caused by a thermal contribution originating from sufficient stellar winds heating the surrounding gas, we studied the spectral index where possible.
We generally find a steeper spectral index, as expected for a thermal origin: $\alpha\approx-2$ evaluated from an in-band analysis on 17 December, 2011, where we {expect} AT~2000ch to {be less bright,} and 13 February, 2012, where we expect AT~2000ch to be highly active (shortly after {maximum brightness}). Values of $\alpha\leq-0.7$ usually indicate a population of ageing electrons. Spectral indices as steep as our values are usually detected far away from the site of origin such as galactic halos, lobes of active galactic nuclei, or diffuse cluster remnants, where electrons had enough time to lose their energy. In terms of compact objects, neutron stars are known to show very steep spectral indices. Here electrons can lose energy very quickly within a small path length due to the strong magnetic field in and close to these objects. So far, it has not been possible to verify very strong magnetic fields for SNe or SN imposters in general. Also taking our findings from the spectral index analysis between the 17 and 18 December, 2011, into account, with $\alpha >-0.9$ (here, we would reach a spectral index indicating a thermal origin with $S_{1.5}\approx60\,GHz$, which could be possible) we find a different approach to be likely: the optical depth of the ejecta significantly increases when the mass-loss rate in a certain time interval reaches sufficient values. In that case, the radio-continuum-emitting electrons cannot leave the ejecta easily and lose energy efficiently during propagation, resulting in steep spectral indices. In the following section, we estimate the mass-loss rate of AT~2000ch and evaluate the optical depth of the ejecta with respect to the free-free emission. We find this to be a dominant factor at least within the first weeks after outburst. Nevertheless, additional synchrotron losses cannot be ruled out and need to be studied in the future.

\subsection{Constraints on the mass-loss rate} \label{sec:massloss}

\renewcommand{\arraystretch}{1.2}
\begin{table*}[ht]
        \caption{Mass-loss rates of AT~2000ch at different episodes derived from optical and radio continuum observations.}
        \centering
        \begin{tabular}{lrrrrrrrrrr}
            date & $L_{\textrm{H}\alpha}$ & $S_\nu$ & $\nu$ & $v_{\textrm{w}}$ & $t$ & $r_{15}$ & $\beta$ & $\dot{M}_\textrm{(\ref{eq:Ofek})}$ & $\dot{M}_\textrm{(\ref{eq:klein})}$ & $\dot{M}_\textrm{(\ref{eq:massloss_radio})}$\\
         (dd.mm.yyyy) & [erg/s/cm$^2$] & [mJy] & [GHz] & [km/s] & [days] & [cm] & & [M$_\odot$/yr] & [M$_\odot$/yr] & [M$_\odot$/yr]\\
                \hline\hline
                01.06.2000$^{*}$ & $2\times10^{-14}$ & & & 1800 & 25 & 0.58 & 8073 & $2.8\times 10^{-5}$ & $<7.7\times 10^{-5}$ & \\
                16.10.2008$^{*}$ & $4\times10^{-14}$ & & & 2300 & 9 & 0.33 & 4593 & $5.1\times 10^{-5}$ & $<1.1\times 10^{-4}$ & \\
                12.01.2016 & $2.7\times10^{-15}$ & & & 2000 & 50 & 1.13 & 15729 & $1.1\times 10^{-5}$ & $<2.8\times 10^{-5}$ & \\
                17.12.2011 & & 53 & 6 & 2500 & 160 & & & & & $<4.3\times 10^{-4}$ \\
                13.02.2012 & & 50 & 6 & 2000 & 15 & & & & & $3.1\times 10^{-6}$\\
                25.03.2012 & & 163 & 1.5 & 2000 & 55 & & & & &$2.0\times 10^{-6}$\\
                24.06.2012 & & 111 & 1.5 & 2500 & 145 & & & & &$<1.9\times 10^{-5}$\\
                01.12.2014 & & $<83$& 1.5 & 2000 & 40 & & & & &$>1.2\times 10^{-6}$\\
                01.12.2014 & & $<5.5$ & 5.5 & 2000 & 40 & & & & &$>3.5\times 10^{-5}$\\
                \hline
        \end{tabular}
        \label{tab:massloss}
        \tablefoot{Columns from left to right: observation date, where the asterisk marks the central date of a consecutive measure, H$\alpha$ luminosity $L_{\textrm{H}\alpha}$, radio flux $S_\nu$, frequency $\nu$, wind velocity $v_\textrm{w}$, elapsed time from latest outburst $t$, radius of the recently ejected material in $10^{15}$\,cm, $\beta$, and the mass-loss rates derived via Equations (\ref{eq:Ofek}), (\ref{eq:klein}), and (\ref{eq:massloss_radio}) denoted in the index.}
\end{table*}

We aim to study the mass-loss rates of AT~2000ch on different timescales after outburst by investigating three individual measurements of the H$\alpha$ luminosity $L_{\textrm{H}\alpha}$ and different radio continuum measurements to give reasonable constraints. A summary of the necessary assumptions and our findings can be found in Table \ref{tab:massloss}. \citet{Wagner2004} report an average luminosity of $L_{\textrm{H}\alpha}\approx2\times10^{-14}$\,erg/s/cm$^2$ for 2000-OT based on the spectra taken between 6 May and 26 June, 2000, \cite{Pastorello2010} measured $L_{\textrm{H}\alpha}\approx4\times10^{-14}$\,erg/s/cm$^2$ as a mean value corresponding to the spectra taken in October, 2008, following 2008-OT, and by measuring the H$\alpha$ brightness corrected for continuum emission taken within the \changes survey on 12 January, 2016,  we find $L_{\textrm{H}\alpha}\approx2.7\times10^{-15}$\,erg/s/cm$^2$. The latest optical observations were performed at the end of 2014; however, based on the periodicity, we assume that AT~2000ch is still within an active phase (magnitude deviations larger than 2\,mag) at $\approx 50$\,days after outburst.
Following \citet{Ofek2013a}, for example, we estimate the corresponding mass-loss rate 
\begin{equation}\label{eq:Ofek}
    \dot{M}= \sqrt{2\times 10^{-39} L_{\textrm{H}\alpha}\, v_{\textrm{w,500}}^2\,\beta^{-1} \,r_{15}}\times 10^{-2} \,\textrm{M}_\odot/\textrm{yr},
\end{equation}
where $L_{\textrm{H}\alpha}$ is given in units of erg/s, the wind speed $v_\textrm{w,500}$ is 500\,km/s, the radius $r_{15}$ is $10^{15}$\,cm corresponding to the radius of the freshly ejected material, and $\beta=(r_1-r)/r$ with $r_1$ is the radius of the star itself.
We estimate the radius of the recently ejected material to be 
\begin{equation}
    r = 3\times10^{15}\, \frac{v_{\textrm{ej}}}{8000\,\textrm{km/s}}\,\left(\frac{t-t_0}{30\,\textrm{days}}\right)^{4/5}\,\textrm{cm}
,\end{equation}
by assuming the measured wind velocity to be the mean velocity of the ejecta $v_{\textrm{ej}}$, which corresponds to the $L_{\textrm{H}\alpha}$ measurements for each individual epoch determined by \citet{Wagner2004} and \citet{Pastorello2010}. For 2016, we approximate the wind velocity by assuming the mean of the pre-stated values. We assume $t_0=0$ to be the time of the latest outburst and $t$ the elapsed time after outburst. Taking the multiple outbursts into account, the surrounding material is expected to become denser with time. Nevertheless, based on estimations of the total radius of the line-emitting region of about $0.2$\,pc \citep{Wagner2004} and $0.25$\,pc \citep{Pastorello2010} from $L_{\textrm{H}\alpha}$ for 2000-OT and 2008-OT, respectively, we expect the freshly ejected material to expand sufficiently quickly into the circumstellar medium before the next outburst takes place and therefore neglect its effect on the determination of $r$.
As we only have a photometric measure for $L_{\textrm{H}\alpha}$ in 2016, we assume the latest velocity measurement of \cite{Pastorello2010} to be a proxy.
Generally, the radius of the freshly ejected material is found to be very small in comparison to the total size of the ejecta, supporting our general assumptions about $r$ (see Table \ref{tab:massloss} for the values).
Reasonable measurements for the stellar size of Eta Car and its companion \citep{Verner,Hillier} let us assume an approximate radius $r_1$ of about $50\times \textrm{R}_\odot$ for AT~2000ch. Even a deviation from the assumed stellar radii by a factor of 10-20 (higher) would result in the same order of mass-loss rate ($10^{-5}$\,M$_\odot$/yr).

Alternatively, we can measure the mass-loss rate based on numerically derived models \citep{Klein}:
\begin{equation} \label{eq:klein}
    \log \frac{L_{\textrm{H}\alpha}}{\textrm{L}_\odot}=2\log\dot{M} - \log\frac{M}{60} + 11.691,
    \quad T_{\textrm{eff}}=30000\,\textrm{K},
\end{equation}
where $M$ denotes the stellar mass, which is assumed to be at least $40\,$M$_\odot$ by \cite{Wagner2004}. We note that these models are only derived for effective temperatures of the central stellar source of between 30000\,K (Eq. \ref{eq:klein}) and 50000\,K \citep[][Eq. 24]{Klein}, where lower temperatures correspond to lower mass-loss rates at fixed stellar mass. \citet{Pastorello2010} estimated an upper limit of the temperature of 10000\,K for AT~2000ch or lower based on SED fits. We therefore expect the derived mass-loss rates to be an upper limit (second last column in Table \ref{tab:massloss}). 
Surprisingly, we find both approaches to agree with a mass-loss rate of several $10^{-5}$\,M$_\odot$/yr, even though several different parameters are assumed, enlarging the uncertainty of the individual measurements. The value derived for 2016 is found to be lower, which can only be partly explained by a larger uncertainty on $L_{\textrm{H}\alpha}$ (photometric measure), indicating a decreased mass-loss rate in 2016.

Considering the mean mass-loss rate of 2000 and 2008 ($\dot{M}\approx4\times10^{-5}$), we evaluate whether the circumstellar medium is expected to be optically thick to radio continuum emission shortly after outburst by investigating the optical depth with respect to the free-free emission \citep{Lang}:
\begin{equation}
 \tau_\textrm{ff}=2.6\times10^{10}\,T_{\textrm{e},4}^{-1.35}\,\nu_{10}^{-2.1}\,v_{\textrm{w},10}^{-2}\,\dot{M}_{0.1}\,r_{15}^{-3}.
\end{equation}
Here, {$T_{\textrm{e},4}$ denotes the temperature of the star in units of 10000\,K, $\nu_{10}$ the measured frequency in 10\,GHz, $v_{\textrm{w},10}$ the wind velocity in 10\,km/s, $\dot{M}_{0.1}$} the mass-loss rate in 0.1\,M$_\odot$/yr, and $r_{15}$ again the radius of the freshly ejected material after outburst in $10^{15}$\,cm. We assume a wind velocity of 2000\,km/s, which corresponds to the mean of the spectroscopically derived velocities from 2000 and 2008 during the active phase after outburst. We also take the mean radius and a temperature of 10000\,K. We find $\tau_\textrm{ff}$ to be {$\gg 1$} and $\gtrsim 1$ for 1.5\,GHz and 6\,GHz within the first 25\,days after outburst, respectively. 
With the increased ejecta radius and reduced mass-loss rate in 2016 we find $\tau_\textrm{ff}$ to reach values of {$\ll 1$} and $\approx 0.3$ for 1.5\,GHz and 6\,GHz, respectively.
We therefore assume the ejected material to be at least optically thick within the first 25\,days after outburst, reducing the brightness of AT~2000ch in radio continuum significantly. However, 50\,days after outburst, $\tau_\textrm{ff}$ has already decreased, becoming a marginal factor (or less significant in the case of the 1.5\,GHz data). Optical depth becomes even less important with temporal distance to the outburst; however, here, the formation of a stellar envelope is not taken into account, which would affect the optical depth even on longer timescales. 

We can further estimate the mass-loss rate of AT~2000ch based on the detected and nondetected (upper limit for $\dot{M}$) radio continuum emission at a certain frequency \citep[adopted from][]{Krauss}:
\begin{align} \label{eq:massloss_radio}
    \dot{M}=\;& 6.46488\times 10^{-10}\times v_\textrm{w}\times \left(\frac{D}{\textrm{Mpc}}\right)^{-8/19} \left(\frac{S_\nu}{mJy}\right)^{-4/19}\nonumber \\
    &\times \left(\frac{\nu}{5\,\textrm{GHz}}\right)^2\left(\frac{t}{10\,\textrm{days}}\right)^2
    \,\textrm{M}_\odot/\textrm{yr},
\end{align}
where we assume equipartition between the ratio of electron to magnetic energy densities, a converted fraction of kinetic to magnetic energy density of 0.1, and a filling factor of 0.5 \citep[simplifying Eq. 5 of][to Eq. \ref{eq:massloss_radio}]{Krauss}. The wind velocity is denoted by $v_\textrm{w}$ given in km/s, the distance $D$ in Mpc, the measured intensity $S_\nu$ in mJy, the frequency $\nu$ is 5\,GHz, and the elapsed time from the latest outburst by $t$ is 10\,days.
We note that a smaller brightness in radio continuum results in a larger mass-loss rate because the influence of the optically thick ejecta with respect to the free-free emission is already considered to reduce the observed emission at a certain mass-loss rate and frequency. However, we  evaluated this effect to be only dominant within the first $~25$\,days at 6\,GHz and $~50$\,days at 1.5\,GHz, ignoring the formation of an envelope around the star, and therefore assume mass-loss rates later during an epoch to provide us with upper limits for the mass-loss rate (see last column in Table \ref{tab:massloss}).

Taking both the $H\alpha$ luminosities and the continuum brightnesses into account, we are able to restrict the mass-loss rate of AT~2000ch to lie between several $10^{-6}$\,M$_\odot$/yr and a few $10^{-5}$\,M$_\odot$/yr. In all models, a constant wind velocity is assumed, which is probably not the case for a stellar object showing a deviation by about several magnitudes within a few days. However, even wind velocities of twice the magnitude assumed here  (which would be quite unlikely as already discussed in Sect. \ref{sec:hist}) would still result in mass-loss rates of a few to several $10^{-5}$\,M$_\odot$/yr.


\section{Summary and conclusions} \label{sec:conc}

In this paper, we characterize the long-term variability of the optical transient and supernova imposter AT~2000ch by extending its optical light curve to a total of two decades, and by investigating its radio continuum emission reaching back to the year 1984. We report the discovery of two outbursts (2013-OT, 2019-OT1), adding to the four known eruptive episodes in the 2000s. These most recent events are consistent with the period of {$\sim 201\pm\,$days} previously observed for 2008-OT, 2009-OT1, and 2009-OT2. In this sense, the variability of AT~2000ch is not at all ``erratic'' , as candidates for SN~2009ip-analogs are often described. Its behavior is, on the contrary, surprisingly consistent over a time-span of two decades. Based on this apparently regular period, we report {nine further significant re-brightening events}, where the sampling of the light curve was insufficient for a clear detection or nondetection of a {re-brightening}. {It is probable that the transient experienced a calmer phase between 2014 and 2018 and over the last two years, while the latest light curve shows only minor changes in brightness and appears less eruptive.}

The individual outbursts themselves do not produce any consistent light-curve shapes, but share common characteristics. Brightness variations of $1-2\,$mag within a few days are commonly observed, as is the presence of a deep minimum following most outbursts, though at different times relative to maximum light. Similarly, the color of AT~2000ch does not follow a consistent pattern. However, we do detect a trend for $r-i$ to be correlated to the r-band light curve, while $g-r$ tends to show an anti-correlation. This leads us to believe that the H$\alpha$ excess in the $r$-band relative to the continuum level is decreased during outburst compared to the quiescent phase.

The latest {four re-brightening events, 2020-RB1, 2020-RB2, 2021-RB1, and 2020-RB2,} are clear exceptions in that their light curves are much smoother than those of all their predecessors, and in that we detect brief episodes of $5-10\,$days with excess emission in the $g$-band, which we interpret as the serendipitous crossings of cavities in the circumstellar envelope over the line of sight. Together, these two observations may be indirect evidence that AT~2000ch is currently transitioning into a phase of relative calm compared to its activity in the past two decades.

We confirm that AT~2000ch is a variable source in the radio regime as well. Its measured radio continuum fluxes during the years 2011 and 2012 clearly exceed the established upper limits in 1994 and 2014. All available radio observations fall into gaps or extremely poorly sampled regions of the optical light curve, meaning that we cannot establish a connection between the activity at optical and radio wavelengths. The measured spectral indices vary from $\alpha > -0.9$  ---which is consistent with anything between purely synchrotron and purely thermal origin--- to $\alpha < -1.76$ and $\alpha < -2.79$, clearly indicating strong energy losses that we can trace back to optically thick ejecta material. The latter scenario is consistent with our mass-loss estimates of several $10^{-6}\,\textrm{M}_{\odot}/\textrm{yr}$ to several $10^{-5}\,\textrm{M}_{\odot}/\textrm{yr}$, which would result in an optically thick circumstellar layer at $1.5\,$GHz and $6\,$GHz for at least $\sim 25$\,days after outburst. 

Based on our radio continuum emission findings of AT~2000ch and the reported upper limit of SN~2009ip, we expect that values of around several tens of $\mu$Jy at 6\,GHz are likely for the class of SN~2009ip analogs at comparable distances (a few Mpc), or generally for SN imposters undergoing tremendous optical brightness variations, producing ejecta that are optically dense to free-free emission. The possible necessity of such a high S/N comparable to those reached within the \changes survey should be taken into account for future studies of comparable sources.


\begin{acknowledgements}
      We thank Andrea Pastorello for the in-depth and fruitful refereeing process. 
      This work has made use of data products from
      the National Radio Astronomy Observatory, a facility of the National Science Foundation operated under cooperative agreement by Associated Universities, Inc. This work made use of images from the Hubble Space Telescope; PropID: 14668; PI: Filippenko.
      This research has made use of the SVO Filter Profile Service (http://svo2.cab.inta-csic.es/theory/fps/) supported from the Spanish MINECO through grant AYA2017-84089. DJB and RJD thank the \textsc{chang-es}-team, in particular J. Irwin and R. Walterbos, for very helpful discussions in an early phase of this project.
\end{acknowledgements}

\bibliographystyle{aa}
\bibliography{literatur}

\appendix
\section{Compiled optical data} \label{optical}

The photometric data used throughout this analysis to study the optical light curve and the color evolution are presented in Table \ref{tab:optical}. Here, we only show the values for the PTF/ZTF measurements and the ones used by amateur astronomers. Detailed information for the data points used for 2000-OT, and 2008-OT to 2009-OT2 can be taken from \cite{Wagner2004} and \cite{Pastorello2010}, respectively.

\renewcommand{\arraystretch}{1.2}
\begin{table*}[ht]
\centering
\caption{\label{tab:optical} Example of compiled data points for AT~2000ch including measurements from PTF, ZTF, and amateur astronomers.}
\begin{tabular}{lcccccccccc}
MJD & Date UTC & Observatory & $r$ & $\sigma_r$ & $g$ & $\sigma_g$ & $i$ & $\sigma_i$ & UCNC & $\sigma_\textrm{UCNC}$ \\
\hline\hline
54905.13308 & 2009-03-15 03:11:38.112 & PTF & 19.82 & 0.17 &  &  &  &  &  &  \\
54905.36183 & 2009-03-15 08:41:02.112 & PTF & 19.95 & 0.20&  &  &  &  &  &  \\
54907.15176 & 2009-03-17 03:38:32.064 & PTF & 19.85 & 0.20&  &  &  &  &  &  \\
... & & & & & & & & & & \\
54911.20442 & 2009-03-21 04:54:21.888 & PTF &  &  & 21.13 & 0.26 &  &  &  &  \\
54911.30429 & 2009-03-21 07:18:10.656 & PTF &  &  & 21.06 & 0.23 &  &  &  &  \\
54916.27868 & 2009-03-26 06:41:17.952 & PTF &  &  & 21.07 & 0.26 &  &  &  &  \\
... & & & & & & & & & &\\
\hline
\end{tabular}
\tablefoot{Columns from left to right: observational date in $MJD$, Date in UTC, Observatory, $r$-band magnitude and uncertainty, $g$-band magnitude and uncertainty, $i$-band magnitude and uncertainty, and unfiltered/clear filter/not corrected magnitudes and uncertainty. The full table can be accessed in electronic form at the CDS via anonymous ftp to cdsarc.u-strasbg.fr}
\end{table*}

\end{document}